\documentclass[prd,aps, preprint,11pt,nofootinbib]{revtex4-1}
\usepackage{geometry}                
\usepackage{graphicx}
\usepackage{amssymb}
\usepackage{mathtools}
\usepackage{epstopdf}
\usepackage{verbatim}
\usepackage{color}
\usepackage{slashed}
\usepackage{bbold}
\usepackage{fullpage}
\usepackage{float}
\usepackage[title]{appendix}
\usepackage[caption=false]{subfig}
\usepackage{footnote}
\usepackage[colorlinks, citecolor=red, linkcolor=blue]{hyperref}

\newcommand{\xg}{x_{\tilde{g}}}
\newcommand{\xw}{x_{\tilde{W}}}
\newcommand{\xb}{x_{\tilde{B}}}
\newcommand{\xmu}{x_{\mu}}

\newcommand{\mq}{\tilde{m}_{q}}
\newcommand{\ml}{\tilde{m}_{\ell}}

\newcommand{\ALP}{A^L_{\gamma -p.}}
\newcommand{\ARP}{A^R_{\gamma -p.}}
\newcommand{\ALD}{A^L_{dip.}}
\newcommand{\ARD}{A^R_{dip.}}

\begin{document}
\setlength{\baselineskip}{0.22in}
\preprint{MCTP-16-09}
\title{ \hspace{0.5in}\\  \hspace {0.5in} \\Impact of Future Lepton Flavor Violation Measurements \\in the Minimal Supersymmetric Standard Model}
\author{Sebastian~A.~R. Ellis and Aaron Pierce} \affiliation{Michigan Center for Theoretical Physics (MCTP), Department of Physics, University of Michigan \\Ann Arbor, MI 48109 USA\\}
\date{\today \hspace{.1cm} \\ }

\vspace{1cm}

\begin{abstract}
Working within the context of the minimal supersymmetric standard model, we compare current bounds from quark flavor changing processes with current and upcoming bounds on lepton flavor violation.  We assume supersymmetry breaking approximately respects CP invariance. Under the further assumption that flavor violating insertions in the quark and lepton scalar masses are comparable, we explore when lepton flavor violation provides the strongest probe of new physics.  We quote results both for spectra with all superpartners near the TeV scale and where scalars are multi-TeV. Constraints from quark flavor changing neutral currents are in many cases already stronger than those expected from future lepton flavor violation bounds, but large regions of parameter space remain where the latter could provide a discovery mode for supersymmetry.
\end{abstract}
\maketitle

\section{Introduction}
\label{Introduction.SEC}

Lepton flavor violation (LFV) and quark flavor changing neutral currents (FCNCs) are powerful probes of new physics, reaching scales well beyond those  accessible at present colliders. A significant effort is underway to improve sensitivity to rare LFV processes such as $\mu \rightarrow e \gamma$ and $\mu$ to e conversion (see Table \ref{Limits.TAB}).  However, for example, the neutral kaon mass difference places strong bounds on flavor violation in the quark sector, and in some models LFV and quark FCNCs are related to one another.   It is interesting to explore under what conditions new LFV experiments will be the most sensitive probe of new physics, superseding limits from the quark sector. We discuss this question in the context of the Minimal Supersymmetric Standard Model (MSSM). 

Many studies of flavor violation within the  MSSM exist, see e.g. \cite{Ellis:1981ts,  Gabbiani:1988rb, Hisano:1995cp, Gabbiani:1996hi,  Arganda:2005ji, Raidal:2008jk, Altmannshofer:2009ne, Isidori:2010kg, Altmannshofer:2013lfa, Arana-Catania:2013ggc,Arana-Catania:2014ooa} for overviews. Indeed, most of the calculations of the rare processes we explore here have appeared elsewhere in the literature. Our focus will be a comparison between LFV and quark FCNCs, trying to get a feel for the relative power of these constraints.

Supersymmetry (SUSY) breaking scalar masses can receive contributions from operators of the form
\begin{align}
K \sim \frac{\kappa_{ij}}{M^2} X_a^\dagger X_a \Phi^\dagger_{i}\Phi_{j}
\end{align}
in the K\"ahler potential.  Here $\Phi$ are MSSM superfields with generation indices $i,j,$ and $X_a$ are fields associated with the breaking of SUSY 
with non-vanishing $F_X$, and $M$ is associated with the mediation scale of SUSY breaking. Such operators can induce off-diagonal terms in the scalar mass matrices, given by $m_{ij}^2 = \kappa_{ij}\langle F_{X_a} \rangle^2 / M^2$.  These terms are a source of flavor violation beyond the Standard Model. The size and form of these off-diagonal contributions depend on the particulars of the UV theory that induces this non-renomalizable operator. It is possible  the SUSY breaking respects a Grand Unified Theory (GUT)  structure, in which case the quark and lepton flavor violation can be related. However, even in this case quark and lepton superfields residing in different representations may feel SUSY breaking differently. For example, in an $SU(5)$ GUT, since the left-handed (LH) lepton superfields reside in the $\bar{\mathbf{5}}$ while the LH quark superfields reside in the $\mathbf{10}$, this leads to the possibility of a mismatch between contributions to LFV and quark FCNCs. (See for example \cite{Ciuchini:2007ha} and discussion in \cite{Altmannshofer:2013lfa}.)

It is also possible that off-diagonal mass terms for squarks and sleptons are {\it a priori} unrelated.  
 Indeed, even if initial flavor violation is related by a symmetry, a mismatch between squark and slepton off-diagonal mass terms may arise once neutrino masses are incorporated into the theory. The inclusion of neutrino Yukawa couplings may lead to sizeable entries in the left-handed slepton mass matrix due to Renormalisation Group Equation (RGE) running from the GUT scale down to the right-handed neutrino scale \cite{Hisano:1997tc, Casas:2001sr, Ellis:2001xt, Ellis:2002fe,Antusch:2006vw, Arganda:2007jw}. Such models naturally lead to non-zero LFV while not contributing to quark FCNCs. This approach has been considered in various contexts, including $SO(10)$ \cite{Casas:2001sr, Mohapatra:2006gs} and $SU(5)$ GUT models \cite{Hisano:1997tc}. The size of these effects are model dependent, but can be large. But even in the case where the quark and lepton flavor violation are decoupled, it is of interest to understand just how different the allowed flavor violation is, consistent with current and upcoming experiments.  

\begin{table}[t]
\centering
\begin{tabular}{| c | c | c |}
\hline 
\textbf{Observable} & \textbf{Exp. Measurement} & \textbf{SM prediction}\\
\hline \hline
$\Delta m_K$ & $(3.484\pm0.006) \times 10^{-12}$ MeV \cite{Agashe:2014kda} & $(3.19\pm0.41$(stat.)$\pm0.96$(sys.)$)\times10^{-12}$ MeV \cite{Bai:2014cva} \\
$\Delta m_{B_d}$& $(3.337\pm0.033) \times 10^{-10}$ MeV \cite{Agashe:2014kda} & $(3.48\pm0.52)\times 10^{-10}$ MeV \cite{Artuso:2015swg}\\
$\sin{2\beta_{d}}$ & $0.682 \pm 0.019$ \cite{Amhis:2014hma} & $0.748^{+0.030}_{-0.032}$ \cite{Charles:2004jd} \\
$\Delta m_{B_s}$ & $(1.1691\pm0.0014) \times 10^{-8}$ MeV \cite{Agashe:2014kda} & $(1.2\pm0.18)\times 10^{-8}$ MeV \cite{Artuso:2015swg} \\
$\sin{2\beta_s}$ & $-0.015 \pm 0.035$ \cite{Amhis:2014hma} & $-0.03761^{+0.00073}_{-0.00082}$ \cite{Charles:2004jd} \\
\hline
\hline 
\textbf{Observable} & \textbf{Current Limit (90\% C.L.)} & \textbf{Future sensitivity (90\% C.L.)}\\
\hline \hline
BR($\mu \to e \gamma$) & $4.2 \times 10^{-13}$  \cite{TheMEG:2016wtm} & $6 \times 10^{-14}$  \cite{Baldini:2013ke} \\
BR($\tau \to e \gamma$) & $3.3 \times 10^{-8}$  \cite{Aubert:2009ag} & $10^{-9}$ \cite{Hayasaka:2013dsa} \\
BR($\tau \to \mu \gamma$) & $4.4 \times 10^{-8}$  \cite{Aubert:2009ag}& $10^{-9}$  \cite{Hayasaka:2013dsa} \\
BR($\mu \to e$)$_{Au}$ & $7.0 \times 10^{-13}$   \cite{Bertl:2006up} & \\
BR($\mu \to e$)$_{Al}$ &  & $10^{-16}$  \cite{Abrams:2012er} \\
BR($\mu \to 3e$) &  $1.0 \times 10^{-12}$ \cite{Bellgardt:1987du}& $10^{-16}$  \cite{Blondel:2013ia, Berger:2014vba} \\

\hline
\end{tabular}
\caption{The experimental measurements and SM predictions for quark observables and the current and future sensitivities of lepton flavor violating processes. Long distance effects in $\Delta m_K$ are difficult to quantify. The quoted SM  $\Delta m_K$ value is a recent Lattice QCD calculation \cite{Bai:2014cva} which uses unphysical values for the pion, kaon and charm quark masses, and as such should not be taken as precise. So, in our numerical work we allow the SUSY contribution to fully saturate the experimental value.}
\label{Limits.TAB}
\end{table}

New phases in the SUSY breaking parameters would contribute to CP-violating processes, such as $\epsilon_K$. If the phases are ${\mathcal O}$(1), extremely strong bounds exist, forcing scalars to be in the PeV regime \cite{Altmannshofer:2013lfa}. It is possible that searches for electric dipole moments (EDMs) could eventually provide constraints competitive with those from $\epsilon_K$, a possibility that has been studied recently in,  e.g. \cite{Demir:2003js, Pospelov:2005pr, Giudice:2005rz,  Ellis:2008zy, McKeen:2013dma, Altmannshofer:2013lfa, Ellis:2014tea}.  However, it is possible a mechanism renders the phases in SUSY breaking parameters small.   Moreover, LFV measurements such as $\mu \rightarrow e \gamma$ are CP-conserving, so a true ``apples to apples" comparison is  with  CP-conserving observables in the quark sector.  In this analysis we will restrict ourselves to the assumption that all phases are zero (or at least negligibly small).   In the kaon sector, for example, the limits from $\Delta M_{K}$ supersede those from $\epsilon_K$ for phases $\lesssim 10^{-2}$.

In this work we consider two scenarios and discuss the interplay between quark FCNCs and LFV in each. In the first, we use the observed Higgs boson mass of 125 GeV as motivation to consider scalar masses that may be (much) heavier than a few TeV, and could fall in the 10's of TeV to even a PeV range \cite{Kane:2011kj, Giudice:2011cg, Draper:2013oza, Arvanitaki:2012ps}.  Additionally, having heavy scalars allows for off-diagonal masses to be relatively large, potentially up to $\mathcal{O}(1)$ of the diagonal masses, thus lessening the need for a mechanism to suppress flavor violation.  At the same time, the observed abundance of dark matter (DM)  indicates either gaugino masses $M_i$ or the Higgsino mass parameter $\mu$ should be near the TeV scale (see, for example, \cite{Wells:2003tf, Pierce:2004mk, Giudice:2004tc}). So, in the first scenario, we imagine a modest hierarchy between the fermionic and scalar superpartners.  In the second scenario, we consider the possibility that all superpartners lie close to the TeV scale. 

In Sections \ref{QFV.SEC} and \ref{LFV.SEC}, we review the structure of the effective Hamiltonians which contribute to quark FCNCs and LFV in the MSSM. In the quark sector, our primary focus is on meson mixing. For LFV we discuss $\ell_{j} \rightarrow \ell_{i} \gamma$ decays and $\mu$ to $e$ conversion.  (We comment briefly on the $b \rightarrow s \gamma$ transition in Sec.~\ref{bsg.SEC}).  We discuss the parametric dependencies of the various operators entering the effective Hamiltonian for each process we consider, and comment on what parameters are most important in what regimes.  We discuss the dependence of both quark FCNCs and LFV on Left-Left ($LL$), Right-Right ($RR$) and $LR$ mixing.  A goal of these sections is to highlight which insertions are most constrained and how this may differ between the quark and lepton sector, an issue which we quantify further in Section \ref{deltaOlogy.SEC}. In Section \ref{xOlogy.SEC} we analyse in more depth how the various gaugino masses $M_i$ and the $\mu$-term impact the strength of quark FCNC constraints relative to LFV bounds. The relative power of LFV and quark FCNCs is summarized in Figs. \ref{GauginoDominationRatio.FIG} -- \ref{TauEDeltaOlogyLargeXBdLRnonZero.FIG}, which represent the main results of this paper. Finally, in Section \ref{Conc.SEC} we summarise the results of our analysis, and comment on the implications. 

\section{Anatomy of quark FCNC processes}
\label{QFV.SEC}
In this section we review contributions to quark flavor violating observables.  In the kaon sector, since we concentrate on CP-conserving new physics, our focus is on $\Delta M_{K}$.  In the $B$ sector, even if new physics contributions are CP-conserving, measurements of CP-violating quantities such as $\sin 2 \beta_d$ are relevant.  We review our treatment  of B-mixing in Sec.~\ref{NumBmeson.SEC}. We briefly comment on $\Delta F = 1$ constraints in \ref{bsg.SEC}.

\subsection{$\Delta F = 2$ transitions}

\begin{figure}
\includegraphics[scale=0.6]{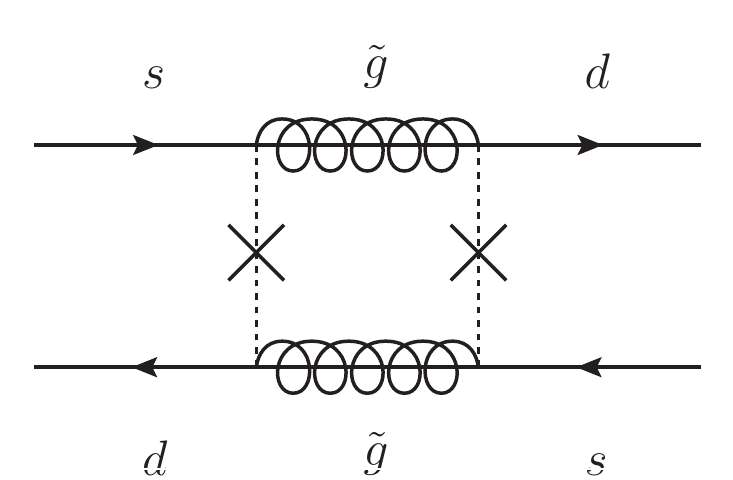}
\caption{Typical kaon mixing diagram induced by SUSY. The crosses represent flavor-violating mass insertions.}
\label{Kmix.FIG}
\end{figure}

The dominant SUSY contribution to meson oscillations is typically gluino-squark box diagrams.\footnote{For large values of the ratio of the vacuum expectation values of the two Higgs doublets $\tan {\beta}$, an additional heavy Higgs-mediated contribution to meson oscillations (see, e.g., \cite{Altmannshofer:2009ne}) may be relevant ($\tan \beta \sim 50$ for $m_A \sim \mq$).} In these processes, one may use the mass insertion approximation for sufficiently small off-diagonal elements in the squark mass matrix, with these insertions appearing on the internal squark lines, shown as crosses in Fig. \ref{Kmix.FIG} for kaon oscillation.  We take the squark mass-squared matrix to be given by 
\begin{align}
\mathcal{M}_{\tilde{q}}^2 = \begin{pmatrix} \mq^2(1 + \delta^{ij}_{LL}) &  \mq^2(\delta^{ij}_{LR}) \\  \mq^2(\delta^{ij}_{RL}) & \mq^2(1 + \delta^{ij}_{RR}) \end{pmatrix},
\end{align}
where the indices $i,j = 1,2,3$ run over generations. An analogous convention is used for sleptons.

The interaction can be described by the corresponding effective Hamiltonian
\begin{align}
\mathcal{H}_{eff} = \sum^5_{i=1} C_i Q_i +\sum^3_{i=1} \tilde{C}_i \tilde{Q}_i + h.c.
\end{align}
where the $C_i$ are the Wilson coefficients for the dimension-6 operators $Q_i$
\begin{align}
\nonumber Q_1 = (\bar{d}_L^\alpha \gamma_\mu &s_L^\alpha)(\bar{d}_L^\beta \gamma_\mu s_L^\beta),~~
Q_2 = (\bar{d}_R^\alpha s_L^\alpha)(\bar{d}_R^\beta s_L^\beta),~~Q_3 = (\bar{d}_R^\alpha s_L^\beta)(\bar{d}_R^\beta  s_L^\alpha),\\ 
&Q_4 = (\bar{d}_R^\alpha s_L^\alpha)(\bar{d}_L^\beta s_R^\beta),~~~~~~Q_5 = (\bar{d}_R^\alpha s_L^\beta)(\bar{d}_L^\beta  s_R^\alpha)
\end{align}
and $\tilde{Q}_i$ given by interchanging $L\leftrightarrow R$ for $i=1,2,3$. For the numerical values of the hadronic matrix elements $\langle \bar{K}_0 | Q_i | K_0 \rangle$ we use the values for the bag factors $B_i(2$ GeV$)$ from \cite{Bertone:2012cu}, the lattice result for $f_K$ from \cite{Bazavov:2010hj}, and the reported kaon mass $m_K$ from \cite{Agashe:2014kda}. Meanwhile for the B-meson hadronic matrix elements, we use the values for the bag factors $B_i(m_b)$ and the lattice results for $f_B,~f_{B_s}$ from \cite{Carrasco:2013zta}, and the reported $B$-meson masses from \cite{Agashe:2014kda}. Expressions for the Wilson coefficients 
including the Leading Order QCD corrections \cite{Bagger:1997gg} are reproduced in Appendix \ref{WilsonCoeffs.APP}. 

\begin{figure}
\centering
\includegraphics[width=0.9\textwidth]{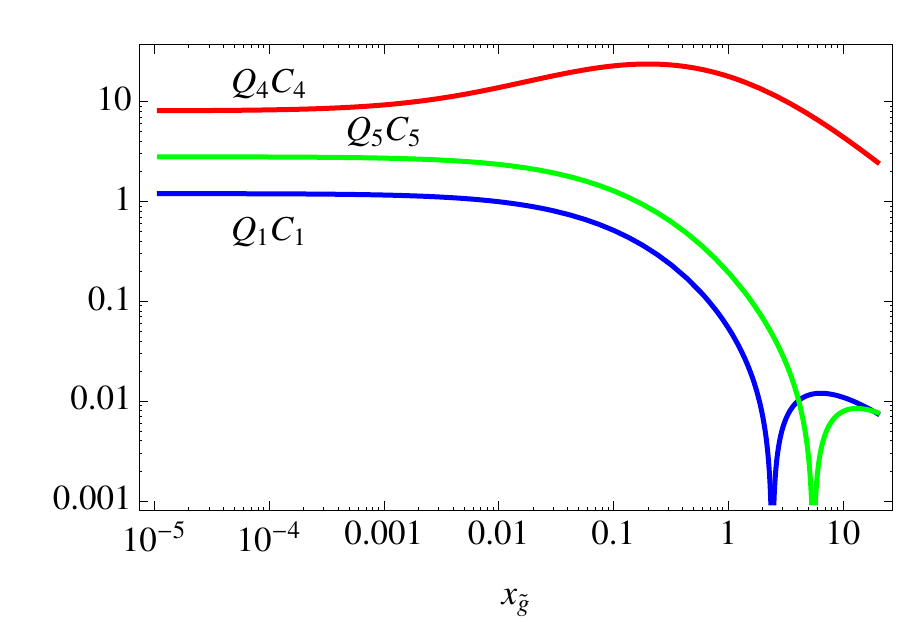}
\caption{The products $C_i(\mu)Q_i$ for kaon oscillations, for insertions $\delta_{LL}=\delta_{RR}= 0.3$, $\delta_{LR}=\delta_{RL}=0$, and $\mq = 20$ TeV (we set $\mu=m_c$). Shown here are $C_1Q_1$ (blue), $C_4Q_4$ (red) and $C_5Q_5$ (green), demonstrating the domination of $C_4Q_4$ for all values of $\xg$. Not shown are $C_2Q_2$ and $C_3Q_3$, which depend only on LR insertions, set to zero here. In any case, these are expected to be subdominant, see text. The numerical values for the $Q_i$ are obtained as described in the text. The relative importance of the $C_i Q_i$ is the same for $B$-meson oscillations.}
\label{Operators.FIG}
\end{figure}

In Fig. \ref{Operators.FIG}, we display the contribution to meson mixing assuming that $\delta_{LL}= \delta_{RR}$.  $\delta_{LR}$ is set to zero -- in any case its contribution is expected to be subdominant, see Eq.~(\ref{largeA.EQN}) below. In both the $ \frac{m_{\tilde{g}}^2}{\mq^2} \equiv x_{\tilde{g}} \ll 1$ and $x_{\tilde{g}} \simeq 1$ regions, for equal sized insertions, the contribution to $\Delta F=2$ processes is dominated by the operator $Q_4$ with coefficient $C_4$. Notably, this dominant operator depends  on the product $\delta_{LL}\times\delta_{RR}$ (rather than $\delta_{LL}^2$ or $\delta_{RR}^2$), so can be varied relative to the others.  As we will see, the relative size of $\delta_{LL}$ and $\delta_{RR}$ will impact the relative strength of  of the quark flavor violation and LFV probes.

$LR$ insertions are not expected to be relevant for  $\Delta F=2$ transitions for large ($\gtrsim$ TeV) squark masses. The $LR$ insertions arise due to off-diagonal terms in the scalar trilinear couplings $A_{ij}$ and have the form 
\begin{align}
\delta^{ij}_{LR} \sim \frac{m_q A^{ij}}{\mq^2}.
\end{align}
The result of the quark mass suppression is that $A$-terms must be very large to affect meson mixing:
\begin{align}
\label{largeA.EQN}
\frac{A^{12}}{\mq} \gtrsim 170 \,  \frac{  \mq}{\text{TeV}} \hspace{.5in} \frac{A^{13}}{\mq} \gtrsim 5 \,  \frac{  \mq}{\text{TeV}}
\hspace{.5in}
\frac{A^{23}}{\mq} \gtrsim 50 \,  \frac{  \mq}{\text{TeV}}. 
\end{align}
Such large $A$-terms would not be expected unless the SUSY-breaking spurion were charged under the flavor symmetry, a possibility which we do not consider further.

While $LR$ insertions are unlikely to be relevant for meson mixing as described above, they are potentially relevant for the $\Delta F=1$ transition of $b \to s \gamma$ (which we discuss later in Section \ref{bsg.SEC}).

\subsection{Treatment of constraints from B-meson observables}
\label{NumBmeson.SEC}

B-meson mixing provides a total of four constraints on the new SUSY contributions. 
For each of the $B_q$ mesons ($q= d, s$) there are the measured mass difference $\Delta m_{B_q}$ as well as  the measurement of the CP violation in the mixing.

In the $B_d$ sector, the observed CP violation in mixing is given by:
\begin{align}
\sin{2\phi_d} = \frac{\sin{2\beta_d}+r_d\sin{\theta_{d}}}{C_{B_d}},
\label{SpsiKS.EQ}
\end{align}
 where $r_d = \frac{|\langle \bar{B}_d|\mathcal{H}_{eff}^{SUSY}| B_d\rangle|}{|\langle \bar{B}_d|\mathcal{H}_{eff}^{SM}| B_d\rangle|}$, $\theta_d$ is a potential new CP-violating phase,
which we take to vanish, and \begin{align}
C_{B_d} = \left(1 + r_d^2 + r_d \cos(2\beta_d - \theta_d) \right)^{1/2}.
\end{align}
Here, $\sin{2\beta_d}$ is the SM prediction, for which we take the latest (Summer 2015) CKMfitter collaboration global fit \cite{Charles:2004jd}\footnote{When constraining the SUSY contribution, we use the global fit as the central value for $S_{\psi K_{s}}$ rather than the directly experimentally measured value. They agree within $2\sigma$.} Similarly in the $B_s$ sector, we use the latest value of $\sin{2\beta_s}$ from the CKMfitter collaboration global fit, and an expression for $\sin{2\phi_s}$ analogous to Eq.~(\ref{SpsiKS.EQ}), with the expectation that the SM prediction is $\sin{2\phi_s} = \sin{2\beta_s}$.  We then calculate the $\chi^2$ values of the combined constraints from the mass difference $\Delta m_{B_q}$ and $\sin{2\beta_q}$ to find the excluded regions in our various plots.

While the experimental precision on both $\sin{2\phi_d}$ and $\sin{2\phi_s}$
is expected to improve \cite{Aushev:2010bq}, improvements in theoretical precision are less easy to forecast. If the expected experimental improvement is matched by theory, this will result in $\mathcal{O}$(1) modifications of the bounds on the allowed $\delta$. In our numerical results, we show the expected improvement assuming the theoretical precision increases by a factor of two. 

\subsection{Treatment of $\Delta F = 1$ transitions}
\label{bsg.SEC}

The $\Delta F=1$ decay of $b \to s \gamma$ is known to impose strong constraints on the $2-3$ sector for TeV-scale superpartners (see for example \cite{Barbieri:1993av,Degrassi:2000qf,Carena:2000uj,Altmannshofer:2009ne}). Particularly when imposing constraints on LR mass insertions, it is necessary to include the results from $b \to s \gamma$ to obtain
the constraints on quark $2-3$ transitions. Constraints on $LL$ and $RR$ insertions can also be derived, and are also relevant. Our procedure for calculating the branching ratio is the following: we take the leading contributions to the operators $C_7,~C_8,~\tilde{C}_7,~\tilde{C}_8$ from heavy Higgs boson and gluino diagrams from \cite{Altmannshofer:2009ne}, and use the expression in \cite{Freitas:2008vh, Altmannshofer:2012ks} to calculate the branching ratio for generic new physics contributions to the above listed operators. We assume that the heavy Higgs bosons are degenerate with the squarks and sleptons. We then impose that the branching ratio be within the 90\% confidence interval given the latest experimental results \cite{Amhis:2014hma}, and the theoretical estimate for the branching ratio at NNLO in the SM \cite{Czakon:2015exa, Misiak:2015xwa}. 
For simplicity, we assume vanishing flavor violation in the up squark sector (which affects potential chargino diagrams, which are usually subdominant in any case).  For heavy Higgs boson masses comparable to squark masses, we find the charged Higgs boson diagram to be smaller than, but not negligibly small compared with the gluino contribution, when $\delta$ is near its experimentally allowed value. We note that the sign of the product $M_{\tilde{g}} A^{23}$ which appears in the gluino diagram is physical.  

In the future, sensitivity of the High Luminosity LHC to flavor changing top quark decays, $t \to h q$ ($q= u,~c$), where $h$ is the Higgs boson, is expected to reach BR($t \to h q$)$\lesssim 2 \times 10^{-4}$ \cite{Agashe:2013hma,ATL-PHYS-PUB-2013-012} with 3 ab$^{-1}$. Recent studies (see for example \cite{Dedes:2014asa} and references therein) indicate that for typical regions of SUSY parameter space, the future sensitivity will be insufficient to probe these rare decays in the MSSM. For this reason we do not compare here the top quark FCNC with the relevant LFV process, $h \to \tau \mu$. This LFV Higgs boson decay has been studied in the context of the MSSM in, for example, \cite{Arganda:2015uca, Aloni:2015wvn}.

\section{Anatomy of LFV processes}
\label{LFV.SEC}
In this section we review supersymmetric contributions to the processes $\ell_i \to \ell_j \gamma$ and $\mu \to e$ conversion in nuclei. We discuss what contributions dominate in what regimes and comment on the dependence on the gaugino masses and $\mu$.

\subsection{$\ell_i \to \ell_j \gamma$}
\label{MuEGam.SEC}

The branching ratio of $\ell_i \to \ell_j \gamma$ is

\begin{align}
\text{BR}(\ell_i \to \ell_j \gamma) = \frac{48\pi^3 \alpha_{em}}{G_F^2}\left( |A_L|^2 + |A_R|^2 \right),
\end{align}
where the amplitudes $A_{L,R}$ are the coefficients of higher-dimensional operators in the effective Hamiltonian
\begin{align}
\mathcal{H}_{eff} = e \frac{m_{\ell_i}}{2}\left( A_L \bar{\ell_j} \sigma^{\mu \nu} P_L \ell_i + A_R \bar{\ell_j} \sigma^{\mu \nu} P_R \ell_i \right) F_{\mu \nu}.
\end{align}

The dominant contribution to $A_L$ arises from Wino loops \cite{Altmannshofer:2013lfa}
\begin{align}
\label{LFVWinoLoop.EQ}
A^{\tilde{W}}_L = \frac{\alpha_2}{4\pi} \frac{1}{\ml^2} \delta_{LL}^{\ell_i \ell_j} \left[ -\frac{1}{8} g_1 (\xw) + g_2(\xw, x_\mu) + \text{sgn}(\mu M_2)\sqrt{\xw x_\mu } t _{\beta} g_3 (\xw, x_\mu)\right],
\end{align}
where the $g_i$ are loop functions given in Appendix \ref{LFmuEgam.APP}, and $\xw, (x_\mu) \equiv \frac{m_{\tilde{W}}^2}{\ml^2},~ \left( \frac{\mu^2}{\ml^2}\right)$. We have abbreviated $\tan{\beta}$ as $t_{\beta}$.   If the sign of $\mu M_2$ is positive (negative), $A^{\tilde{W}}_L$ exhibits destructive (constructive) interference. We will refer to each of these cases in the following analysis.

There are additional contributions to $A_L$ and $A_R$ due to a Bino loop \cite{Paradisi:2005fk, Altmannshofer:2009ne}
\begin{align}
A^{\tilde{B}}_L &\supset \frac{\alpha_1}{4\pi} \frac{1}{\ml^2} \delta_{LL}^{\ell_i \ell_j} \text{sgn}(\mu M_1)\sqrt{\xb x_\mu} t_{\beta} \left[f_{3n} (\xb) + \frac{f_{2n}(\xb, x_\mu)}{x_\mu  - \xb}\right],\\
A^{\tilde{B}}_R & \supset \frac{\alpha_1}{4\pi} \frac{1}{\ml^2} \delta_{RR}^{\ell_i \ell_j} \text{sgn}(\mu M_1) \sqrt{\xb x_\mu} t_{\beta}  \left[ f_{3n} (\xb) - \frac{2 f_{2n}(\xb, x_\mu)}{x_\mu  - \xb}\right],
\label{BinoLFV.EQ}
\end{align}
with the $f_{2,3 n}$ are loop functions given in Appendix \ref{LFmuEgam.APP} and $\xb \equiv \frac{m_{\tilde{B}}^2}{\ml^2}$.

While the above contributions to $A_R$ and $A_L$ apply to all $\ell_i \to \ell_j \gamma$ processes, there is an additional diagram which gives an important contribution for $\mu \to e \gamma$ only, arising due to a Bino loop with two flavor changing insertions combined with a flavor-conserving $LR$ insertion on an internal stau line \cite{Paradisi:2005fk, Altmannshofer:2013lfa}. The flavor-conserving insertion results in an enhancement of $m_\tau / m_\mu$: 
\begin{align}
A^{\tilde{B}}_R \supset \frac{\alpha_1}{4\pi} \left( \frac{m_\tau}{m_\mu}\right)\text{sgn}(\mu M_1) \frac{\sqrt{\xb x_\mu} t_{\beta}}{\ml^2} f_{4n}(\xb) \delta^{\mu \tau}_{LL} \delta^{\tau e}_{RR}\ ,
\label{BinoLoop.EQ}
\end{align}
where $f_{4n}(\xb)$ is a loop function that can be found in Appendix \ref{LFmuEgam.APP}. The analogous expression for $A_L$ is found by taking Eq. (\ref{BinoLoop.EQ}) and interchanging the $LL$ and $RR$ insertions. This diagram is of particular interest if a flavor symmetry suppresses $1-2$ insertions, since Eq. (\ref{BinoLoop.EQ}) only depends on $1-3$ and $2-3$ insertions. 

In Section \ref{QFV.SEC} we saw that meson mixing did not put meaningful constraints on off-diagonal trilinear terms even for TeV scale scalars. In contrast, the LR mixing contributions to LFV may be non-negligible.
Consider the contribution to radiative lepton decay arising from a Bino loop, reproduced below \cite{Paradisi:2005fk, Altmannshofer:2009ne}
\begin{align}
A^{\tilde{B}}_{L} & \supset \frac{\alpha_1}{2\pi} \frac{\delta_{RL}^{\ell_i \ell_j}}{\ml} \frac{\sqrt{\xb}}{m_\mu} f_{2n}(\xb)\ ,
\label{LR.EQ}
\end{align}
with $A^{\tilde{B}}_{R}$ given by the $\delta_{LR}$ insertion. For $\xb \sim 1$, we see that this is only suppressed by one power of $\ml$. Since $\delta_{LR}^{ij}$ arises due to terms of the form
\begin{align}
\delta^{ij}_{LR}\simeq \frac{m_f A^{ij}}{\ml^2}\ ,
\end{align}
we can use these expressions to constrain the ratio of $A^{ij}/\ml$ for a given value of $\ml$.
\begin{figure}
\centering
\includegraphics[width=0.7\textwidth]{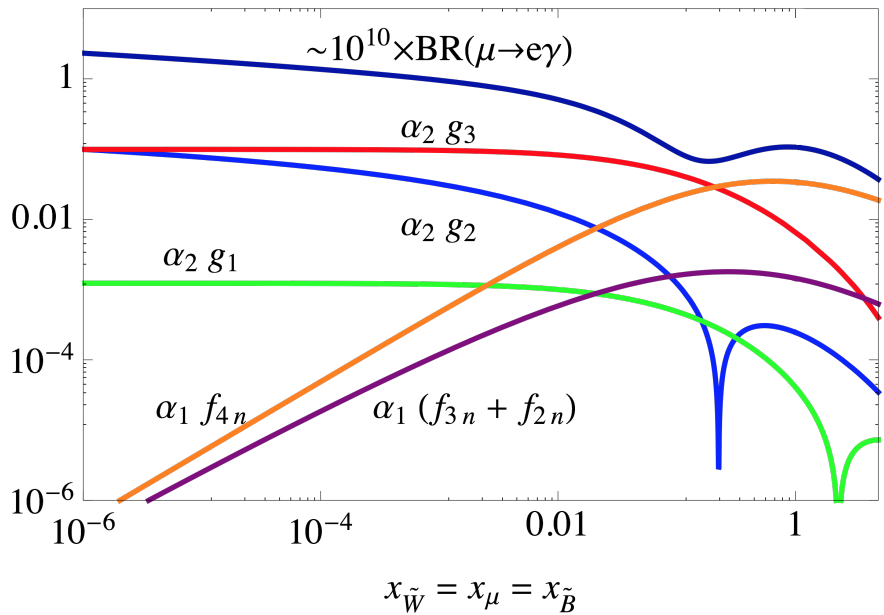}
\caption{The relative importance of various operators to the branching ratio, as well as the total branching ratio scaled up. This shows that for small $x$ (with all being set equal), the loop functions $g_2$ and $g_3$ dominate, while for larger values the Bino loop functions $f_{2n},~f_{3n}$, and $f_{4n}$ become important. We set $\tilde{m}_{{\ell}}= 20$ TeV, and we set $\delta^{\mu e}_{LL}=\delta^{\mu e}_{RR} = \delta^{\mu \tau}_{LL}\delta^{\tau e}_{RR} = 0.3$ so that the effective $\delta^{\mu e}$ is the same for each operator. $\tan \beta =10$.}
\label{OpDepMu.FIG}
\end{figure}

In Fig. \ref{OpDepMu.FIG} we show the relative contributions to $\mu \rightarrow e \gamma$ (arbitrary units) for comparable insertions:  $\delta^{\mu e}_{LL}=\delta^{\mu e}_{RR} = \delta^{\mu \tau}_{LL}\delta^{\tau e}_{RR} = 0.3$. The $g_{i}$ and $f_{i}$ correspond to the loop functions introduced in Eqs.~(\ref{LFVWinoLoop.EQ})--(\ref{BinoLoop.EQ}). $\tan \beta$ is set to 10.  The dominant contributions to $\mu \to e \gamma$ are from the Wino-Higgsino mixing diagrams, denoted by $g_2$ and $g_3$, at small $x_i$. Since both of these only depend on $\delta_{LL}$ (see Eqn.~(\ref{LFVWinoLoop.EQ})), at small $x_i$, $\mu \to e \gamma$ will place constraints on $\delta_{LL}$, but not $\delta_{RR}$. As $x_i$ approaches 1, the Bino contributions porportional  to $f_{in}$ can become important. The dominant operator is that with the $LR$ flavor-conserving insertion, as long as $\delta^{\mu \tau}\delta^{\tau e}$ is not too suppressed relative to the single $\delta^{\mu e}$ insertion. 

In Fig. \ref{OpDepMu.FIG}, we see that the largest branching ratio of $\mu \to e \gamma$ is obtained in the small $x_{\tilde{W}},~x_\mu,~\xb$ regime. While the figure shown sets $\tan \beta=10$, since the dominant contributions are proportional to $g_3$ (small $x$) and $f_{4n}$ (large $x$) both of which are also proportional to $\tan \beta$, the scaling is straightforward.   The statement was also found to apply for maximising the branching ratios of $\tau \to \mu \gamma$ and $\tau \to e \gamma$.  This is to be contrasted with Fig.~\ref{Operators.FIG} where small $x$  did not enhance the meson mixing.  Thus, we expect LFV to be a relatively powerful probe in the  small $x$ regime. Given the non-trivial $x_i$ dependence, however, we will give a more detailed study of  the  dependence on combinations of $\xb, \xw$ and $x_\mu$ in Section \ref{xOlogy.SEC}.

\subsection{$\mu \to e$ conversion in Nuclei}

We decompose the contributions to $\mu \to e$ conversion.The branching ratio is given by
\begin{align}
\nonumber \text{BR}(\mu \to e)_N &= \bigg\{\bigg|\frac{1}{4}e A_L^* D+2(2g_{L,V}^u+g_{L,V}^d)V^{(p)}+2(g_{L,V}^u+2g_{L,V}^d)V^{(n)}\bigg|^2 \\&+ \bigg|\frac{1}{4}e A_R^* D+2(2g_{R,V}^u+g_{R,V}^d)V^{(p)}+2(g_{R,V}^u+2g_{R,V}^d)V^{(n)}\bigg|^2 \bigg\}\frac{1}{\omega_{capture}},
\label{Conv.EQN}
\end{align}
where $\omega_{capture}$ is the muon capture rate of the nucleus. The $A_{L(R)}$ are the same dipole coefficients that were given in Section \ref{MuEGam.SEC}, and $g^{u,d}_{L(R),V}$ are the penguin- and box-type Wilson coefficients coupling to up or down-type quarks. The terms $D$, $V^{(p)}$ and $V^{(n)}$ are overlap integrals calculated in \cite{Kitano:2002mt} whose values are presented in Appendix \ref{LFconversion.APP} for convenience. 

At $\xw \sim 1$, the branching ratio for $\mu \to e$ conversion is dominated by the dipole contributions $A_{L,R}$. In this limit there is a simple relation between the $\mu \to e \gamma$ branching ratio and that of $\mu \to e$ conversion, namely: 
\begin{align}
\text{BR}(\mu \to e)_N &\simeq \frac{G_F^2 D^2}{192 \pi^2 \omega_{capture}} \text{BR}(\mu \to e \gamma) \sim \begin{cases} \frac{\alpha_{em}}{3}\text{BR}(\mu \to e \gamma), \text{~when N is Aluminium,}\\ \frac{\alpha_{em}}{2}\text{BR}(\mu \to e \gamma), \text{~when N is Gold}.\end{cases}
\end{align}
This will apply to our analysis in the case of TeV scale scalars.
Given the future experimental improvements on measuring both $\mu\to e$ conversion and $\mu \to e \gamma$ (see Tab. \ref{Limits.TAB}),  in the case of dipole domination, conversion can impose limits on LFV insertions
comparable to those from $\mu \to e \gamma$.

The Wilson coefficients $g^{u,d}_{L(R),V}$ can be decomposed into the box-, $\gamma$-penguin and $Z$-penguin diagram contributions as
\begin{align}
g_{L(R)V}^q = g_{L(R)V}^{q,box} + g_{L(R)V}^{q,\gamma} + g_{L(R)V}^{q, Z}.
\end{align}
Wino loops give the dominant contributions to the the $g_{LV}^q$.  
Since the operators corresponding to these coefficients become important relative to the dipole contribution at small $x_i$, we present here the leading contributions in that regime \cite{Altmannshofer:2013lfa}.
\begin{align}
5g_{LV}^{u,box}=g_{LV}^{d,box} = \frac{g_2^4}{(4\pi)^2 \mq^2}\delta^{\mu e}_{LL}\frac{5}{4} f\left(\frac{\ml^2}{\mq^2}\right),
\end{align}

\begin{align}
g_{LV}^{u,\gamma-peng.}=-2g_{LV}^{d,\gamma-peng.} &= \frac{-2 e^2 g_2^2}{3(4\pi)^2 \ml^2}\delta^{\mu e}_{LL} f_{\gamma,L} (\xw) \\ &\simeq \frac{-2 e^2 g_2^2}{(4\pi)^2 \ml^2}\delta^{\mu e}_{LL}\left\{ \frac{1}{4} + \frac{1}{9} \log\left(\xw \right)\right\},
\end{align}
where the second line is in the limit of small $\xb, \xw \ll 1$.
\begin{align}
\nonumber g_{LV}^{u,Z-peng.}&=\frac{-\left(1-\frac{4}{3}\sin^2\theta_W\right)}{\left(1-\frac{8}{3}\sin^2\theta_W\right)}g_{LV}^{d,Z-peng.} \\ \nonumber&= \frac{-g_2^4}{(4\pi)^2 \ml^2}\delta^{\mu e}_{LL}\frac{1}{16}\left(1-\frac{8}{3}\sin^2\theta_W\right)\\&\times \left\{\cos^2\beta f_1\left(\xw,x_\mu \right) + \sin^2\beta f_2\left(\xw,x_\mu \right) + \text{sgn}(\mu M_2)\sqrt{\xw x_\mu}\sin\beta \cos \beta f_3\left(\xw,x_\mu \right) \right\},
\label{LHZpeng.EQ}
\end{align}
where $f(x),~f_{\gamma,L}(x) ,~f_1(x),~f_2(x)$ and $f_3(x)$ are loop functions given in Appendix \ref{LFconversion.APP}.

The contributions proportional to $\delta_{RR}^{\mu e}$ can also be derived, and are presented here in the mass insertion approximation\footnote{Complete expressions for both $LL$ and $RR$ contributions in the mass eigenstate basis can be found in \cite{Arganda:2007jw}.}, to our knowledge, for the first time. Here, Bino exchange dominates. 
In the small $x_i$ limit, the box diagrams give
\begin{align}
g_{RV}^{u,box}=g_{RV}^{d,box} = \frac{g_1^4}{(4\pi)^2 \mq^2}\delta^{\mu e}_{RR}\frac{1}{4} f\left(\frac{\ml^2}{\mq^2}\right),
\end{align}
while the $\gamma$-penguin diagrams contribute
\begin{align}
g_{RV}^{u,\gamma-peng.}=-2g_{RV}^{d,\gamma-peng.} &=\frac{-2 e^2 g_1^2}{3(4\pi)^2 \ml^2}\delta^{\mu e}_{RR} f_{\gamma,R} (\xb) \\ &\simeq \frac{-2 e^2 g_1^2}{(4\pi)^2 \ml^2}\delta^{\mu e}_{RR}\left(\frac{1}{4}\right),
\end{align}
where $f_{\gamma,R} (x)$ is a loop function given in Appendix \ref{LFconversion.APP}. In the final line we have taken the $\xb \to 0$ limit.
The $Z$-penguin diagrams give
\begin{align}
g_{RV}^{u,Z-peng.}&=\frac{-\left(1-\frac{4}{3}\sin^2\theta_W\right)}{\left(1-\frac{8}{3}\sin^2\theta_W\right)}g_{RV}^{d,Z-peng.} =\frac{-g_1^4}{(4\pi)^2 \ml^2} \frac{1}{4}\left(1-\frac{8}{3}\sin^2\theta_W\right)\delta^{\mu e}_{RR} \cos{2\beta} f_{Z,R}(\xb,\xmu).
\end{align}
where $f_{Z,R}(\xb,\xmu)$ is a loop function given in Appendix \ref{LFconversion.APP}.

\begin{figure}
\centering
\includegraphics[width=0.8\textwidth]{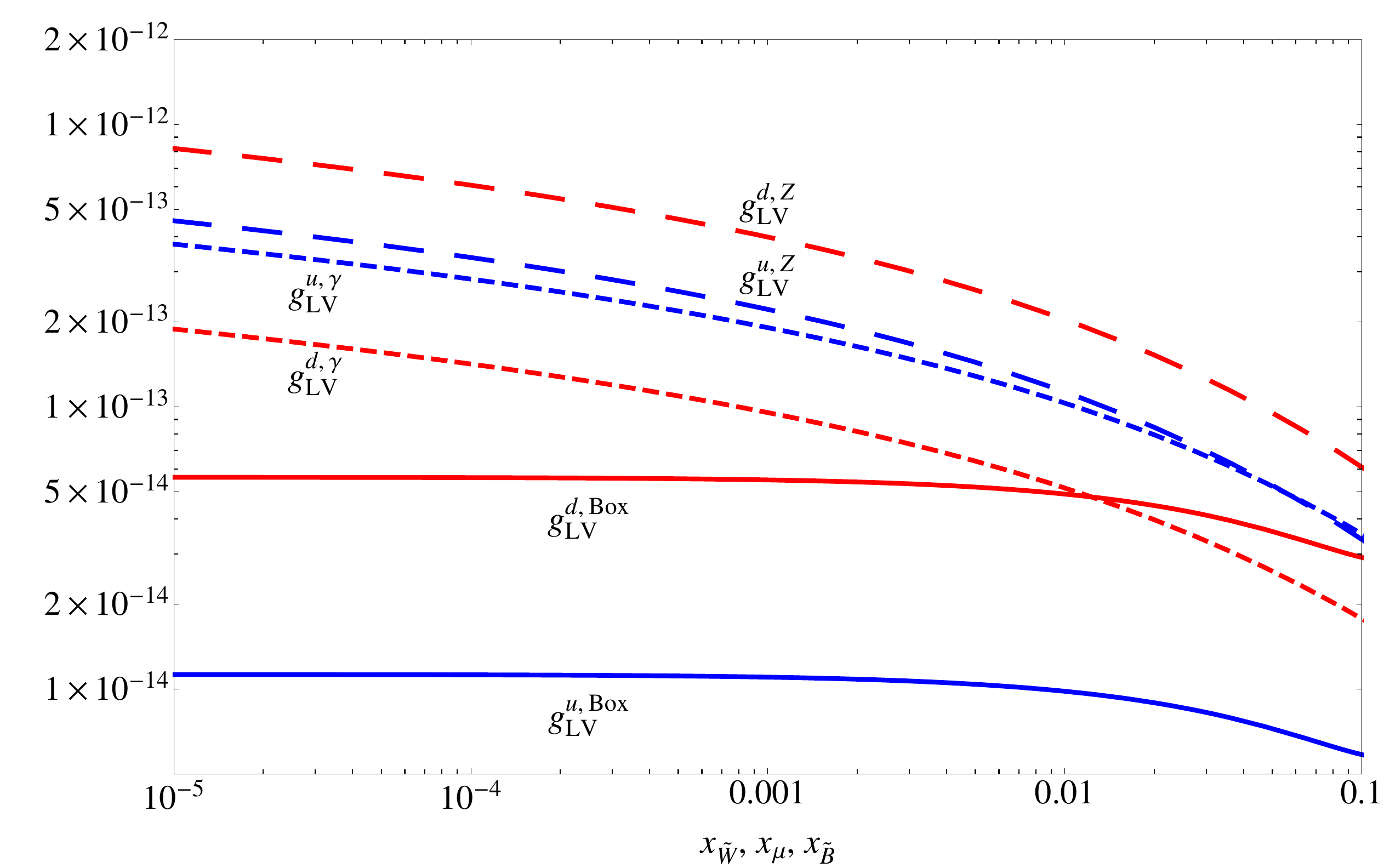}
\caption{The relative importance of various non-dipole operators to the branching ratio as a function of $x_{\tilde{W}},~x_\mu$. We have taken $\ml=\mq=20$ TeV, $t_{\beta}=10$, and $\delta_{LL}=\delta_{RR} = 0.3$. Not shown are the various RH non-dipole operators. They achieve a maximum of $8\times10^{-15}$ for the Bino d-quark Z-penguin diagrams (at $x_i \sim 10^{-5}$) and a maximum of $4\times10^{-15}$ for the Bino u-quark $\gamma$-penguin diagrams (at $x_i \sim 0.1$). 
}
\label{OpDepMuCon.FIG}
\end{figure}

In Fig. \ref{OpDepMuCon.FIG} we show the dependence of non-dipole operators on a common $x_i$. We see the branching ratio of $\mu \to e$ conversion is dominated by the $\gamma/Z$-penguin diagrams for small  $x_i$. 

\subsubsection{Interference between dipole and non-dipole operators in $\mu \to e$ conversion}
\label{ConInt.SEC}

We now review interference effects exhibited in $\mu \to e$ conversion. Most importantly, there is interference between the dipole operators and the non-dipole operators listed above. 
The physical sign $\text{sgn}(\mu M_i)$, where $i$=$1,2$ appears in Eqs.~(\ref{LFVWinoLoop.EQ} -- \ref{BinoLoop.EQ}) in the dipole operators, and in Eq.~(\ref{LHZpeng.EQ}) in the non-dipole operators. While in the Wino Z-penguin operator, Eq.~(\ref{LHZpeng.EQ}), it has only a small effect on the overall size of the contribution, in the dipole operator of Eqs. (\ref{LFVWinoLoop.EQ} -- \ref{BinoLoop.EQ}) it not only changes the size, but also the sign of these contributions relative to the sum of the non-dipole operators. The result is that if $\text{sgn}(\mu M_i) = -~(+)$ the branching ratio of $\mu \to e$ conversion exhibits constructive (destructive) interference.

\begin{figure}[t]
\centering
\includegraphics[width=0.7\textwidth]{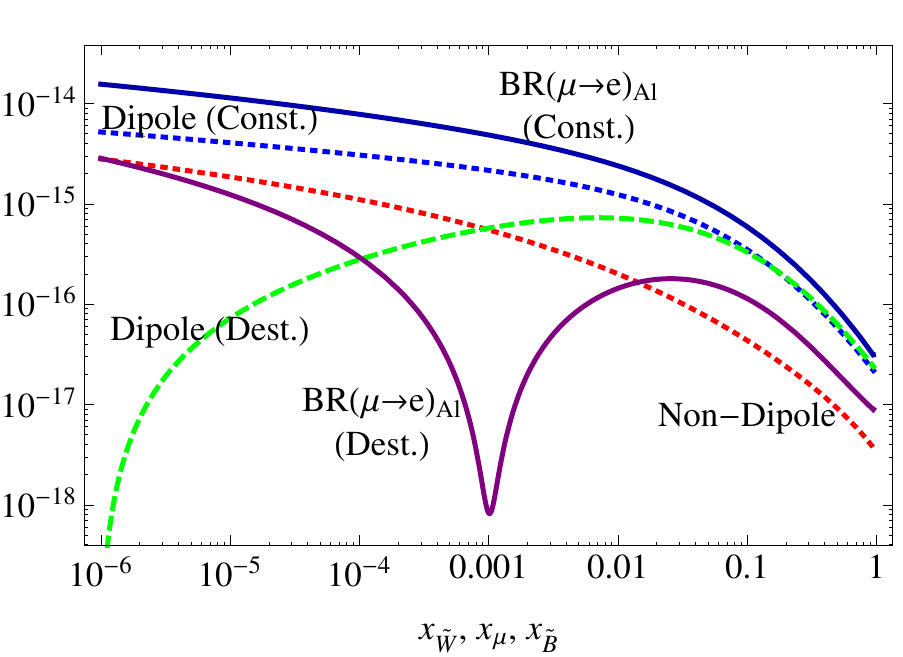}
\caption{The interference of dipole and non-dipole operators as a function of $x_{\tilde{W}},~x_\mu,~ \xb$. We have taken $\ml=\mq=20$ TeV, $t_{\beta}=10$, and $\delta_{LL}=\delta_{RR} = 0.1$. The blue (green) dotted line shows the constructive (destructive) contribution from the dipole operators, while the red dotted line shows the contribution from the non-dipole operators. The dark blue line shows the constructive branching ratio, while the purple line shows the destructive branching ratio.}
\label{MuConInterf.FIG}
\end{figure}

At large values of $x_i$ the dipole operators dominate, and the interference effects are lessened. At smaller values of $x_i$ however, the dipole and non-dipole operators both contribute, and indeed, there is a region where the LH dipole and LH non-dipole parts cancel exactly.  In this case, the branching ratio for $\mu \to e$ conversion is given by the RH contributions, which are themselves dominated by the non-dipole parts in this regime. This is shown in Fig. \ref{MuConInterf.FIG} for $\ml = 20$ TeV, but  a similar cancellation is robust for all values of $\ml$.

\subsection{Rare $\ell_i \to 3\ell_j$ decays }
\label{mu3e.SEC}

We now examine the operators that contribute to the rare decays of $\ell_i \to 3 \ell_j$. In our numerical analysis we restrict ourselves to the decay $\mu \to 3e$, but analytic results apply to rare tau decays as well. We concentrate on $\mu \to 3e$ because of the expected improvement in sensitivity from the Mu3e experiment \cite{Blondel:2013ia, Berger:2014vba}, which aims to probe BR$(\mu \to 3e) \lesssim 10^{-16}$. We do not consider $\tau \to 3 e (\mu)$ decays in our numerical analysis, as the expected future sensitivity is not much greater than that of $\tau \to e(\mu) \gamma$ \cite{Aushev:2010bq}.

The branching ratio of $\ell_i \to 3\ell_j$ is given by \cite{Hisano:1995cp, Arganda:2005ji} as
\begin{align}
\nonumber \text{BR}(\ell_i \to 3\ell_j) \simeq \frac{6\pi^2 \alpha_{em}^2}{G_F^2} \Bigg\{ & |\ALP|^2 + |\ARP|^2 - 2 \left( \ALP (\ARD)^*+ \ALD (\ARP)^* +h.c.\right) 
\\ \nonumber& + \left( \frac{16}{3} \log \frac{m_\mu}{m_e} - \frac{22}{3} \right)\left( |\ALD|^2 + |\ARD|^2\right) 
\\ \nonumber&+ \frac{1}{6} \left( |B_1^L|^2 + |B_1^R|^2\right) + \frac{1}{3} \left( |B_2^L|^2 + |B_2^R|^2\right) 
\\ \nonumber&+ \frac{1}{24} \left( |B_3^L|^2 + |B_3^R|^2\right) + 6 \left( |B_4^L|^2 + |B_4^R|^2\right) 
\\ \nonumber &-\frac{1}{2} \left( B_3^L (B_4^L)^* + B_3^R (B_4^R)^* + h.c.\right) 
\end{align}
\begin{align}
\nonumber &+\frac{1}{3} \left( \ALP (B_1^L)^* + \ARP (B_1^R)^* + \ALP (B_2^L)^* + \ARP (B_2^L)^* + h.c. \right) 
\\ \nonumber &-\frac{2}{3} \left( \ARD (B_1^L)^* + \ALD (B_1^R)^* +\ALD (B_2^R)^* + \ARD (B_2^L)^* + h.c. \right)
\\ \nonumber &+\frac{1}{3}\Big[ 2 \left( |F_{LL}|^2 +|F_{RR}|^2\right) + |F_{LR}|^2 +|F_{RL}|^2
\\ \nonumber &+ \left( B_1^L (F_{LL})^* + B_1^R (F_{RR})^* + B_2^L (F_{LR})^* + B_2^R (F_{RL})^* + h.c. \right) 
\\ \nonumber &+2 \left( \ALP (F_{LL})^* + \ARP (F_{RR})^* + h.c. \right) + \left( \ALP (F_{LR})^* + \ARP (F_{RL})^* + h.c.\right)
\\ & - 4 \left( \ARD (F_{LL})^* + \ALD (F_{RR})^* + h.c. \right) - 2 \left( \ALD (F_{RL})^* + \ALD (F_{LR})^* + h.c.\right)
\Big]
\Bigg\} \ ,
\label{mu3eBR.EQ}
\end{align}
where $A_{dip.}^{L,R}$ are the dipole operator coefficients from from Eqs. (\ref{LFVWinoLoop.EQ}) - (\ref{BinoLoop.EQ}) above, $A_{\gamma -p.}^{L,R}$ are the photo-penguin operator coefficients, the $B_i^{L,R}$ are from box-type operators and the $F_{LL,RR,LR,RL}$ are from $Z$-penguin operators, as defined in \cite{Arganda:2005ji}.

Typically at moderate and low $\tan \beta$, the $\mu \to e \gamma$ dipole operators dominate the $\mu \to 3e$ decay rate \cite{Arganda:2005ji, Altmannshofer:2013lfa}, in large part due to the appearance of the $\log m_\mu / m_e$ in the second line of Eq. (\ref{mu3eBR.EQ}) above.\footnote{This logarithm arises due to the phase space integration of the final state fermions-- there is a infrared singularity cutoff by the electron mass.} There exists then a fairly simple relation between the two branching ratios:
\begin{align}
\frac{\text{BR}(\mu \to 3 e)}{\text{BR}(\mu \to e \gamma)} \simeq \frac{\alpha_{em}}{3\pi} \left( 2\log \frac{m_\mu}{m_e} - \frac{11}{4}\right) \simeq 6.1 \times 10^{-3} \ .
\end{align}

\begin{figure}
\centering
\includegraphics[width=0.7\textwidth]{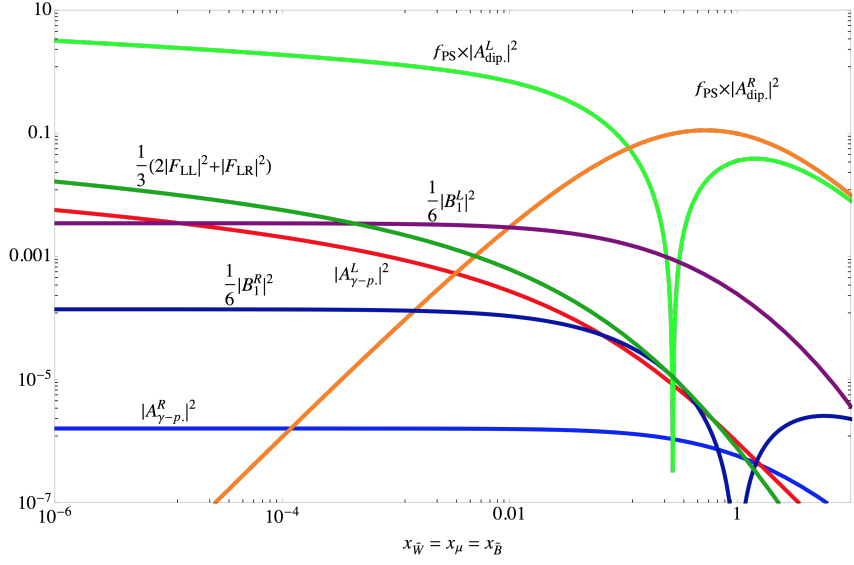}
\caption{The relative importance of various operators to the branching ratio of $\mu \to 3 e$, in arbitrary units. We have scaled each one by the appropriate numerical factor ($f_{PS} \equiv \left(\frac{16}{3} \log \frac{m_\mu}{m_e} - \frac{22}{3} \right)$ contains the IR logarithm induced by integration over phase space). This shows that for all values of $x$ (with all being set equal), the dipole coefficents dominate. We set $\tilde{m}_{{\ell}}= 20$ TeV, and we set $\delta^{\mu e}_{LL}=\delta^{\mu e}_{RR} = \delta^{\mu \tau}_{LL}\delta^{\tau e}_{RR} = 0.3$ so that the effective $\delta^{\mu e}$ is the same for each operator. $\tan \beta =10$. }
\label{OpDepMu3E.FIG}
\end{figure}

As can be seen in Fig. \ref{OpDepMu3E.FIG}, the dipole operators, enhanced by the phase space factor, greatly dominate over the other operators that contribute to the branching ratio in all regions of $x_i$ parameter space. In our analysis we include the numerical contributions from the other operators, which are given in the mass insertion approximation in Appendix \ref{Mu3e.APP}.

\section{Dependence on Fermionic Superpartner Masses}
\label{xOlogy.SEC}

In Section \ref{LFV.SEC} above, we saw that there is non-trivial dependence of LFV observables on $\xb,~\xw,~x_\mu$ and $t_\beta$. Additionally, the dependence on $\xg$ of quark sector observables was shown in Fig. \ref{Operators.FIG}. In this section we examine in detail how the LFV constraints compare with the quark FCNC constraints as a function of various combinations of gaugino masses and $\mu$. With this aim in mind, we show the ratio of squark to slepton mass 
\begin{equation} R_{ij}  \equiv \mq/ \ml,
\end{equation}
 for which constraints derived from transitions between generations $i$ and $j$ are equally strong from the quark and lepton sectors. We investigate this ratio as a function of various  $x_i$.    Because in this section we set $\delta_{LR} =0$,   the behavior of the various transitions always goes as $\delta/\tilde{m}^2$.  Thus, our results in terms of $R_{ij}$ with fixed $\delta$ can be reinterpreted as ratios of $\sqrt{\delta_\ell / \delta_q}$ for equal sfermion masses.

Before discussing the relative power of different measurements,  we first want to determine what $x_{i}$ affect our observables most.  We first examine the dependence of meson oscillation observables on $\xg$.  This can be gleaned by studying Fig. \ref{Operators.FIG}.   There is $\mathcal{O}(1)$ variation between small $\xg$ and $\xg \sim 1$, while for $\xg \gg 1$ the variation becomes important. Since we restrict ourselves to either the situation where $\xg \ll 1$ or $\xg \sim 1$, the relative power of LFV and quark FCNC observables with respect to $\xg$ is at most  $\mathcal{O}(1)$.
In addition, the dependence of the LFV observables on $\xb$ in the regions we consider ($\xb \ll 1 \to \xb \sim 1$) is only slight. Only if one has large $\xmu \gtrsim 1$ as well as large $\xb \gtrsim 1$ does the variation become appreciable. As such, varying $\xb$ does not allow one to change the relative power of the LFV and quark FCNC observables much. Therefore, we concentrate on the effect of varying $\xmu$ and $\xw$.

We now move to quantify the relative power of quark and LFV constraints by solving for $R$ in several cases. 
We first specify $\xg$.
We then  find the squark mass which saturates the bound from meson mixing.
Similarly, once we fix $x_{\mu}$, $\xb$ and $\xw$ we can find the corresponding slepton mass which saturates the current limits on the processes $\mu \to e \gamma$, $\tau \to e \gamma$ and $\tau \to \mu \gamma$. We also consider the combinations which saturate the future sensitivity to $\mu \to e \gamma$, $\mu \to e$ conversion in Aluminium and $\mu \to 3e$. Combining these two results yields $R$.
Results shown in this section assume constructive interference as defined for the LFV processes in Section \ref{LFV.SEC}. Were we to examine the case of destructive interference, when comparing $\ell_i \to \ell_j \gamma$ with quark FCNCs, we would find qualitatively similar behavior of $R_{ij}$ for values of $x_i \lesssim 1$.  $R$ is increased by at most a factor of 2. For large values of $x_i$, the interference effect is lessened. When comparing $\mu \to e$ conversion however, interference effects can be important, as discussed in Section \ref{ConInt.SEC}. If we were to examine destructive interference, $R$ would become very large near $x_i \sim 10^{-3}$.

We study two separate regimes, one where we fix $\xb=\xw=\xg\sim1$, which corresponds to TeV-scale physics, and one where we fix $\xb=\xw=\xg\sim10^{-3}$, corresponding to heavy scalars, but with $\mathcal{O}$(TeV) gauginos. We then allow only $\xmu$ to vary, primarily because its variation captures most of the important effects.  We have already argued that $\xb$ and $\xg$'s effects are easily understood.   In principle, we could have shown the variation with respect to $\xw$, but it follows approximately the same pattern as varying $\xmu$. This can be understood by considering Fig. \ref{OpDepMu.FIG}. In the small $x_i$ regime, the dominant contribution to the $\mu \to e \gamma$ transition arises due to the LH Wino-Higgsino mixing diagrams with loop functions $g_2 (\xw, \xmu)$ and $g_3 (\xw, \xmu)$. In the small $x_i$ limit, these functions are approximately
\begin{align}
g_2(\xw, \xmu) &\sim \frac{\xw \log \xw}{\xmu - \xw}+\frac{\xmu \log \xmu}{\xw - \xmu}, \\
g_3(\xw, \xmu) &\sim \frac{ \log \xw}{\xmu - \xw}+\frac{ \log \xmu}{\xw - \xmu},
\label{smallXLFV.EQ}
\end{align}
so that the behavior as a function of $\xmu$ and $\xw$ is the same. Therefore varying one while keeping the other fixed is enough to illustrate the general behavior.

\begin{figure}
\centering
\subfloat[$\xb = \xw = \xg \sim 1$.]{
\includegraphics[width=0.46\textwidth]{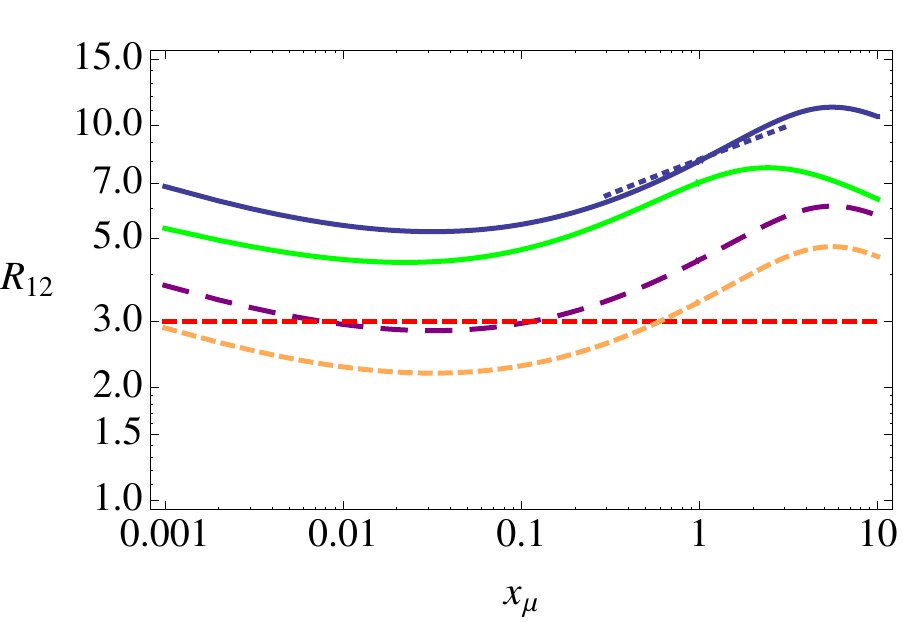}
\label{LargeX.FIG}
}
\hspace{0.1 cm}
\subfloat[$\xb = \xw = \xg \sim 10^{-3}$.]{
\includegraphics[width=0.46\textwidth]{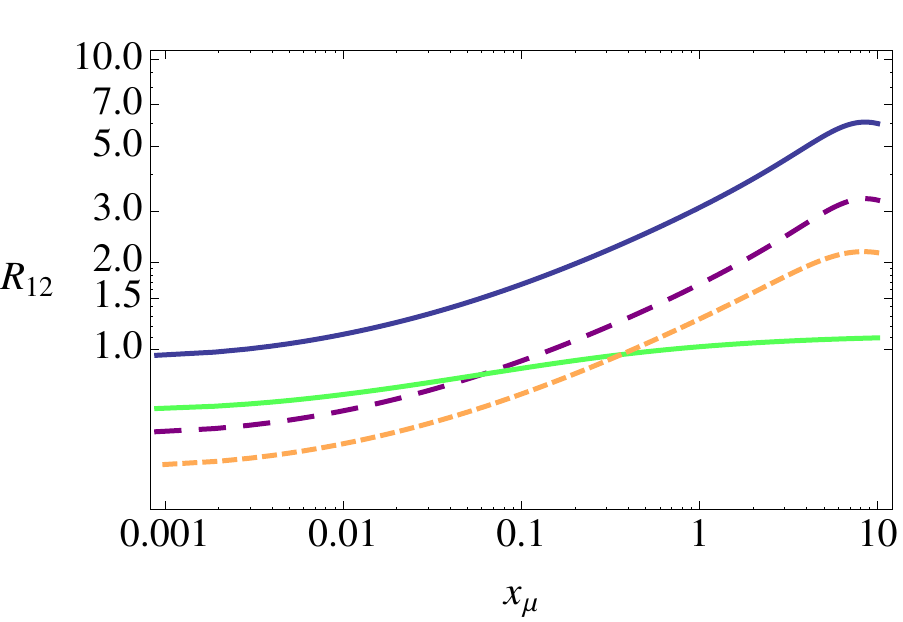}
\label{SmallX.FIG}
}
\caption{These figures show $x_\mu$ vs.~the ratio $R_{12}$ of squark to slepton mass that saturates the experimental bounds from kaon oscillations and rare $\mu$ decays.
We set $\delta_{LL,RR} = 0.1$, $\delta_{LR}=0$ and $t_{\beta}=10$. In the left figure,  $\xb = \xw =1$, while in the right figure $\xb = \xw =10^{-3}$. The solid blue line is the current $\mu \to e \gamma$ constraint, the dashed purple line is the future $\mu \to e \gamma$ sensitivity, the solid green line is the future sensitivity to $\mu \to e $ conversion in Aluminium, and the dashed orange line corresponds to future $\mu \to 3 e$ sensitivity. The dotted blue line corresponds to the function for $R_{12}$ given in Eq. (\ref{R12semAn.EQ}). The dashed red line is the ratio of $\mq/\ml$ obtained by running from the GUT scale to the low scale given initial conditions for a universal gaugino mass $m_{1/2}(M_{GUT}) = 3$ TeV and universal scalar mass $m_0(M_{GUT}) = 0.5$ TeV.}
\label{GauginoDominationRatio.FIG}
\end{figure}

\begin{figure}
\centering
\subfloat[$\xb = \xw = \xg \sim 1$. ]{
\includegraphics[width=0.465\textwidth]{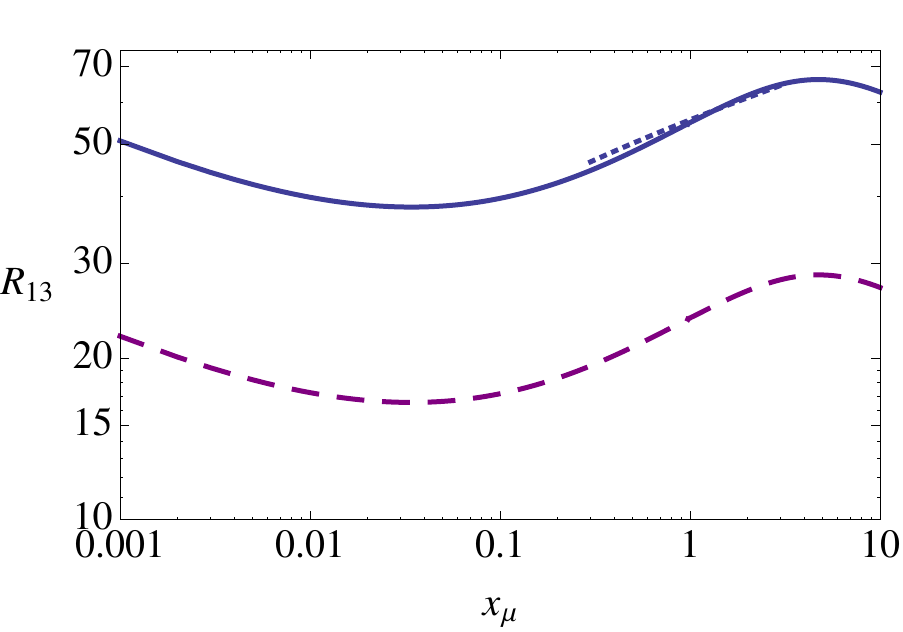}

\label{LargeXBd.FIG}}
\hspace{0.5 cm}
\subfloat[ $\xb = \xw = \xg \sim 10^{-3}$. ]{
\includegraphics[width=0.465\textwidth]{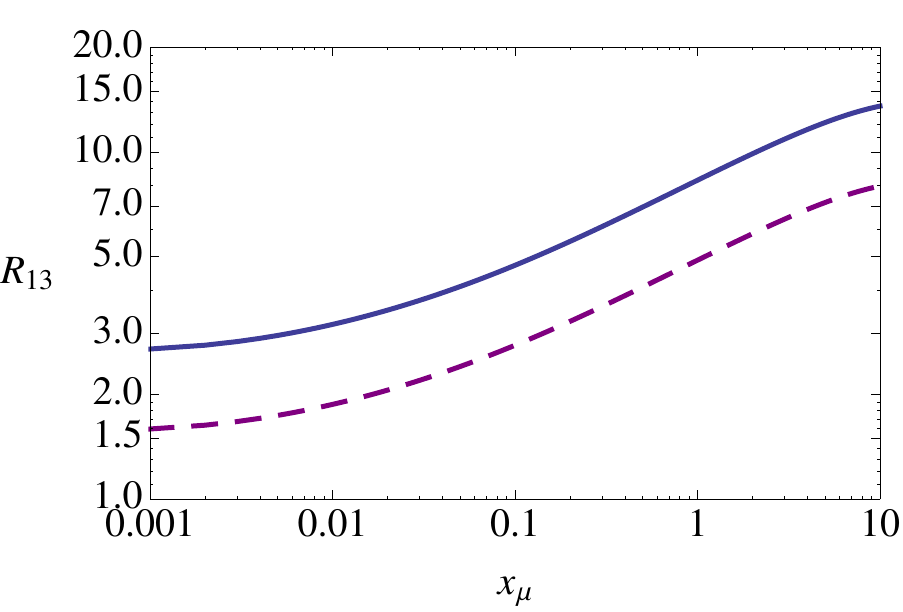}
\label{SmallXBd.FIG}}
\vspace{-.5 cm}
\subfloat[$\xb = \xw = \xg \sim 1$.]{
\includegraphics[width=0.465\textwidth]{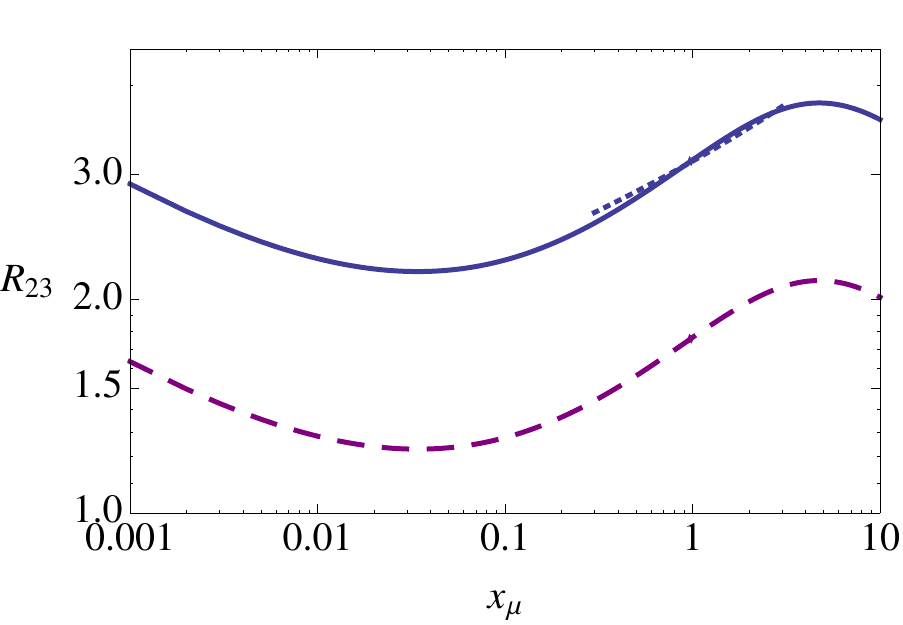}
\label{LargeXBs.FIG}}
\hspace{0.5 cm}
\subfloat[ $\xb = \xw = \xg \sim 10^{-3}$. ]{
\includegraphics[width=0.465\textwidth]{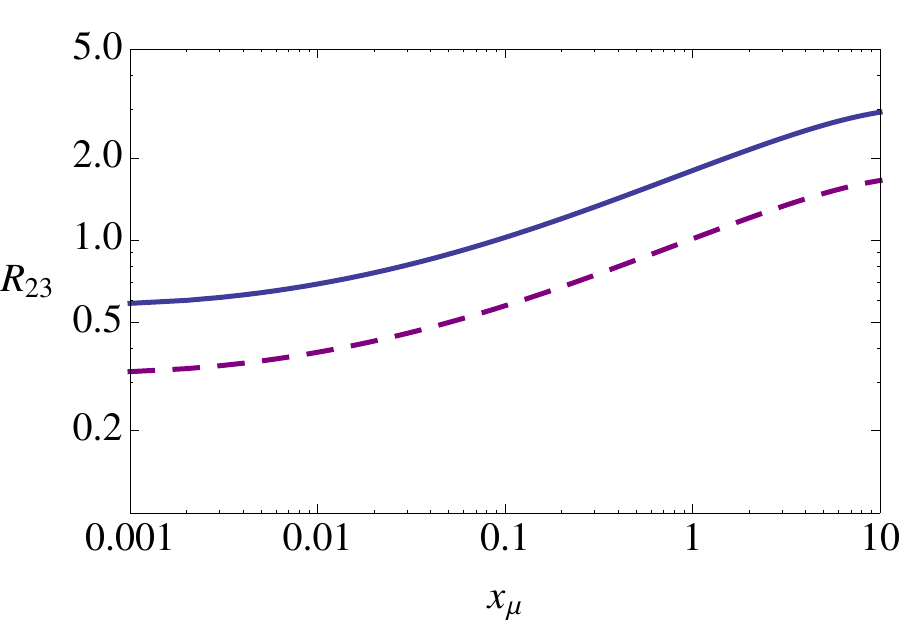}
\label{SmallXBs.FIG}}
\caption{The upper (lower) figures show $x_\mu$ vs. the ratio $R_{13}~(R_{23})$ of squark to slepton mass that saturates the current experimental bounds from $B_d$ ($B_s$) meson oscillations and $\tau \to e \gamma$ ($\tau \to \mu \gamma$). We set $\delta_{LL,RR} = 0.1$, $\delta_{LR}=0$ and $t_{\beta}=10$. In the left figure,  $\xb = \xw =1$, while in the right figure $\xb = \xw =10^{-3}$. The solid blue line is calculated using the current $\tau \to e \gamma$ ($\tau \to \mu \gamma$) constraint, while the dashed purple line uses the future $\tau \to e \gamma$ ($\tau \to \mu \gamma$) sensitivity. The dotted blue lines corresponds to the functions for $R_{13}$ (upper left) and $R_{23}$ (lower left) given in Eqs. (\ref{R13semAn.EQ}) and (\ref{R23semAn.EQ}) respectively.}
\label{GauginoDominationRatioBd.FIG}
\end{figure}

In the large $x_i \sim 1$ regime, there is more complicated dependence on various contributions to the $\mu \to e \gamma$ amplitude. We see from both Fig. \ref{OpDepMu.FIG} and Fig. \ref{LargeX.FIG} that the region $0.3 \lesssim \xmu,\xw \lesssim 3$ is where most variation occurs. This is also true for $1-3$ and $2-3$ transitions, as can be seen in Figs. \ref{LargeXBd.FIG} and \ref{LargeXBs.FIG}. 
In this regime (the region $0.3 \lesssim \xmu,\xw \lesssim 3$), we find the following functions 
\begin{align}
\label{R12semAn.EQ}
R_{12} &\simeq 7.2 + \sqrt{0.85 \, \xw \xmu} + \log \xw + \log \xmu, \\
\label{R13semAn.EQ}
R_{13}  &\simeq 54 + \sqrt{2.5 \, \xw \xmu} + 7(\log \xw + \log \xmu), \\ 
\label{R23semAn.EQ}
R_{23} &\simeq 2.6 + \sqrt{0.28 \, \xw \xmu} + 0.2(\log \xw + \log \xmu),
\end{align}
capture this behavior accurately to within $\lesssim 6 \%$.  We choose this particular functional form because it closely matches the functional form of the full expressions (found in Appendix \ref{LFmuEgam.APP}), with a small number of parameters.

The situation where $\xb  \sim \xw \sim \xg \sim 1$ at the low scale could be realized with a GUT-scale boundary condition such that a universal scalar mass $m_0$ is small compared  with high scale gaugino masses $m_{1/2}$. In this case, one arrives at a low-energy spectrum where the slepton masses are dominated by the Wino mass, and the squark masses are dominated by the gluino mass. We find that with these boundary conditions, the low scale squark to slepton mass ratio is fixed, and is approximately $\mq/\ml \sim 3$. This is shown by the dotted red line in Fig. \ref{LargeX.FIG}. Given the constraints from the running on $\xb,~\xg,~\xw$, the only free parameters in this case are $\mu$ and $t_{\beta}$. We show in Fig. \ref{LargeX.FIG} how the squark to slepton mass ratio varies as a function of $\xmu$ for the $1-2$ sector, and in Figs. \ref{LargeXBd.FIG} and \ref{LargeXBs.FIG} for the $1-3$ and $2-3$ sectors respectively. 

We show the results for the ``heavy scalar case" $\xb = \xw = \xg \sim 10^{-3}$ in Fig. \ref{SmallX.FIG} for the $1-2$ sector, and in Figs. \ref{SmallXBd.FIG} and \ref{SmallXBs.FIG} for the $1-3$ and $2-3$ sectors respectively. This situation could arise for example if the boundary conditions at the GUT-scale are such that a universal gaugino mass $m_{1/2}$ is suppressed relative to $m_0$.

We notice that smaller $\xmu$  increases the relative strength of the LFV probes for all transitions.  Moreover, $R$ decreases as a function of increasing $t_\beta$, i.e. LFV becomes relatively powerful at large $t_\beta$.   For the current constraints from $\mu \to e \gamma$, $R_{12}$ decreases from $R_{12} \approx20$ for $t_\beta =2$ to $R_{12} \approx 6$ for $t_\beta =20$ when $x_i\sim1$, and from $R_{12} \approx1.6$ for $t_\beta =2$ to $R_{12} \approx 0.7$ for $t_\beta =20$ when $x_i\sim10^{-3}$. 

The constraint from $\mu \to e$ conversion shows a decrease from $R_{12} \approx 10$ for $t_\beta =2$ to $R_{12} \approx 3$ for $t_\beta =20$ for $x_i\sim1$, and from $R_{12} \sim 0.7$ for $t_\beta =2$ to $R_{12} \sim 0.5$ for $t_\beta =20$ when $x_i\sim10^{-3}$.
For all transitions, being in the small $\xb,~\xw,~\xg$ regime results in significant increases in the relative strength of LFV observables relative to quark FCNC observables. The increase is a factor of a few for the $1-2$ and $1-3$ transitions, and up to an order of magnitude for the $2-3$ transitions. In the $1-2$ and $2-3$ transitions, we see that at small values of $x_i$ the ratio drops below 1, meaning that the LFV constraints become stronger than those from the meson oscillation observables. In the $1-3$ sector however, the ratio does not drop below 1, a result that is echoed in Section \ref{deltaOlogy.SEC}, where we see that $B_d$ meson oscillations are a stronger constraint than $\tau \to e \gamma$ in much of the $\delta$ parameter space also. We also observe that in Fig. \ref{SmallX.FIG} for small $\xb,~\xw,~\xg$, $\mu \to e$ conversion in the future goes from being a weaker constraint than the future sensitivity of $\mu \to e \gamma$ at small $\xmu$, to being the stronger constraint for $\xmu \gtrsim 0.04$.

From our results in this section, we can see that having small $x_i$ results in a relative strengthening of the LFV constraints for all transitions.
Nevertheless, varying $x_i$ over 3 orders of magnitude typically only results in variation of $R$ by $\mathcal{O}$(few), and at most an order of magnitude.

\section{Constraints on $\delta$}
\label{deltaOlogy.SEC}

In this section we examine constraints on the flavor off-diagonal mass insertions. We show results for $1-2$ transitions (comparing $\mu \to e \gamma$, $\mu \to e$ conversion and $\mu \to 3e$ with $\Delta m_K$), and also for $1-3$ and $2-3$ transitions (comparing $\tau \to e(\mu) \gamma$ with $B_{d(s)}$ meson mixing). 
These analyses summarize the relative sensitivity of quark and lepton flavor violation probes now and into the future.

It is also of interest to connect these results to GUT constructions and or textures.
If SUSY breaking respects, e.g., an $SU(5)$ GUT symmetry, particles residing within a 
${\bf \bar{5}}$ or a ${\bf 10}$ may share a common soft mass. Inspired by this relation, we define
\begin{align}
\delta^{\tilde{\ell}_i}_{LL}=\delta^{\tilde{d}_i}_{RR}&\equiv \delta_{\bar{\mathbf{5}}}, \\
\nonumber \delta^{\tilde{\ell}_i}_{RR}=\delta^{\tilde{u}_i}_{RR}=\delta^{\tilde{q}_i}_{LL}&\equiv \delta_{\mathbf{10}}. 
\label{Deltas.EQ}
\end{align}
We will comment on how our results can be rephrased in this language below.

In subsection \ref{LRzeroSmallX.SEC} we consider the situation where the $LR$ insertions are zero, and all the mass-squared ratios $x_i \sim \mathcal{O}(10^{-2} -10^{-3})$, indicative of a significant but modest hierarchy between sfermions and fermionic superpartners.  In this region of parameter space we are interested in the possibility that gauginos and the $\mu$ parameter are all around the TeV scale, but sfermions are much heavier, in the tens to hundreds of TeV, akin to models of split or mini-split SUSY \cite{Wells:2003tf, Giudice:2004tc, ArkaniHamed:2004fb, ArkaniHamed:2004yi, Wells:2004di, Acharya:2007rc, Acharya:2008zi, Arvanitaki:2012ps}. There is strong motivation for such models, and their implications for flavor physics have been considered before \cite{ArkaniHamed:2004yi, Giudice:2005rz, Altmannshofer:2013lfa, Moroi:2013sfa, Ellis:2014tea, Ellis:2015dra}. 
For even smaller $x$, for fixed sfermion mass, the $\Delta F=2$  is essentially unchanged.  
The LFV BRs increase logarithmically as you go to smaller $x_i$ (see Eq.~(\ref{smallXLFV.EQ})). However, too small $x_i$ will result in too small gaugino masses and $\mu$ unless the sfermion mass is raised.

In subsection \ref{LRzeroLargeX.SEC} we consider the case where $LR$ insertions are zero, but the $x_i = 1$. Finally in subsection \ref{LRnonzeroLargeX.SEC} we analyse the case where $LR$ insertions are non-zero, and the $x_i = 1$. In the latter two sections with $x_i =1$, the region of parameter space we consider is one where once again gauginos and the $\mu$ parameter are around the TeV scale, motivated both by naturalness and by dark matter considerations. Therefore we set the sfermion masses to also be at the TeV scale.

\subsection{$\delta_{LR} = 0$, $x$ small}
\label{LRzeroSmallX.SEC}

\begin{figure}
\centering
\subfloat[Constructive interference in LFV processes.]{
\includegraphics[width=0.49\textwidth]{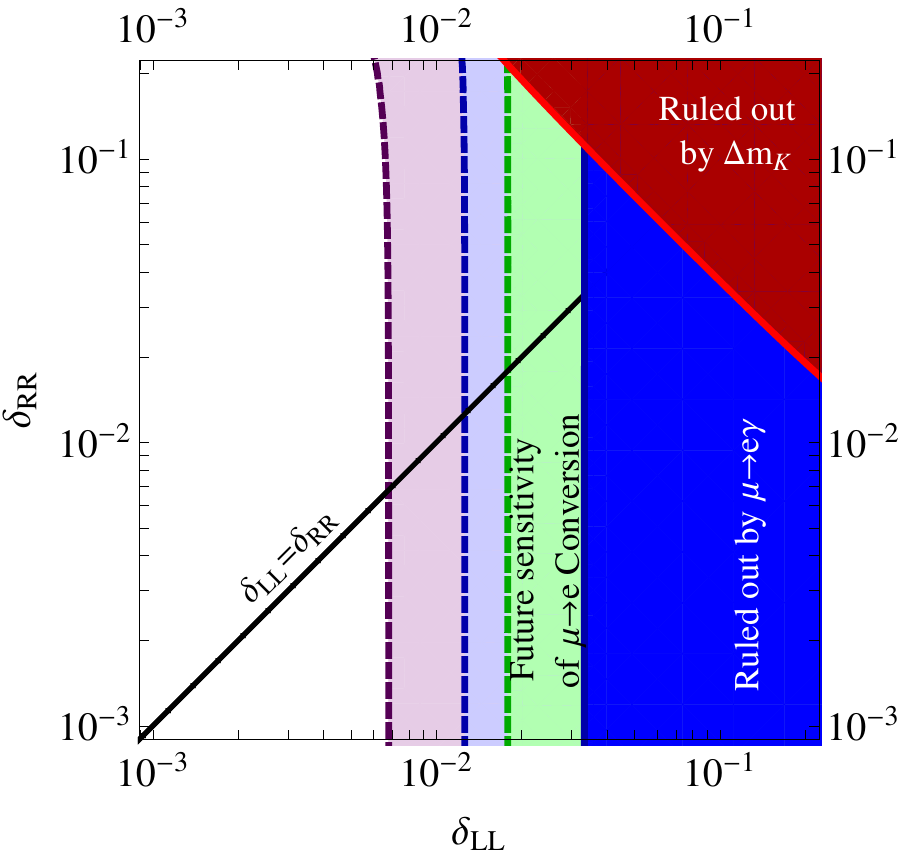} 
}
\subfloat[Destructive interference in LFV processes.]{
\includegraphics[width=0.49\textwidth]{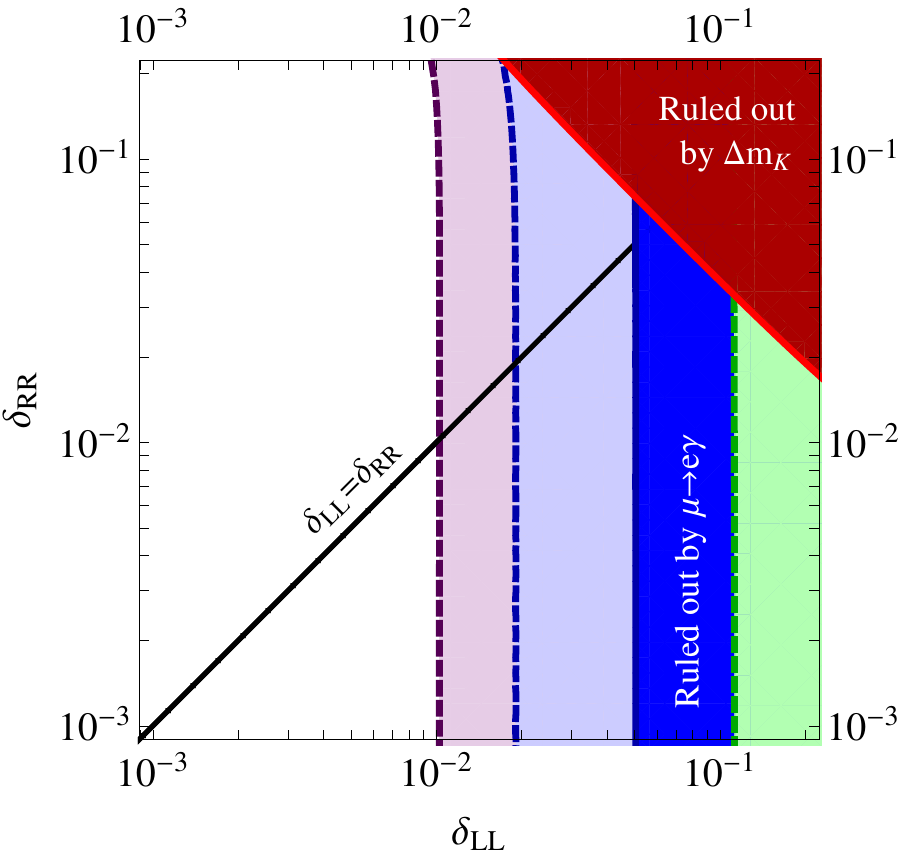}
}
\caption{$\delta_{LL}$ vs $\delta_{RR}$ plots for $1-2$ insertions.  These plots compare constraints from $\Delta m_K$ (red), BR($\mu \to e \gamma$) (current (future) in dark (light) blue), $\mu\to e$ conversion (green) and BR($\mu \to 3e$) (purple). All regions correspond to the measured (projected) limits at 90\% C.L. We have set $\mq=\ml = 20$ TeV, $x_{\tilde{g}}=x_\mu = x_{\tilde{W}} = x_{\tilde{B}} = 5\times 10^{-3}$, and $t_\beta = 10$.}
\label{MuEDeltaOlogy.FIG}
\end{figure}

\begin{figure}
\centering
\subfloat[Constructive interference in LFV, $1-3$ transitions]{
\includegraphics[width=0.49\textwidth]{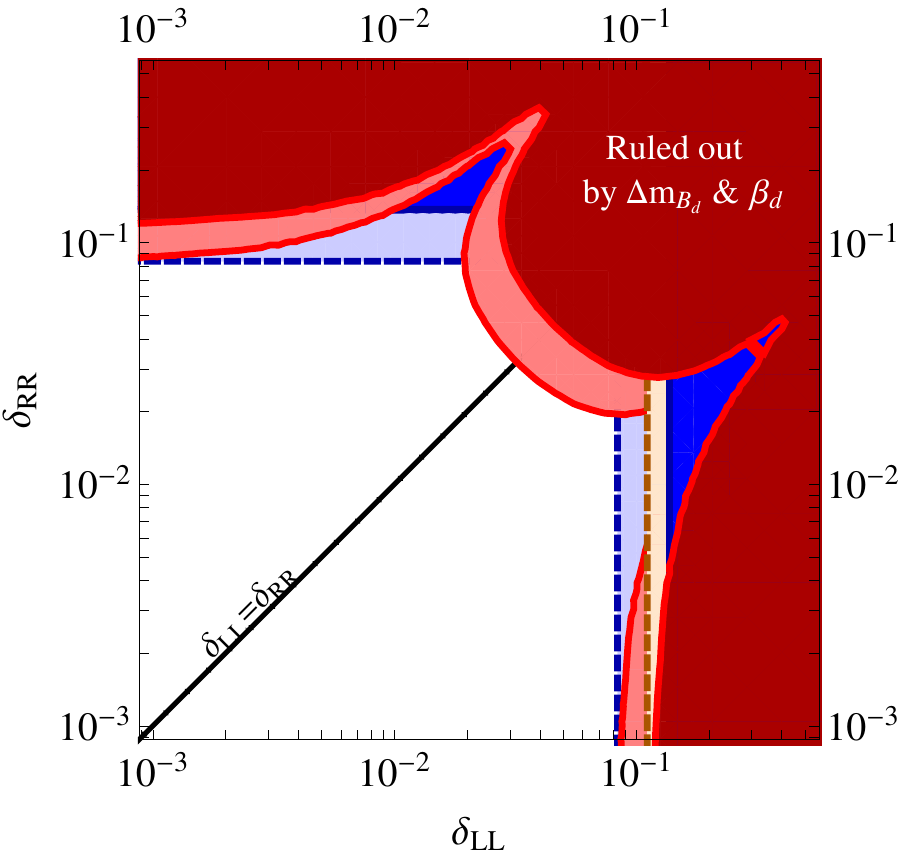} 
\label{TauEDeltaOlogy.FIG}
}
\subfloat[Destructive interference in LFV, $1-3$ transitions]{
\includegraphics[width=0.49\textwidth]{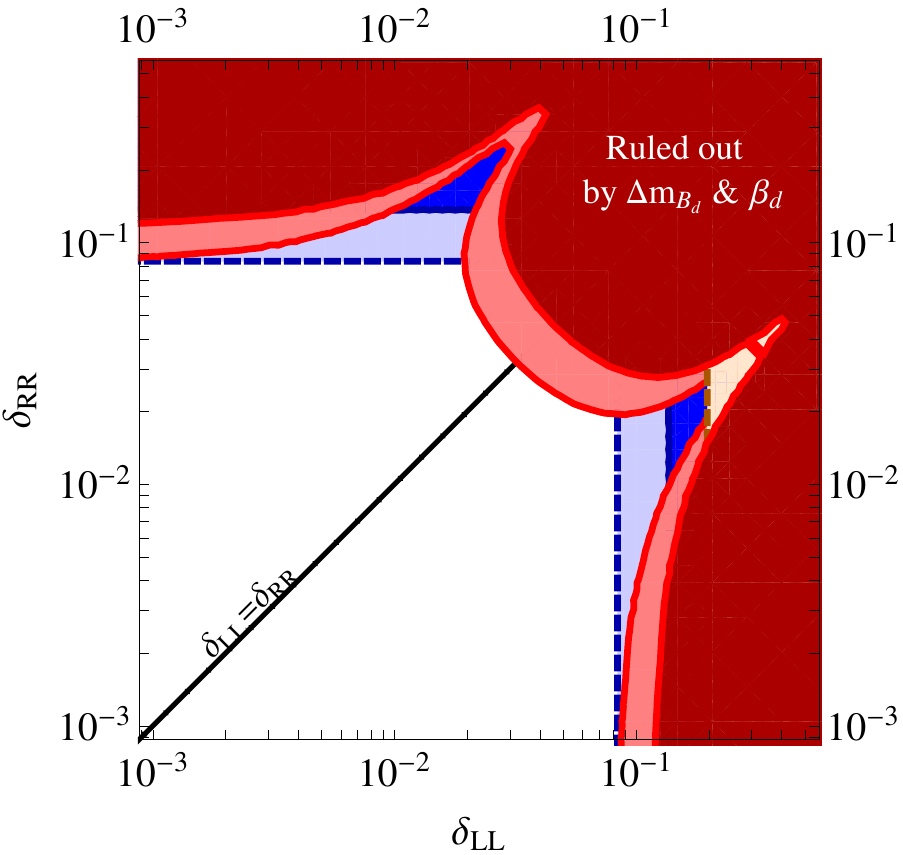}
\label{TauEDeltaOlogyDest.FIG}
}

\hspace{1cm}
\subfloat[Constructive interference in LFV, $2-3$ transitions]{
\includegraphics[width=0.49\textwidth]{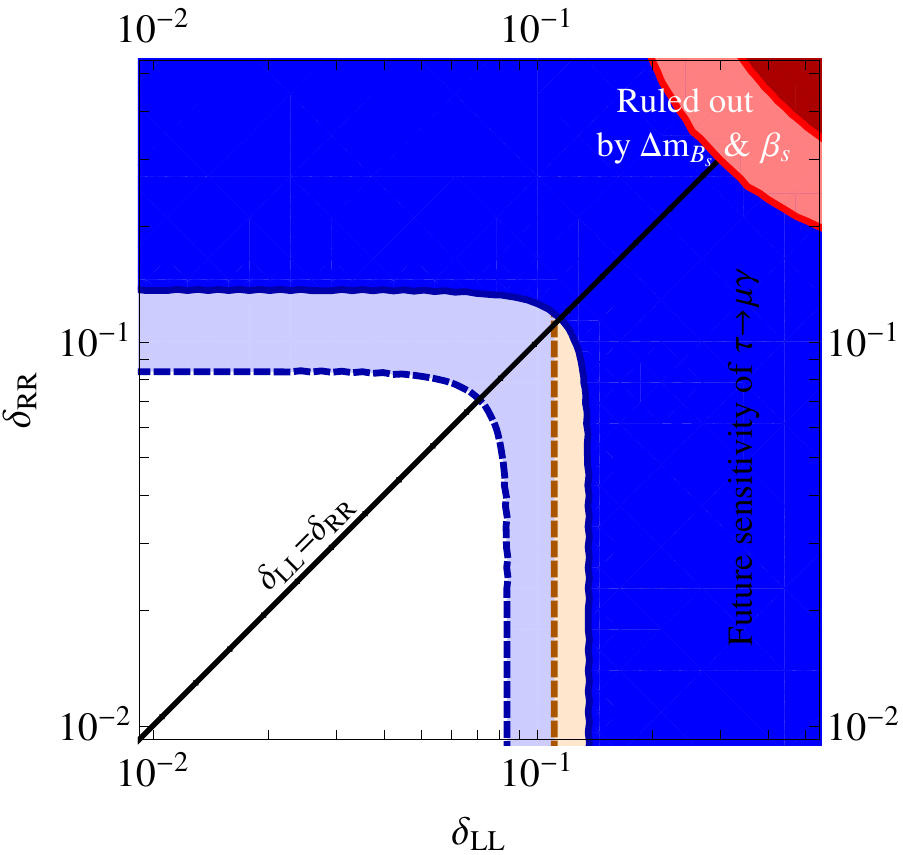} 
\label{TauMuDeltaOlogy.FIG}
}
\subfloat[Destructive interference in LFV, $2-3$ transitions]{
\includegraphics[width=0.49\textwidth]{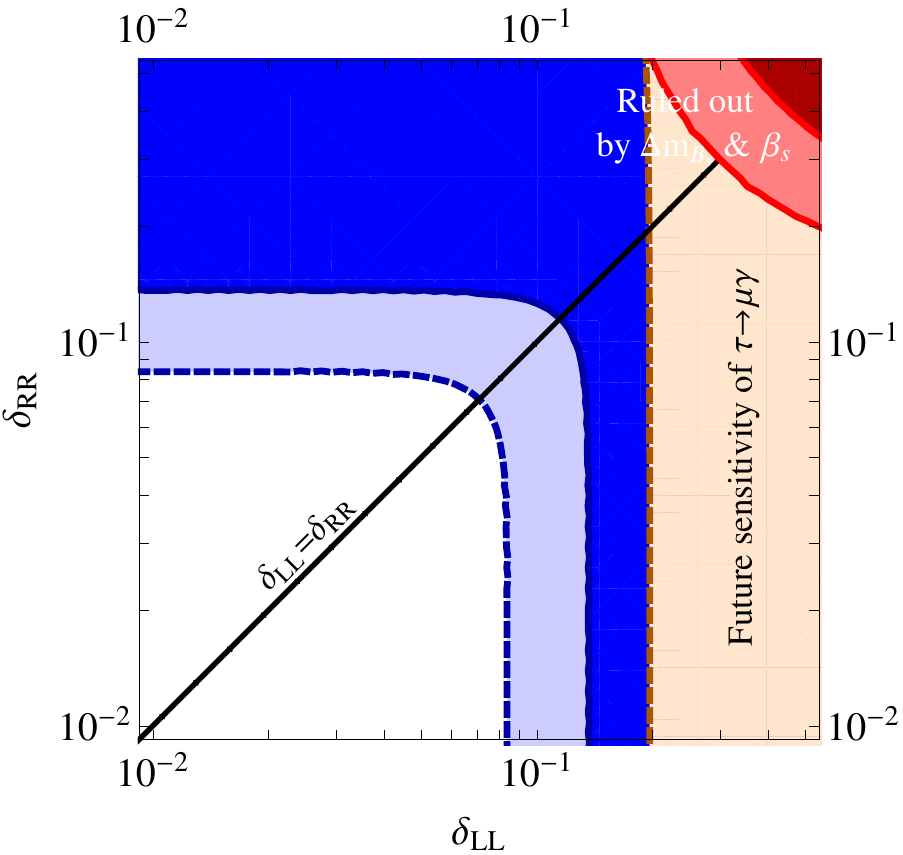}
\label{TauMuDeltaOlogyDest.FIG}
}
\caption{$\delta_{LL}$ vs $\delta_{RR}$ plots for $1-3$ (upper) and $2-3$ (lower) insertions.  These plots compare constraints from $\Delta m_{B_{d}}$ and $\beta_d$, BR($\tau \to e \gamma$) for the upper plots, and from $\Delta m_{B_{s}}$ and $\beta_s$, BR($\tau \to \mu \gamma$).  The dark red regions are already excluded, and the light red shows the potential future reach  with a factor of two improvement. The light orange region shows the future sensitivity of $\tau \to e/\mu \gamma$. We have set $\mq=\ml = 5$ TeV, $x_{\tilde{g}}=x_\mu = x_{\tilde{W}} = x_{\tilde{B}} \simeq 0.04$, and $t_\beta = 10$.  Also shown is a dark blue region excluded by $\mu \to e \gamma$ making the further assumption that  $\delta^{13}_{LL,RR} = \delta^{23}_{RR,LL}$. The light blue is the future sensitivity given the same assumption. All regions shown are excluded at 90\% C.L.}
\label{TauDeltaOlogy.FIG}

\end{figure}

In this subsection we consider
$x_i \ll 1$, corresponding to a scenario where there is a  hierarchy between scalar and fermionic superpartners.
In Figs.~\ref{MuEDeltaOlogy.FIG} and \ref{TauDeltaOlogy.FIG}, we display regions excluded by quark and lepton FCNC limits in the $\delta_{LL}$, $\delta_{RR}$ plane. 
Note, meson mixing constraints are symmetric under $\delta_{LL} \leftrightarrow \delta_{RR}$, so these figures can be reinterpreted as  $\delta_{RR} \leftrightarrow \delta_{\mathbf{10}}$ and $\delta_{LL} \leftrightarrow \delta_{\bar{\mathbf{5}}}$. In our numerical work, we have set $\delta$'s in the up sector to vanish, but flavor violation in the up sector is in any case subdominant for the meson mixing considered here.

In Fig.~\ref{MuEDeltaOlogy.FIG}, which corresponds to $1-2$ transitions, we have chosen $\mq=\ml = 20$ TeV and $x_i = 5\times 10^{-3}$, while for the $1-3$ (Figs.~\ref{TauEDeltaOlogy.FIG},~\ref{TauEDeltaOlogyDest.FIG}) and $2-3$  transitions (Fig.~\ref{TauMuDeltaOlogy.FIG},~\ref{TauMuDeltaOlogyDest.FIG}),  we have chosen $\mq=\ml = 5$ TeV and $x_i = 0.04$. The sfermion masses and value of the common $x_i$ for $1-2$ transitions is chosen so as to avoid falling into the region where the destructive interference in $\mu \to e$ conversion is most important, around $x_i \sim 10^{-3}$ (see Section \ref{ConInt.SEC} for further discussion). The smaller sfermion masses and larger $x_i$ for $1-3$ and $2-3$ transitions are chosen so that useful constraints can be shown, and to comply with limits from Run I of the LHC on gluino masses, respectively.  While we have chosen to show plots for particular sfermion mass assignments, the corresponding limits on $\delta_{LL,RR}$ will scale with the masses according to the expressions given in the quark FCNC/LFV anatomy sections \ref{QFV.SEC}, \ref{LFV.SEC} above. 
We discuss each of these figures now in turn.

In Fig. \ref{MuEDeltaOlogy.FIG}, 
we display limits from  $\Delta m_K$, $\mu \to e \gamma$ , $\mu \to 3e$ and $\mu \to e$ conversion. For muon conversion we use the future experimental sensitivity in Aluminium, shown in Table \ref{Limits.TAB} above. We see from Fig. \ref{MuEDeltaOlogy.FIG} a) that in the case of constructive interference for the LFV processes (see discussion below Eqn.~\ref{LFVWinoLoop.EQ}), $\mu \to e \gamma$ is already a stronger constraint than $\Delta m_K$.
Also in this case, the future sensitivity of $\mu \to e \gamma$ will be superior to that of $\mu \to e$ conversion. In the case of destructive interference the current constraint from $\mu \to e \gamma$ is stronger than the future sensitivity of $\mu \to e$ conversion (see Fig. \ref{MuEDeltaOlogy.FIG} b)). This is due to $\mu \to e$ conversion experiencing large interference between the dipole and non-dipole operators in the region near $x_i \sim 10^{-3}$, while the interference in $\mu \to e \gamma$, arising only within the dipole operators, is less pronounced. For both constructive and destructive interference, the strongest constraint will be from the improvement on $\mu \to e \gamma$, and eventually from the planned Mu3e experiment. 

In Fig.~\ref{TauDeltaOlogy.FIG} we assume $\delta^{12}$ is negligibly small, and as such only $\delta^{13}$ and $\delta^{23}$ processes will be relevant. We choose $x_i = 0.04$ in these plots so that the gauginos and $\mu$ are all at 1 TeV.
We compare $\tau \to e \gamma$ with $\Delta m_{B_d}$, $\beta_d$ in Figs. \ref{TauEDeltaOlogy.FIG} and \ref{TauEDeltaOlogyDest.FIG}, and $\tau \to \mu \gamma$ with $\Delta m_{B_s}$, $\beta_s$ in Figs. \ref{TauMuDeltaOlogy.FIG} and \ref{TauMuDeltaOlogyDest.FIG}. In both sets of plots we also consider the possibility that $\mu \to e \gamma$ can provide a constraint on $\delta^{13}\delta^{23}$ due to the LR flavor-conserving insertion in the $m_\tau / m_\mu$ enhanced Bino loop contribution from Section \ref{MuEGam.SEC}. We compare the possible constraint from $\mu \to e \gamma$ under the assumption that $\delta^{13}_{LL,RR} = \delta^{23}_{RR,LL}$, but $\delta^{12} = 0$.   In this case the only operator contributing to $\mu \to e \gamma$ is the $LR$ flavor-conserving Bino operator from section \ref{LFV.SEC}. Presented are results for both constructive and destructive interference between operators contributing to the rare $\tau$ decays.

For the B-meson observables, the central region near $\delta_{LL}=\delta_{RR}$ is dominated by the operator $Q_4$, as defined in Section \ref{QFV.SEC}, while the extended regions at small $\delta_{LL}$ and $\delta_{RR}$ are dominated by the operators $Q_1$ and $\tilde{Q}_1$ respectively. The two regions where constraints from $B_d$-mesons are weaker occur due to cancellation between $Q_1/\tilde{Q}_1$ and $Q_4$. 

In Fig.~\ref{TauEDeltaOlogy.FIG} we observe that in the future $\tau \to e \gamma$ has the potential to be a stronger constraint (for constructive interference) on $\delta_{LL}$ than the current bound from ${B_d}$ mixing for small $\delta_{RR} \lesssim 2 \times 10^{-2}$. However, if the constraints from $B_d$ mixing improve by a factor of two this will reduce the region where $\tau \to e \gamma$ has the potential to be a stronger constraint to $5 \times 10^{-3} \lesssim \delta_{RR} \lesssim 2 \times 10^{-2}$. If $\delta_{LL} \geq \delta_{RR}$, the constraints from $B_d$ mixing will remain the strongest. In the case of destructive interference, the future $\tau \to e \gamma$ constraint will only be stronger in a small region where the contributions from the quark FCNC operators $Q_1$ and $Q_4$ cancel. The current constraints on $\delta_{LL}$ from $\mu \to e \gamma$ under the assumptions stipulated above are weaker (stronger) than the future constraints from $\tau \to e \gamma$ in the case of constructive (destructive interference).   

Meanwhile from Fig.~\ref{TauMuDeltaOlogy.FIG} we see that the constraints from $B_s$ mixing are currently stronger than those from LFV in all regions of $\delta$-space. In the future however, the constraints on $\delta_{LL}$ from $\tau \to \mu \gamma$ will become stronger for all values of $\delta_{RR}$ in the case of constructive interference, and for $\delta_{RR}\lesssim 0.3$ in the case of destructive interference. The constraints on $\delta_{RR}$ from $\mu \to e \gamma$ apply only if $\delta^{13}_{LL,RR} = \delta^{23}_{RR,LL}$. Under this assumption, the current constraints from $\mu \to e \gamma$ are always stronger than those from $B_s$ meson mixing, and stronger (weaker) than the future constraints from $\tau \to \mu \gamma$ for constructive (destructive) interference.  Additionally, $\mu \to e \gamma$ places constraints on $\delta^{23}_{RR}$.  Limints from  $\tau \to \mu \gamma$ one this insertion are very weak (not visible on this plot).  For both constructive and destructive interference, the future sensitivity of $\mu \to e \gamma$ under the stipulated assumptions is greater than the future sensitivity of $\tau \to \mu \gamma$. 

Note that $\mu \to 3e$ can also constrain both $1-3$ and $2-3$ transitions in the same way as $\mu \to e \gamma$, since the same dipole operators dominate both decays. We do not include these constraints in Fig. \ref{TauDeltaOlogy.FIG}, as they can be inferred from the relevant constraints in Fig. \ref{MuEDeltaOlogy.FIG}.

To summarize,  for small $x_i \ll 1$ (heavy scalars), LFV observables either currently or will provide stronger constraints on left-handed flavor violation than the quark sector for both $1-2$ and $2-3$ transitions. In the case of $1-3$ transitions however, the constraints from $B_d$ meson mixing will remain comparable to or stronger than those from LFV observables in most of the parameter space.  We also wish to re-emphasize the potential of $\mu \to e \gamma$ to provide constraints on $1-3$ and $2-3$ transitions due to the $LR$ flavor conserving operator from section \ref{MuEGam.SEC}.

\subsection{$\delta_{LR} = 0$, $x= 1$}
\label{LRzeroLargeX.SEC}

In this subsection we consider the situation when $x_i = 1$. We consider a common superpartner mass $M_{SUSY} = 1$ TeV in this section.

\begin{figure}
\centering
\subfloat[Constructive interference for $\mu \to e \gamma$.]{
\includegraphics[width=0.49\textwidth]{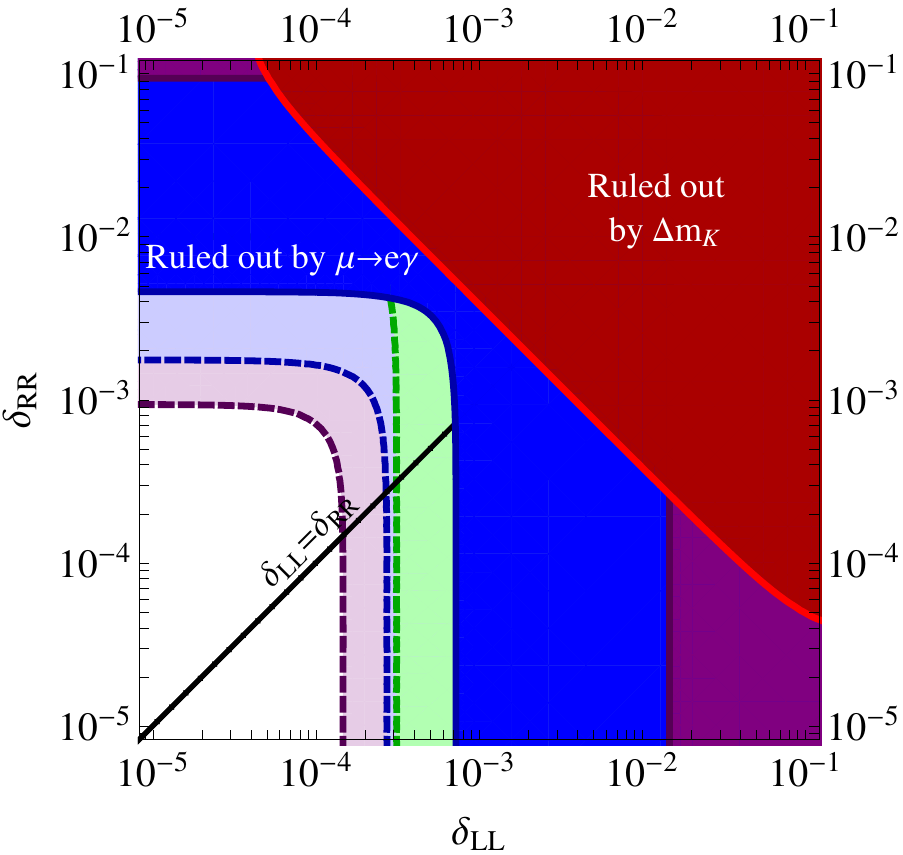} 
\label{LargeXCon.FIG}
}
\subfloat[Destructive interference for $\mu \to e \gamma$.]{
\includegraphics[width=0.49\textwidth]{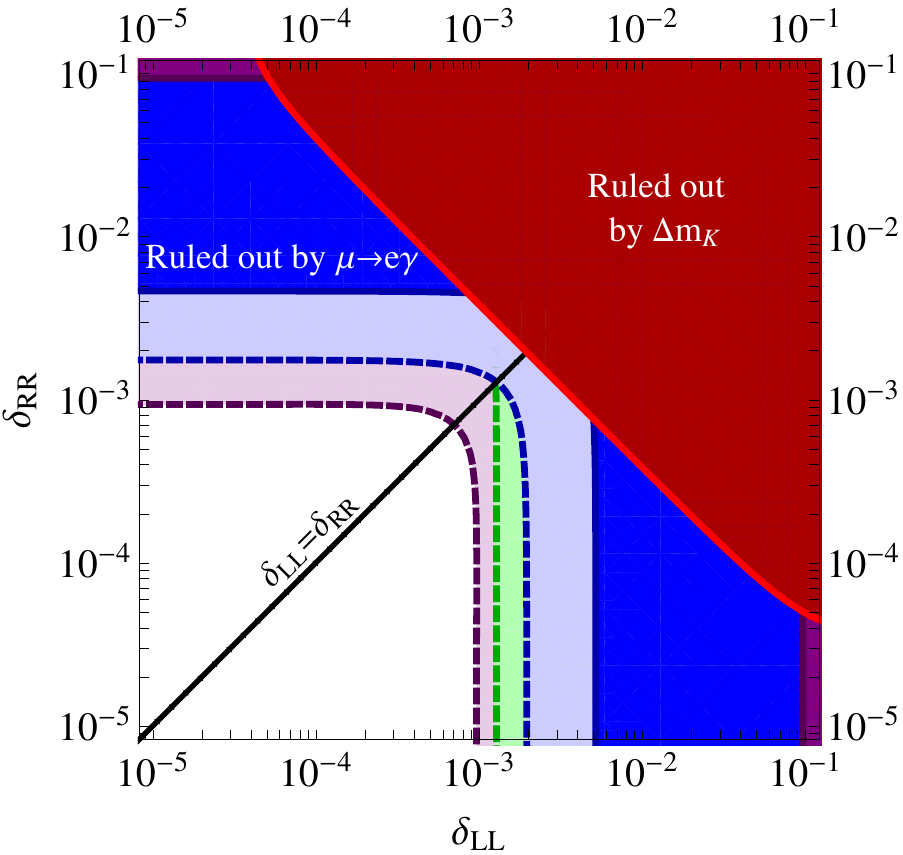}
\label{LargeXDest.FIG}
}
\caption{$\delta_{LL}$ vs $\delta_{RR}$ plots for $1-2$ insertions.  These plots compare the current and future constraints from $\Delta m_K$, BR($\mu \to e \gamma$), BR($\mu \to 3e$) and $\mu\to e$ conversion. All regions correspond to the measured (projected) limits at 90\% C.L. . We have set $\mq=\ml = 1$ TeV, $x_{\tilde{g}}=x_\mu = x_{\tilde{W}} = x_{\tilde{B}} = 1$, and $t_\beta = 10$.}
\label{MuEDeltaOlogyLRZeroLargeX.FIG}
\end{figure}

In Fig.~\ref{MuEDeltaOlogyLRZeroLargeX.FIG} we compare the current and future bounds on $\delta$ in the $1-2$ sector  from  LFV processes and $\Delta m_K$. We consider both  constructive (a) and  destructive (b) interference in the LFV processes. We observe from Fig. \ref{LargeXCon.FIG} that $\mu \to e \gamma$ is currently a stronger constraint than $\Delta m_K$ when there is constructive interference.  The future sensitivity of $\mu \to e \gamma$ will be greater than that of $\mu \to e$ conversion for constructive interference. On the other hand, from Fig. \ref{LargeXDest.FIG} (destructive interference), we see that $\Delta m_K$ is currently a stronger constraint than $\mu \to e \gamma$ if $\delta_{LL}=\delta_{RR}$, and $\mu \to 3e$ will become the strongest constraint in the future. The constraints on $\delta_{RR}$ currently are strongest from $\mu \to e \gamma$, but in the future will be strongest from $\mu \to 3e$. Currently, $\mu \to e \gamma$ also dominates the constraint on $\delta_{LL}$ for constructive interference. For destructive interference  $\delta_{LL}$ will be most strongly constrained by $\mu \to 3e$, which is slightly stronger than $\mu \to e$ conversion. This is in contrast with the situation at small $x_i$, where we saw that the constraint from $\mu \to e$ conversion would be weak in the case of destructive interference. As can be understood by examining Fig. \ref{MuConInterf.FIG}, this is due to the interference at large $x_i \sim1$ not being as pronounced as at small $x \sim \mathcal{O}(\text{few}) \times 10^{-3}$.

\begin{figure}
\centering
\subfloat[Constructive interference in LFV, $1-3$ transitions]{
\includegraphics[width=0.49\textwidth]{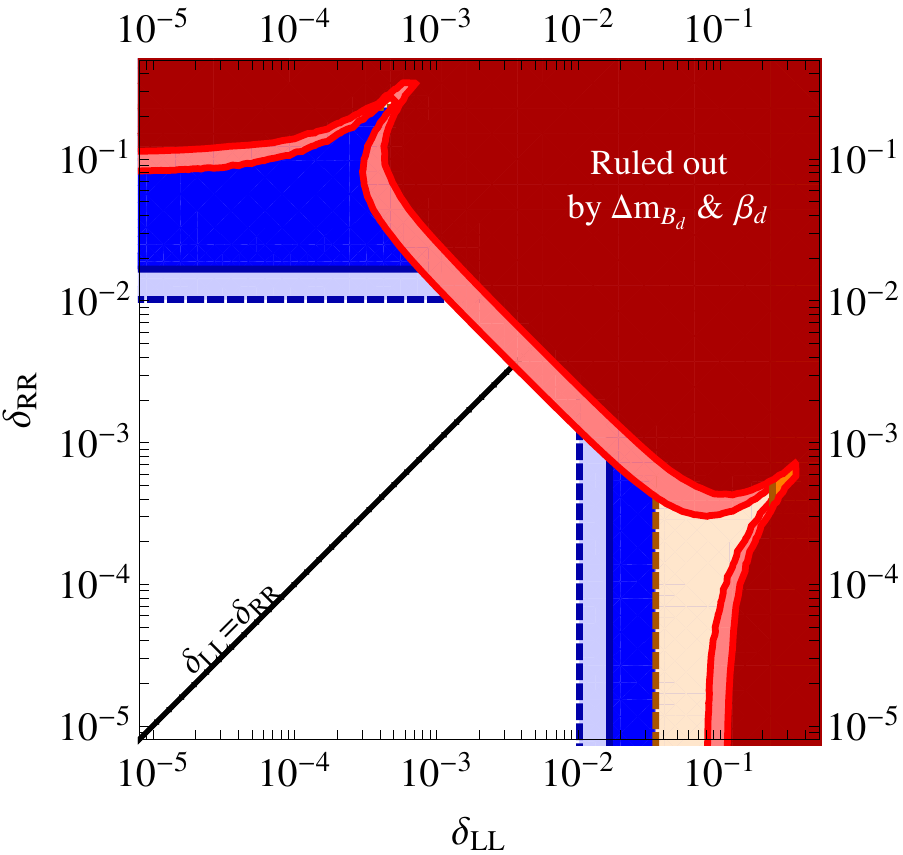} 

\label{TauEDeltaOlogyLargeXa.FIG}}
\subfloat[Destructive interference in LFV, $1-3$ transitions]{
\includegraphics[width=0.49\textwidth]{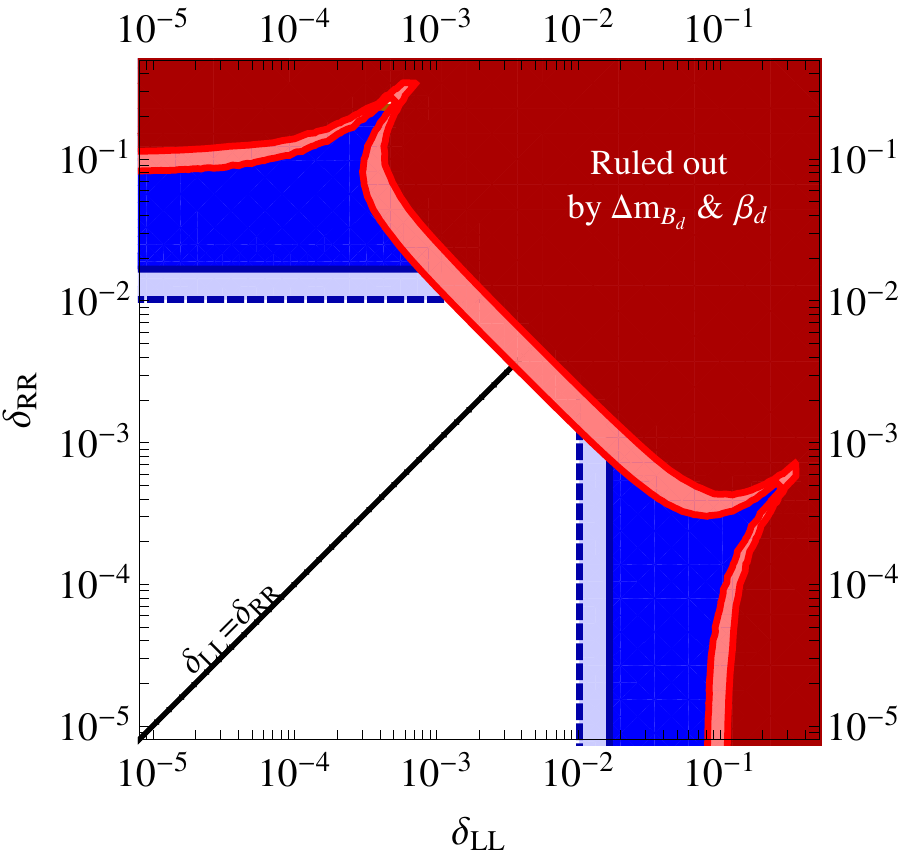}
\label{TauEDeltaOlogyLargeXb.FIG}
}
\vspace{0.1cm}
\subfloat[Constructive interference in LFV, $2-3$ transitions]{
\includegraphics[width=0.49\textwidth]{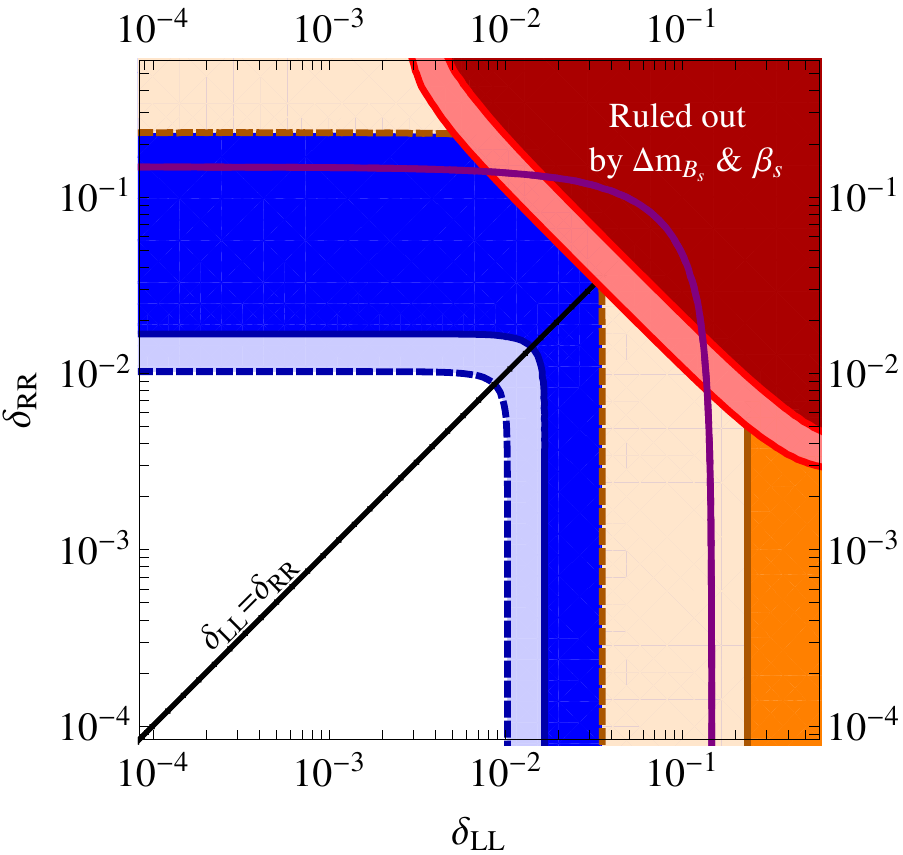} 
\label{TauMuDeltaOlogyLargeXa.FIG}
}
\subfloat[Destructive interference in LFV, $2-3$ transitions]{
\includegraphics[width=0.49\textwidth]{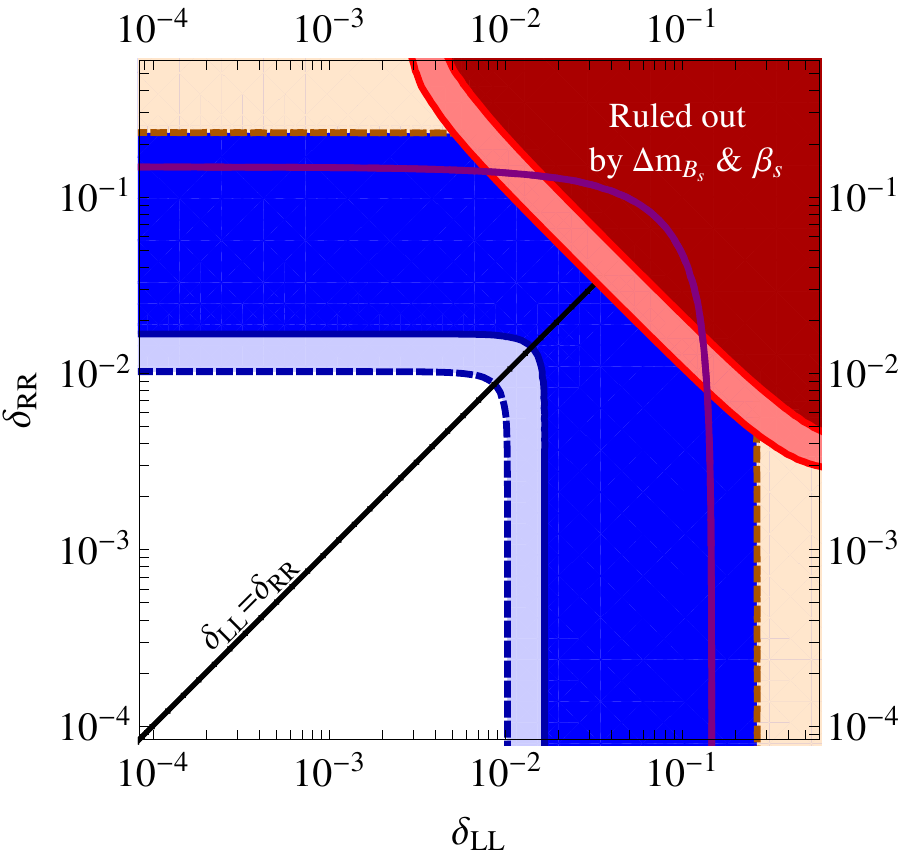}
\label{TauMuDeltaOlogyLargeXb.FIG}
}
\caption{$\delta_{LL}$ vs $\delta_{RR}$ plots for $1-3$ (upper) and $2-3$ (lower) insertions. The upper plots compare the current constraints from $\Delta m_{B_{d}}$, $\beta_d$, $\tau \to e \gamma$ on $\delta^{13}$. The lower plots compare constraints from $\Delta m_{B_{s}}$, $\beta_s$, $b \to s \gamma$, $\tau \to \mu \gamma$. The dark red regions are excluded by $B$ meson mixing, the light red is a potential factor of two improvement. The light orange region shows the future sensitivity of $\tau \to \mu \gamma$. The purple line shows the current limits from $b \to s \gamma$. Also shown is a dark blue region excluded by $\mu \to e \gamma$ assuming $\delta^{13}_{LL,RR} = \delta^{23}_{RR,LL}$. The light blue is the future sensitivity given the same assumption. We have set $\mq=\ml = 1$ TeV, $x_{\tilde{g}}=x_\mu = x_{\tilde{W}} = x_{\tilde{B}} \simeq 1$, and $t_\beta = 10$.}
\label{TauEDeltaOlogyLargeX.FIG}
\end{figure}

Turning now to $1-3$ transitions, we see from Figs. \ref{TauEDeltaOlogyLargeXa.FIG} and \ref{TauEDeltaOlogyLargeXb.FIG} that if $\delta^{12}=0$, the bound from $B_d$ mixing will remain a stronger constraint than $\tau \to e \gamma$ in a large region of parameter space. This result is largely independent of whether interference in the leptonic observable is constructive or destructive, and in the destructive case $\tau \to e \gamma$ will not improve on the $B_d$ mixing bound at all. As in the previous section, we compare the possible constraint from $\mu \to e \gamma$ under the assumption that $\delta^{13}_{LL,RR} = \delta^{23}_{RR,LL}$. If this assumption is correct, $\mu \to e \gamma$ is already a stronger probe than the future sensitivity of $\tau \to e \gamma$ in all of the parameter space shown, regardless of interference. However, despite improvements in $\mu \to e \gamma$, the sensitivity will not be competitive with the constraints from $B_d$ meson mixing near the line of $\delta_{LL}=\delta_{RR}$.

Finally, we perform the same analysis for $\tau \to \mu \gamma$, comparing with bounds from $B_s$ mixing. From  Fig. \ref{TauMuDeltaOlogyLargeXa.FIG}, there is a region where $\tau \to \mu \gamma$ already provides the strongest constraint on $1-3$ mixing in the case of constructive interference.  In the future, such a region will exist for destructive interference as well as seen in Fig. \ref{TauMuDeltaOlogyLargeXb.FIG}.
Additionally, we note  $\mu \to e \gamma$ (again, with the added assumption $\delta^{13}_{LL,RR} = \delta^{23}_{RR,LL}$) is already a stronger probe than both of the other observables in all of the parameter space, and will remain so into the future. 
If this assumption does not hold, then we note that $b \to s \gamma$, shown by the purple lines, is currently the strongest constraint on $\delta_{RR}$ for small $\delta_{LL}$ regardless of the sign of the product $M_{\tilde{g}}A^{23}$, which appears in the gluino diagrams contributing to the amplitude.  
The future sensitivity of $\tau \to \mu \gamma$ will improve on these constraints on $\delta_{LL}$ only if there is constructive interference in the $\tau$ decay amplitude. It will not however improve on the constraints on $\delta_{RR}$, but rather will have comparable sensitivity.

Note that $\mu \to 3e$ can also constrain both $1-3$ and $2-3$ transitions in the same way as $\mu \to e \gamma$, since the same dipole operators dominate both decays. We do not include these constraints in Fig. \ref{TauEDeltaOlogyLargeX.FIG}, as they can be inferred from the relevant constraints in Fig. \ref{MuEDeltaOlogyLRZeroLargeX.FIG}.

\subsection{$\delta_{LR} \neq 0$, $x=1$}
\label{LRnonzeroLargeX.SEC}

For TeV-scale superpartner masses, the factor $m_f / \tilde{m}$ in the LR insertions is small, but not negligibly so. We cannot assume that $\delta_{LR}=0$ as we had done when the superpartners were of $\mathcal{O}$(10) TeV. So, in the $x \sim 1$ case, given a particular $\tilde{m}$, using the known SM fermion mass, we relax the $\delta_{LR}=0$ assumption.  Indeed, we place limits on the ratio $A^{ij}/\tilde{m}$. In this subsection we assume $\delta_{RR} = \delta_{LL} \equiv \delta$.

\begin{figure}
\centering
\subfloat[Constructive interference for $\mu \to e \gamma$.]{
\includegraphics[width=0.49\textwidth]{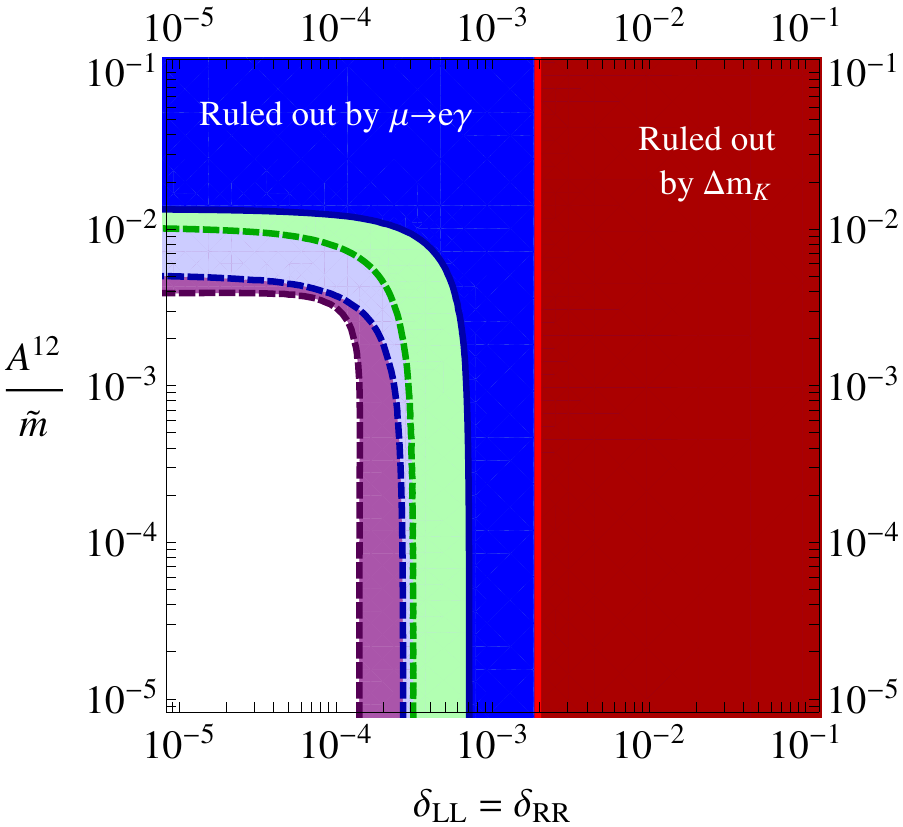} 
}
\subfloat[Destructive interference for $\mu \to e \gamma$.]{
\includegraphics[width=0.49\textwidth]{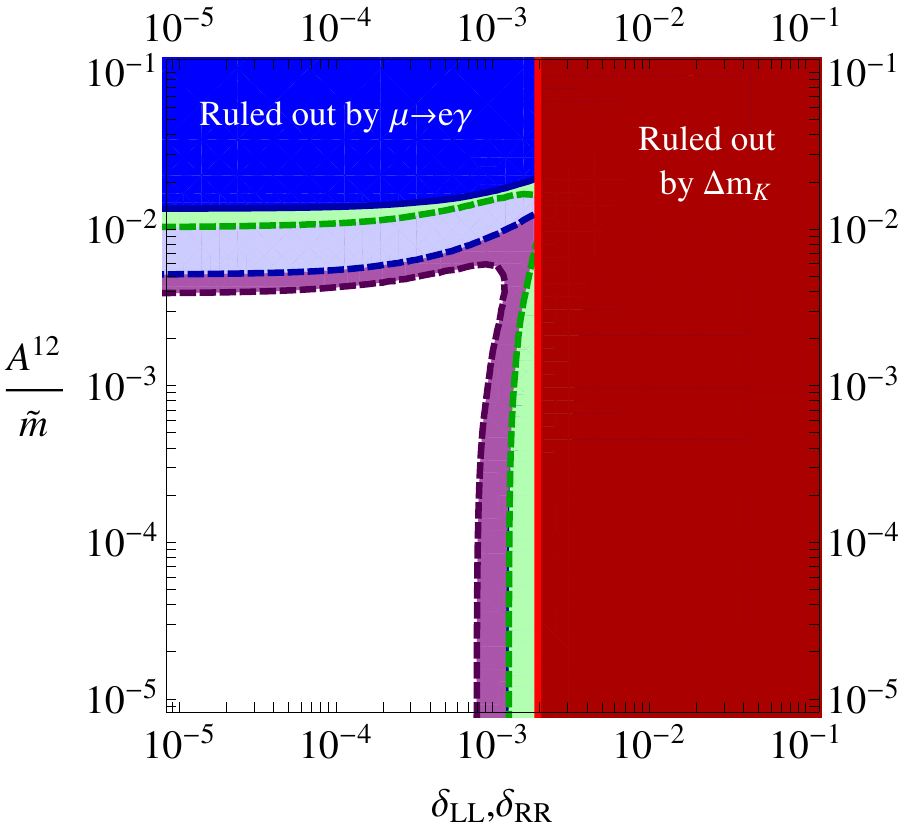}
}
\caption{$\delta_{LL},\delta_{RR}$ vs $A^{12} / \tilde{m}$ plots comparing the constraints from $\Delta m_K$ (red), the current (future) limit on BR($\mu \to e \gamma$) (dark (light) blue), the future sensitivity of $\mu\to e$ conversion (light green) and the future sensitivity of BR($\mu \to 3e$) (purple), all at the 90\% C.L.. We have set $\mq=\ml = 1$ TeV, $x_{\tilde{g}}=x_\mu = x_{\tilde{W}} = x_{\tilde{B}} = 1$, and $t_\beta = 10$.}
\label{MuEDeltaOlogyLRnonZeroSmallX.FIG}
\end{figure}

We see from Fig. \ref{MuEDeltaOlogyLRnonZeroSmallX.FIG} that in the case of constructive interference $\mu \to e \gamma$ places stronger constraints on the size of $A^{12}/ \tilde{m}$ than $\Delta m_K$ in all regions of parameter space, and large regions if there is destructive interference. Note in the case of constructive interference, $\mu \to e \gamma$ is already constraining $A^{12} \lesssim 10^{-2} \tilde{m}$ for TeV-scale SUSY masses. This is also true when there is destructive interference except for a sliver of parameter space near the $A^{12}/\tilde{m} = \delta_{LL,RR}$ line, where the interference is most pronounced. Eventually, $\mu \to 3e$ will be the strongest constraint on $A^{12}/\tilde{m}$, although only slightly improving on the future $\mu \to e \gamma$ constraint.

\begin{figure}
\centering
\subfloat[Constructive interference for $\tau\to e \gamma$.]{
\includegraphics[width=0.49\textwidth]{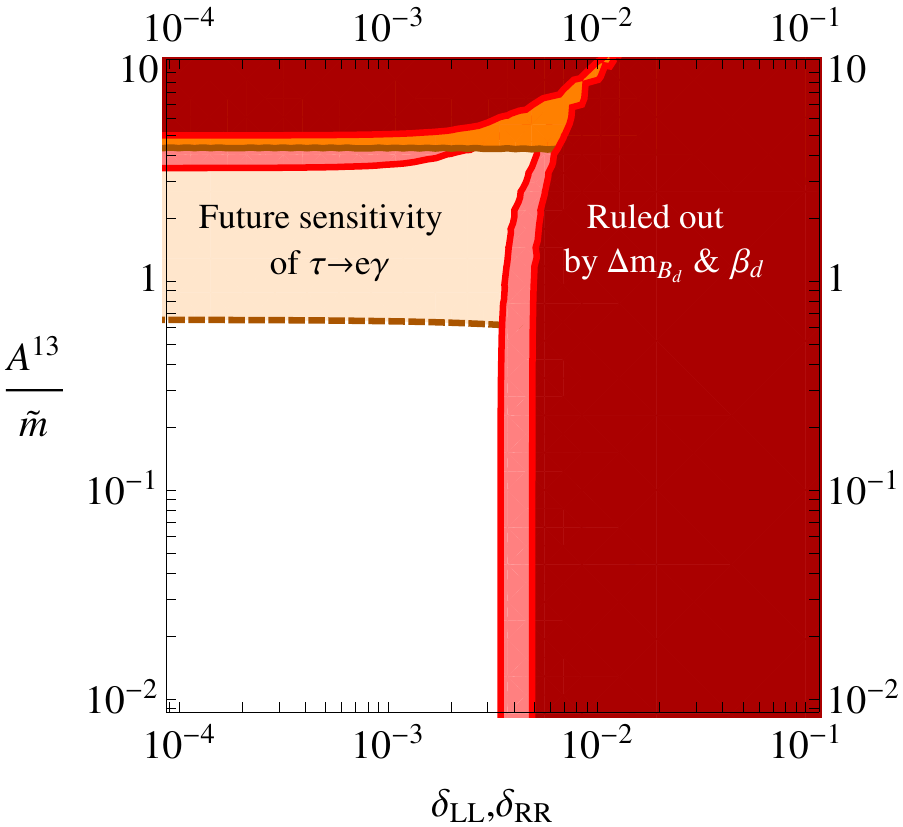} 
\label{TauEDeltaOlogyLargeXBdLRnonZeroa.FIG}
}
\subfloat[Destructive interference for $\tau\to e \gamma$.]{
\includegraphics[width=0.49\textwidth]{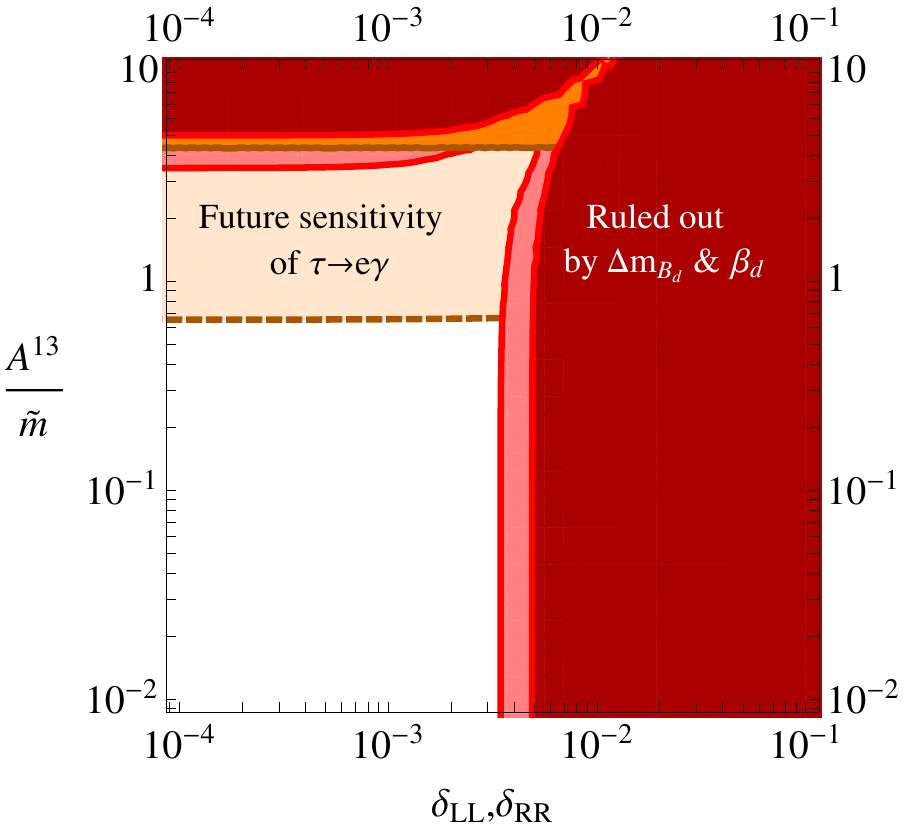}
\label{TauEDeltaOlogyLargeXBdLRnonZerob.FIG}
}
\vspace{.01cm}
\subfloat[Constructive interference for $\tau\to \mu \gamma$.]{
\includegraphics[width=0.49\textwidth]{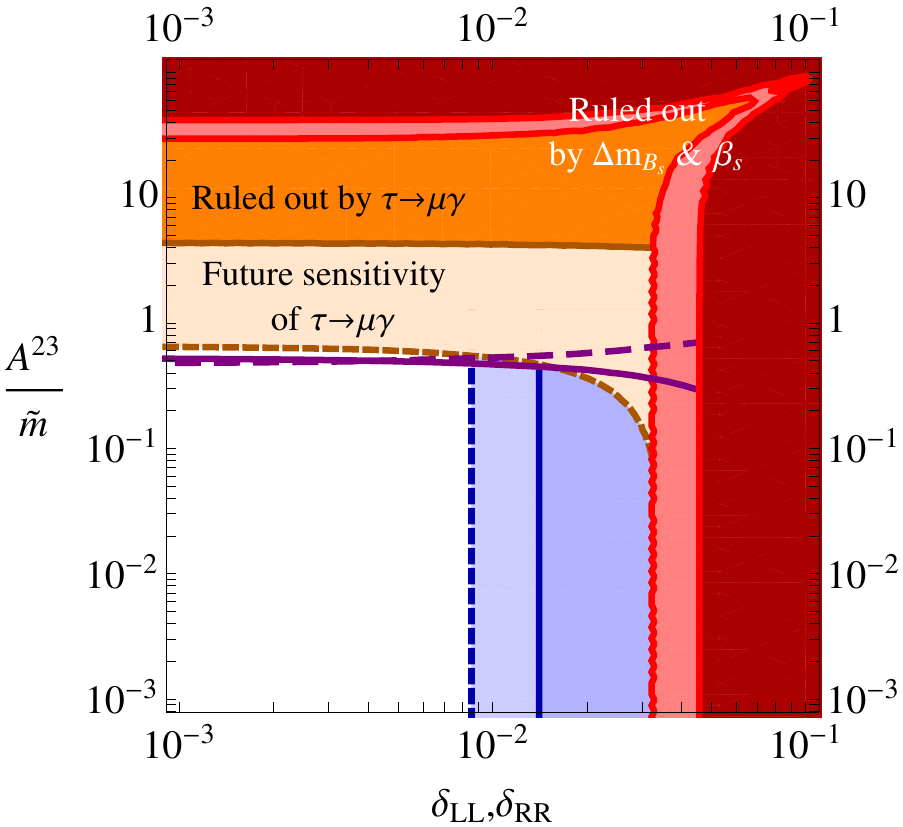} 
\label{TauMuDeltaOlogyLargeXBsLRnonZeroa.FIG}
}
\subfloat[Destructive interference for $\tau\to \mu \gamma$.]{
\includegraphics[width=0.49\textwidth]{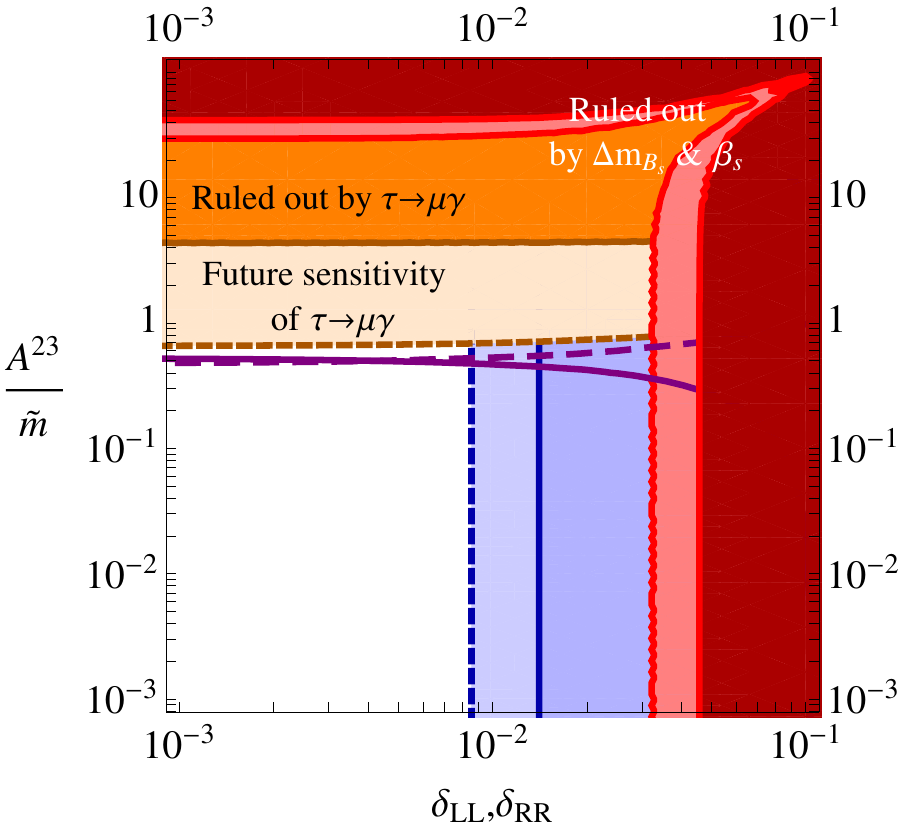}
\label{TauMuDeltaOlogyLargeXBsLRnonZerob.FIG}
}
\caption{$\delta_{LL},\delta_{RR}$ vs. $A^{13} / \tilde{m}$ (upper) and $\delta_{LL},\delta_{RR}$ vs. $A^{23} / \tilde{m}$ (lower) plots.  In the upper plots, we compare the current and future constraints from $\Delta m_{B_{d}}$, $\beta_d$, $\tau \to e \gamma$. The lower plots compare the current and future constraints from $\Delta m_{B_{s}}$, $\beta_s$, $b \to s \gamma$, $\tau \to \mu \gamma$. The dark red region is excluded by meson mixing at  90 \% C.L., and the light red assumes a factor of two improvement. The solid (dashed) purple line shows the limit from $b \to s \gamma$ in the case of constructive (destructive) interference. The dark (light) orange region shows the current (future) sensitivity of $\tau \to e \gamma$  (top) and $\tau \to \mu \gamma$ (bottom). We have set $\mq=\ml = 1$ TeV, $x_{\tilde{g}}=x_\mu = x_{\tilde{W}} = x_{\tilde{B}} \simeq 1$, and $t_\beta = 10$. In both panels, the dark (light)  blue gives a current (future) exclusion from $\mu \to e \gamma$ assuming $\delta^{13}=\delta^{23}$.}
\label{TauEDeltaOlogyLargeXBdLRnonZero.FIG}
\end{figure}

In the $1-3$ sector, we find from Figs. \ref{TauEDeltaOlogyLargeXBdLRnonZeroa.FIG} and \ref{TauEDeltaOlogyLargeXBdLRnonZerob.FIG} that $B_d$ mixing imposes a stronger constraint than $\tau \to e \gamma$ in large regions of parameter space. However, for small $\delta_{LL,RR} \lesssim 2 \times 10^{-3}$, we find that $\tau \to e \gamma$, both in the case of constructive and destructive interference, provides a stronger constraint than $\Delta m_{B_d}$ and $S_{\psi K_s}$ on $A^{13}/\tilde{m}$. We see that while currently the limit is weak:  $A^{13} \lesssim 4 \tilde{m}$, in the future $\tau \to e \gamma$ will be sensitive up to $A^{13} \lesssim 0.6 \tilde{m}$. Additionally we note that under the assumption that $\delta^{13} = \delta^{23}$, $\mu \to e \gamma$ does not improve the constraints on $\delta_{LL, RR}$. 

Similarly for the $2-3$ sector, we see from Figs. \ref{TauMuDeltaOlogyLargeXBsLRnonZeroa.FIG} and \ref{TauMuDeltaOlogyLargeXBsLRnonZerob.FIG} that currently $\tau \to \mu \gamma$ imposes a stronger constraint on $A^{23}/\tilde{m}$ than $B_s$ for $\delta_{LL,RR} \lesssim 3 \times 10^{-2}$ only. In this region the current constraint from $B_s$ mixing gives $A^{23}/\tilde{m} \lesssim 50$, improving to $A^{23}/\tilde{m} \lesssim 40$, while the current constraint from $\tau \to \mu \gamma$ yields $A^{23}/\tilde{m} \lesssim 4$ for both constructive and destructive interference. The future sensitivity of $\tau \to \mu \gamma$ will constrain $A^{23}/\tilde{m} \lesssim 0.6$ for $\delta_{LL,RR} \lesssim 10^{-2}$. However, the strongest constraint comes from $b \to s \gamma$, which bounds $A^{23}/\tilde{m} \lesssim 0.5$ for small $\delta_{LL,RR}$, both in the case of constructive (sgn($m_{\tilde{g}}A^{23}=+$) and destructive (sgn($m_{\tilde{g}}A^{23}=-$)) interference. As in Figs. \ref{TauMuDeltaOlogyLargeXa.FIG} and \ref{TauMuDeltaOlogyLargeXb.FIG}, we see that for $\delta_{LL}=\delta_{RR}$, if $\delta^{13} = \delta^{23}$, $\mu \to e \gamma$ can provide a stronger constraint than both $\tau \to \mu \gamma$ and $B_s$ mixing.

Again, $\mu \to 3e$ can constrain both $1-3$ and $2-3$ transitions in the same way as $\mu \to e \gamma$. We do not include these constraints in Fig. \ref{TauEDeltaOlogyLargeXBdLRnonZero.FIG}, as they can be inferred from the relevant constraints in Fig. \ref{MuEDeltaOlogyLRnonZeroSmallX.FIG}.

\section{Summary}
\label{Conc.SEC}
 
We have analysed various quark- and lepton-flavor violating processes in the absence of new CP violating phases. While the absence of  new CP violating phases is a strong assumption, because LFV measurements are CP conserving, in some ways it provides for the most direct comparison between the two sectors. In general, relaxing this assumption will strengthen  -- considerably in the case of the $1-2$ sector --  the bounds on quark flavor violation.  

In the case of heavy scalars, a scenario well motivated by the observed Higgs boson mass, LFV is a particularly powerful probe on $LL$ flavor violation. In the $1-2$ sector, improvements on bounds on $\mu \rightarrow e \gamma$, $\mu \to 3e$ and $\mu-e$ conversion will probe new parameter space, even accounting for comparable flavor violation in the quark sector.  Similarly, again for $\delta_{LL}$, $\tau \rightarrow \mu \gamma$ is a powerful probe.  $\tau \rightarrow e \gamma$, on the other hand, does not represent as strong a constraint as $B_{d}$ mixing over much of the parameter space (assuming comparable flavor violation the squark and slepton matrices). In an $SU(5)$ GUT context, these bounds can be interpreted as probes of flavor violation in the ${\bf \bar{5}}$ scalar masses.

In the case where all superpartner masses are close to the TeV scale, we obtain similar results on the $LL$ flavor violation.  But in this case, LFV has the opportunity to place limits on $RR$ insertions as well.  These limits can be reinterpreted as probes of  flavor violation in the ${\bf 10}$ scalar masses in an $SU(5)$ GUT. Moreover, for TeV scalars, $LR$ insertions are likely to give important contributions  to LFV observables. Significant bounds already exist on off-diagonal trilinear couplings $A_{ij}$ and these will only strengthen as the experimental sensitivity to LFV improves.

In all, in cases where squark mass matrices are related to slepton mass matrices, quark FCNCs provide a significant constraint. In some areas of parameter space, even improvement of LFV bounds will not make them the most sensitive.  However, there are large swathes of parameter space where LFV has the chance to be a discovery tool.   

\acknowledgments{We would like to thank W.~Altmannshofer for extensive discussions of his previous work, and James Wells for a reading of the manuscript. This work is supported by the U.S. Department of Energy, Office of Science, under grant DE-SC0007859.}

\appendix

\section{Wilson coefficients for $\Delta=2$ processes}
\label{WilsonCoeffs.APP}
\begin{itemize}
\item $x_{\tilde{g}} \simeq 1$
\end{itemize}

At the SUSY scale defined as $M_{SUSY} = \sqrt{m_{\tilde{g}}\mq}$, the squarks and gluinos are integrated out, and the Wilson coefficients are given by \cite{Bagger:1997gg}:

\begin{align}
\nonumber&C_1(M_{SUSY}) = \frac{\alpha^2_s(M_{SUSY})}{216 \mq^2} \left[ (24 x f_6 (x) + 66 \tilde{f}_6 (x))\delta_{LL}^2 \right],\\
\nonumber&C_2(M_{SUSY}) = \frac{\alpha^2_s(M_{SUSY})}{216 \mq^2}  \left[ 204 x f_6 (x) \delta_{RL}^2 \right],\\
\nonumber&C_3(M_{SUSY}) = \frac{\alpha^2_s(M_{SUSY})}{216 \mq^2}  \left[ -36  \tilde{f}_6 (x) \delta_{RL}^2 \right],\\
\nonumber&C_4(M_{SUSY}) = \frac{\alpha^2_s(M_{SUSY})}{216 \mq^2}  \left[ (504 x f_6 (x) - 72 \tilde{f}_6 (x))\delta_{LL} \delta_{RR} - 132 \tilde{f}_6(x) \delta_{LR} \delta_{RL} \right],\\
&C_5(M_{SUSY}) = \frac{\alpha^2_s(M_{SUSY})}{216 \mq^2} \left[  (24 x f_6 (x) +120 \tilde{f}_6 (x))\delta_{LL} \delta_{RR} - 180 \tilde{f}_6(x) \delta_{LR} \delta_{RL} \right]
\end{align}
where the $\delta_{XY}$ are mass insertions, and the loop functions $f_6(x)$ and $\tilde{f}_6(s)$ are given by

\begin{align}
\nonumber f_6(x) = \frac{6(1+3x)\log x + x^3 - 9x^2 -9x + 17}{6(x-1)^5},\\
 \tilde{f}_6(x) = \frac{6x(1+x)\log x - x^3 - 9x^2 + 9x + 1}{3(x-1)^5}
\end{align}
The $\tilde{C}_i$ are obtained by swapping L $\leftrightarrow$ R everywhere for $i=1,2,3$.

At the hadronic scale, the Wilson coefficients are given by

\begin{align}
\nonumber &C_1 (\mu_{had}) = \eta_1 C_1(M_{SUSY}),\\
\nonumber &C_2 (\mu_{had}) = \eta_{22} C_2(M_{SUSY})+\eta_{23} C_3(M_{SUSY}),\\
\nonumber &C_3 (\mu_{had}) = \eta_{32} C_2(M_{SUSY})+\eta_{33} C_3(M_{SUSY}),\\
\nonumber &C_4 (\mu_{had}) = \eta_{4} C_4(M_{SUSY})+\frac{1}{3}(\eta_{4}-\eta_{5}) C_5(M_{SUSY}),\\
 &C_5 (\mu_{had}) = \eta_5 C_5(M_{SUSY})
\end{align}
where
\begin{align}
\nonumber &\eta_1 = \left( \frac{\alpha_s (m_c)}{\alpha_s (\mu_{had})}\right)^{6/27}\left( \frac{\alpha_s (m_b)}{\alpha_s (m_c)}\right)^{6/25}\left( \frac{\alpha_s (m_t)}{\alpha_s (m_b)}\right)^{6/23}\left( \frac{\alpha_s (M_{SUSY})}{\alpha_s (m_t)}\right)^{6/21},\\
\nonumber & \eta_{22} = 0.983 \eta_2 + 0.017 \eta_3, ~~ \eta_{23} = -0.258 \eta_2 + 0.258 \eta_3,\\
\nonumber & \eta_{32} = -0.064 \eta_2 + 0.064 \eta_3, ~~ \eta_{33} = 0.017 \eta_2 + 0.983 \eta_3,\\
& \eta_2 = \eta_1^{-2.42}, ~~ \eta_3 = \eta_1^{2.75}, ~~ \eta_4 = \eta_1^{-4}, ~~ \eta_5 = \eta_1^{1/2}
\end{align}

\begin{itemize}
\item $x_{\tilde{g}} \ll 1$
\end{itemize}

In this case one integrates out the squarks at $M_{SUSY} = \mq$, then run down to the gluino mass scale, at which point the gluino is integrated out before running down to the hadronic scale. The Wilson coefficients at the hadronic scale have been computed to be \cite{Bagger:1997gg}

\begin{align}
\nonumber & C_1(\mu_{had}) = \frac{\alpha^2_s(M_{SUSY})}{216 \mq^2} \left[ -22 \delta_{LL}^2 \kappa_1 \right],\\
\nonumber & \tilde{C}_1(\mu_{had}) = \frac{\alpha^2_s(M_{SUSY})}{216 \mq^2} \left[ -22 \delta_{RR}^2 \kappa_1 \right],\\
\nonumber & C_4(\mu_{had}) = \frac{\alpha^2_s(M_{SUSY})}{216 \mq^2} \left[ \delta_{LL}\delta_{RR} \frac{8}{3}(4\kappa_4 + 5 \kappa_5) + \delta_{LR}\delta_{RL}(64 \kappa_4 - 20 \kappa_5) \right],\\
& C_5(\mu_{had}) = \frac{\alpha^2_s(M_{SUSY})}{216 \mq^2} \left[ \delta_{LL}\delta_{RR}(-40 \kappa_5) + \delta_{LR}\delta_{RL}(60 \kappa_5) \right]
\end{align}
where
\begin{align}
\nonumber &\kappa_1 = \left( \frac{\alpha_s (m_c)}{\alpha_s (\mu_{had})}\right)^{6/27}\left( \frac{\alpha_s (m_b)}{\alpha_s (m_c)}\right)^{6/25}\left( \frac{\alpha_s (m_t)}{\alpha_s (m_b)}\right)^{6/23}\left( \frac{\alpha_s (m_{\tilde{g}})}{\alpha_s (m_t)}\right)^{6/21}\left( \frac{\alpha_s (\mq)}{\alpha_s (m_{\tilde{g}})}\right)^{6/15},\\
&\kappa_4 = \kappa_1^{-4},~~ \kappa_5 = \kappa_1^{1/2}
\end{align}
Note the power of $\left( \frac{\alpha_s (\mq)}{\alpha_s (m_{\tilde{g}})}\right)$ is $6/15$, and not $6/13$ as in \cite{Bagger:1997gg}. This is due to the assumption in \cite{Bagger:1997gg} that the third generation of squarks would be of a similar mass as the gluino. Removing this assumption changes the beta function coefficient.
\begin{itemize}
\item $x_{\tilde{g}} \gg 1$
\end{itemize}

Contrary to the case where the gluino is considerably lighter than the squarks, in this case the gluino is integrated out first at $m_{\tilde{g}}$, then the squarks are integrated out at $\mq$ before evolving down to the hadronic scale. The Wilson coefficients at the hadronic scale are given by

\begin{align}
\nonumber &C_1 (\mu_{had}) = \frac{\alpha^2_s(M_{SUSY})}{216 \mq^2} \left(4 \varepsilon_3^2 \eta'_1 \right) \delta_{LL}^2,\\
\nonumber &C_2 (\mu_{had}) = \frac{\alpha^2_s(M_{SUSY})}{216 \mq^2} \left( \left(\frac{2}{3}(64\varepsilon_1^2 - \varepsilon_2^2)-8\varepsilon_3^2\right) \eta'_{22} + (2\varepsilon_2^2 - 8 \varepsilon_3^2) \eta'_{23}  \right) \delta_{RL}^2,\\
\nonumber &C_3 (\mu_{had}) = \frac{\alpha^2_s(M_{SUSY})}{216 \mq^2} \left( \left(\frac{2}{3}(64\varepsilon_1^2 - \varepsilon_2^2)-8\varepsilon_3^2\right) \eta'_{32} + (2\varepsilon_2^2 - 8 \varepsilon_3^2) \eta'_{33}  \right) \delta_{RL}^2,\\
\nonumber &C_4 (\mu_{had}) = \frac{\alpha^2_s(M_{SUSY})}{216 \mq^2} \left( \frac{4}{3} (64\varepsilon_1^2 \eta'_4 - \varepsilon_2^2\eta'_5)\right) \delta_{LL}\delta_{RR},\\
&C_5 (\mu_{had}) = \frac{\alpha^2_s(M_{SUSY})}{216 \mq^2} \left(4 \varepsilon_2^2 \eta'_5 \right) \delta_{LL} \delta_{RR}
\end{align}
where $\eta'_i$ are the same as the $\eta_i$ in the $x_{\tilde{g}} \simeq 1$ case, and 
\begin{align}
\varepsilon_1 = \left( \frac{\alpha_s(\mq)}{\alpha_s(m_{\tilde{g}})}\right)^{-8/5}, ~~ \varepsilon_2 = \varepsilon_1^{7/16},~~\varepsilon_3 = \varepsilon_1^{3/8}
\end{align}
and $\tilde{C}_i$ are given by interchange of L and R for $i=1,2,3$.

\section{Loop functions for $\ell_i \to \ell_j \gamma$}
\label{LFmuEgam.APP}

We reproduce here the loop functions for the calculation of the branching ratio of $\ell_i \to \ell_j \gamma$.
\begin{align}
g_1(x) &= \frac{1+16x + 7x^2}{(1-x)^4} + \frac{2x(4+7x + x^2)}{(1-x)^5}\log x
\\ g_2(x,y) &=-\frac{11 + 7(x+y) - 54 xy + 11(x^2y + y^2 x) + 7x^2 y^2}{4(1-x)^3(1-y)^3}  \\ \nonumber&+ \frac{x(2+6x+x^2)}{2(1-x)^4(y-x)}\log{x} + \frac{y(2+6y+y^2)}{2(1-y)^4(x-y)}\log{y}
\\ g_3(x,y) &= - \frac{40 -33(x+y) + 11(x^2 + y^2) + 7(x^2 y + y^2 x) - 10x y}{4(1-x)^3(1-y)^3} \\ \nonumber&+ \frac{2+6x + x^2}{2(1-x)^4(y-x)}\log x + \frac{2+6y + y^2}{2(1-y)^4(x-y)}\log y
\end{align}
\begin{align}
f_{2n}(x) &= \frac{-5 x^2 + 4 x + 1 + 2 x (x + 2) \log x}{
 4 (1 - x)^4}
\\ f_{2n}(x,y) &= f_{2n}(x) - f_{2n}(y)
\\ f_{3n}(x) &=\frac{1 + 9 x - 9 x^2 - x^3 + 6 x (x + 1) \log x}{3 (1 - x)^5}
\\ f_{4n}(x) &= \frac{-3 - 44 x + 36 x^2 + 12 x^3 - x^4 - 12 x (3 x + 2) \log x}{
 6 (1 - x)^6}
\end{align}

\section{Loop functions and overlap integrals for $\mu \to e$ conversion in nuclei}
\label{LFconversion.APP}
We reproduce here the loop functions used for the calculation of $\mu \to e$ conversion for convenience:
\begin{align}
f(x) &= \frac{1}{8(1-x)} + \frac{x\log x}{8(1-x)^2},\\
 f_1(x,y) &= \frac{x^3(3-9y) + (y-3)y^2 + x^2(3y-1)(1+4y) + xy(y(13-11y)-4)}{2(1-x)^2(1-y)^2(x-y)^2}\\ \nonumber&+ \frac{x(2x^3 + 2y^2 + 3xy(1+y) -x^2(1+9y))}{(1-x)^3(x-y)^3}\log x \\ \nonumber&+\frac{y^2(y+x(7y-5)-3x^2)}{(1-y)^3(x-y)^3}\log y,
 \end{align}
 \begin{align}
f_2(x,y) &= \frac{x^3(1-3y) + 3(y-3)y^2 + x(y-3)y(y+4) + x^2(y(13-4y)-11)}{2(1-x)^2(1-y)^2(x-y)^2}\\ \nonumber&+ \frac{x(2x^3 + 2y^2 + 3x^2(1+y) -xy(9+y))}{(1-x)^3(y-x)^3}\log x \\ \nonumber&+\frac{y^2(x^2+x(7-5y)-3y)}{(1-y)^3(y-x)^3}\log y,\\
 f_3(x,y) &= -\frac{12(x+y+x^2+y^2+x^2y+y^2x-6xy}{(1-x)^2(1-y)^2(x-y)^2}\\ \nonumber &+\frac{24x(x^2-y)}{(1-x)^3(y-x)^3}\log x + \frac{24y(y^2-x)}{(1-y)^3(x-y)^3}\log y ,
\end{align}
These loop functions take into account the separation of scale between the gauginos, the $\mu-$term and the scalar masses.
Other loop functions used for the calculation of $\mu \to e$ conversion are:
\begin{align}
f_{\gamma,L}(x) &=  \frac{1-6 x + 18 x^2 - 10x^3 - 3x^4 + 12 x^3 \log x}{36(x-1)^5}-\frac{4(7-18 x + 9 x^2 + 2 x^3 + (3-9x^2)\log x)}{36(x-1)^5}\\
f_{\gamma,R}(x) &=  \frac{1-6 x + 18 x^2 - 10x^3 - 3x^4 + 12 x^3 \log x}{9(x-1)^5} \\
f_{Z,R}(x,y) &=\frac{x(x(1+2x)+2(x-1)\sqrt{x}\sqrt{y}-(2+x)y)}{ (x-1)^3 (x - y)^2}\log x \\&\nonumber-\frac{y(y(1+2y)+2(y-1)\sqrt{x}\sqrt{y}-(2+y)x)}{ (y-1)^3 (x - y)^2}\log y \\ \nonumber&+ \frac{y(5+y)+x^2(1+5y)+x(5+y(5y-22))-4\sqrt{x}\sqrt{y}(y-1)(1-x)}{2 (x-1)^2 (x - y) (y-1)^2}
\end{align}

The overlap integrals which appear in Eq. (\ref{Conv.EQN}) were calculated in \cite{Kitano:2002mt}, and are given here for convenience for $_{13}^{27}$Al:
\begin{itemize}
\item $D=0.0357 (m_\mu)^{5/2}$,
\item $V^{(p)} = 0.0159 (m_\mu)^{5/2}$,
\item $V^{(n)} = 0.0169 (m_\mu)^{5/2}$.
\end{itemize}

\section{Subdominant operator coefficients and loop functions for $\ell_i \to 3 \ell_j$ decays}
\label{Mu3e.APP}

As discussed in Section \ref{mu3e.SEC}, the dipole operators dominate the decay $\ell_i \to 3 \ell_j$. Here we list the sub-dominant photo-penguin, box-type and $Z$-penguin contributions. 

The photo-penguin operator coefficients are closely related to those for $\mu \to e$ conversion, and are:
\begin{align}
\ALP = \frac{-g_2^2}{(4\pi)^2 \ml^2} \delta_{LL}^{\ell_i \ell_j} f_{\gamma,L} (\xw) \ , \\ 
\ARP = \frac{-g_2^2}{(4\pi)^2 \ml^2} \delta_{RR}^{\ell_i \ell_j} f_{\gamma,R} (\xb) \ ,
\end{align}
where the $LH$ contributions arise dominantly from Wino exchange, while the $RH$ contributions arise from Bino exchange. The loop functions $f_{\gamma,L(R)}$ can be found in Appendix \ref{LFconversion.APP}, and are the same that appeared in the $\mu \to e$ conversion process.

The box-type operator coefficients arise due to neutralino/chargino and slepton exchange, in various configurations. The Higgs-mediated diagrams which contribute to $B_2$ and $B_3$ are subdominant in the regime of low to moderate $\tan\beta$ considered here, and thus the dominant coefficients are the $B_1^{L,R}$, given by
\begin{align}
e^2 B_1^L = \frac{g_2^4}{(4\pi)^2}\delta_{LL}^{\ell_i \ell_j} f_{Box,L}(\xw) \ , \\
e^2 B_1^R = \frac{g_2^4}{(4\pi)^2}\delta_{RR}^{\ell_i \ell_j} f_{Box,R}(\xw) \ ,
\end{align}
where the loop functions $f_{Box,L(R)}$ are given below.

The $Z$-penguin operator coefficients which give rise to $\ell_i \to 3 \ell_j$ decays are the following:
\begin{align}
\nonumber F_{LL} = &\frac{g_2^2}{(4\pi)^2}\frac{1}{4 \sin^2\theta_W}\delta_{LL}^{\ell_i\ell_j}\left(-\frac{1}{2} + \sin^2\theta_W\right) \\ &\times \left\{\cos^2\beta f_1\left(\xw,x_\mu \right) + \sin^2\beta f_2\left(\xw,x_\mu \right) + \text{sgn}(\mu M_2)\sqrt{\xw x_\mu}\sin\beta \cos \beta f_3\left(\xw,x_\mu \right) \right\} \ ,\\
 F_{LR} =& F_{LL} \times \frac{\sin^2\theta_W}{\left(-\frac{1}{2} + \sin^2\theta_W\right)}  \ ,\\
 F_{RR} = &\frac{g_1^2}{(4\pi)^2 \ml^2}\tan^2 \theta_W \delta^{\ell_i \ell_j}_{RR} \cos{2\beta} f_{Z,R}(\xb,\xmu) \ ,\\
 F_{RL} =& F_{RR} \times \frac{\left(-\frac{1}{2} + \sin^2\theta_W\right)}{\sin^2\theta_W}\ ,
\end{align}
where the loop functions $f_{1,2,3}, ~ f_{Z,R}$ are the same loop functions as in $\mu \to e$ conversion, and are found in Appendix \ref{LFconversion.APP}.

Loop functions which appear in the calculation for $\ell_i \to 3 \ell_j$ are:
\begin{align}
f_{Box,L}(x) = \frac{5+(4-9x) x + 2 x (6+x) \log x}{8(x-1)^3} \\
f_{Box,R}(x) = \frac{1+(4-5x) x + 2 x (2+x) \log x}{8(x-1)^3}
\end{align}

\bibliography{LFVvsQFV}

\begin{thebibliography}{72}%
\makeatletter
\providecommand \@ifxundefined [1]{%
 \@ifx{#1\undefined}
}%
\providecommand \@ifnum [1]{%
 \ifnum #1\expandafter \@firstoftwo
 \else \expandafter \@secondoftwo
 \fi
}%
\providecommand \@ifx [1]{%
 \ifx #1\expandafter \@firstoftwo
 \else \expandafter \@secondoftwo
 \fi
}%
\providecommand \natexlab [1]{#1}%
\providecommand \enquote  [1]{``#1''}%
\providecommand \bibnamefont  [1]{#1}%
\providecommand \bibfnamefont [1]{#1}%
\providecommand \citenamefont [1]{#1}%
\providecommand \href@noop [0]{\@secondoftwo}%
\providecommand \href [0]{\begingroup \@sanitize@url \@href}%
\providecommand \@href[1]{\@@startlink{#1}\@@href}%
\providecommand \@@href[1]{\endgroup#1\@@endlink}%
\providecommand \@sanitize@url [0]{\catcode `\\12\catcode `\$12\catcode
  `\&12\catcode `\#12\catcode `\^12\catcode `\_12\catcode `\%12\relax}%
\providecommand \@@startlink[1]{}%
\providecommand \@@endlink[0]{}%
\providecommand \url  [0]{\begingroup\@sanitize@url \@url }%
\providecommand \@url [1]{\endgroup\@href {#1}{\urlprefix }}%
\providecommand \urlprefix  [0]{URL }%
\providecommand \Eprint [0]{\href }%
\providecommand \doibase [0]{http://dx.doi.org/}%
\providecommand \selectlanguage [0]{\@gobble}%
\providecommand \bibinfo  [0]{\@secondoftwo}%
\providecommand \bibfield  [0]{\@secondoftwo}%
\providecommand \translation [1]{[#1]}%
\providecommand \BibitemOpen [0]{}%
\providecommand \bibitemStop [0]{}%
\providecommand \bibitemNoStop [0]{.\EOS\space}%
\providecommand \EOS [0]{\spacefactor3000\relax}%
\providecommand \BibitemShut  [1]{\csname bibitem#1\endcsname}%
\let\auto@bib@innerbib\@empty
\bibitem [{\citenamefont {Ellis}\ and\ \citenamefont
  {Nanopoulos}(1982)}]{Ellis:1981ts}%
  \BibitemOpen
  \bibfield  {author} {\bibinfo {author} {\bibfnamefont {J.~R.}\ \bibnamefont
  {Ellis}}\ and\ \bibinfo {author} {\bibfnamefont {D.~V.}\ \bibnamefont
  {Nanopoulos}},\ }\href {\doibase 10.1016/0370-2693(82)90948-0} {\bibfield
  {journal} {\bibinfo  {journal} {Phys. Lett.}\ }\textbf {\bibinfo {volume}
  {B110}},\ \bibinfo {pages} {44} (\bibinfo {year} {1982})}\BibitemShut
  {NoStop}%
\bibitem [{\citenamefont {Gabbiani}\ and\ \citenamefont
  {Masiero}(1989)}]{Gabbiani:1988rb}%
  \BibitemOpen
  \bibfield  {author} {\bibinfo {author} {\bibfnamefont {F.}~\bibnamefont
  {Gabbiani}}\ and\ \bibinfo {author} {\bibfnamefont {A.}~\bibnamefont
  {Masiero}},\ }\href {\doibase 10.1016/0550-3213(89)90492-6} {\bibfield
  {journal} {\bibinfo  {journal} {Nucl. Phys.}\ }\textbf {\bibinfo {volume}
  {B322}},\ \bibinfo {pages} {235} (\bibinfo {year} {1989})}\BibitemShut
  {NoStop}%
\bibitem [{\citenamefont {Hisano}\ \emph {et~al.}(1996)\citenamefont {Hisano},
  \citenamefont {Moroi}, \citenamefont {Tobe},\ and\ \citenamefont
  {Yamaguchi}}]{Hisano:1995cp}%
  \BibitemOpen
  \bibfield  {author} {\bibinfo {author} {\bibfnamefont {J.}~\bibnamefont
  {Hisano}}, \bibinfo {author} {\bibfnamefont {T.}~\bibnamefont {Moroi}},
  \bibinfo {author} {\bibfnamefont {K.}~\bibnamefont {Tobe}}, \ and\ \bibinfo
  {author} {\bibfnamefont {M.}~\bibnamefont {Yamaguchi}},\ }\href {\doibase
  10.1103/PhysRevD.53.2442} {\bibfield  {journal} {\bibinfo  {journal} {Phys.
  Rev.}\ }\textbf {\bibinfo {volume} {D53}},\ \bibinfo {pages} {2442} (\bibinfo
  {year} {1996})},\ \Eprint {http://arxiv.org/abs/hep-ph/9510309}
  {arXiv:hep-ph/9510309 [hep-ph]} \BibitemShut {NoStop}%
\bibitem [{\citenamefont {Gabbiani}\ \emph {et~al.}(1996)\citenamefont
  {Gabbiani}, \citenamefont {Gabrielli}, \citenamefont {Masiero},\ and\
  \citenamefont {Silvestrini}}]{Gabbiani:1996hi}%
  \BibitemOpen
  \bibfield  {author} {\bibinfo {author} {\bibfnamefont {F.}~\bibnamefont
  {Gabbiani}}, \bibinfo {author} {\bibfnamefont {E.}~\bibnamefont {Gabrielli}},
  \bibinfo {author} {\bibfnamefont {A.}~\bibnamefont {Masiero}}, \ and\
  \bibinfo {author} {\bibfnamefont {L.}~\bibnamefont {Silvestrini}},\ }\href
  {\doibase 10.1016/0550-3213(96)00390-2} {\bibfield  {journal} {\bibinfo
  {journal} {Nucl. Phys.}\ }\textbf {\bibinfo {volume} {B477}},\ \bibinfo
  {pages} {321} (\bibinfo {year} {1996})},\ \Eprint
  {http://arxiv.org/abs/hep-ph/9604387} {arXiv:hep-ph/9604387 [hep-ph]}
  \BibitemShut {NoStop}%
\bibitem [{\citenamefont {Arganda}\ and\ \citenamefont
  {Herrero}(2006)}]{Arganda:2005ji}%
  \BibitemOpen
  \bibfield  {author} {\bibinfo {author} {\bibfnamefont {E.}~\bibnamefont
  {Arganda}}\ and\ \bibinfo {author} {\bibfnamefont {M.~J.}\ \bibnamefont
  {Herrero}},\ }\href {\doibase 10.1103/PhysRevD.73.055003} {\bibfield
  {journal} {\bibinfo  {journal} {Phys. Rev.}\ }\textbf {\bibinfo {volume}
  {D73}},\ \bibinfo {pages} {055003} (\bibinfo {year} {2006})},\ \Eprint
  {http://arxiv.org/abs/hep-ph/0510405} {arXiv:hep-ph/0510405 [hep-ph]}
  \BibitemShut {NoStop}%
\bibitem [{\citenamefont {Raidal}\ \emph {et~al.}(2008)\citenamefont {Raidal}
  \emph {et~al.}}]{Raidal:2008jk}%
  \BibitemOpen
  \bibfield  {author} {\bibinfo {author} {\bibfnamefont {M.}~\bibnamefont
  {Raidal}} \emph {et~al.},\ }\bibfield  {booktitle} {\emph {\bibinfo
  {booktitle} {{Flavor in the era of the LHC. Proceedings, CERN Workshop,
  Geneva, Switzerland, November 2005-March 2007}}},\ }\href {\doibase
  10.1140/epjc/s10052-008-0715-2} {\bibfield  {journal} {\bibinfo  {journal}
  {Eur. Phys. J.}\ }\textbf {\bibinfo {volume} {C57}},\ \bibinfo {pages} {13}
  (\bibinfo {year} {2008})},\ \Eprint {http://arxiv.org/abs/0801.1826}
  {arXiv:0801.1826 [hep-ph]} \BibitemShut {NoStop}%
\bibitem [{\citenamefont {Altmannshofer}\ \emph {et~al.}(2010)\citenamefont
  {Altmannshofer}, \citenamefont {Buras}, \citenamefont {Gori}, \citenamefont
  {Paradisi},\ and\ \citenamefont {Straub}}]{Altmannshofer:2009ne}%
  \BibitemOpen
  \bibfield  {author} {\bibinfo {author} {\bibfnamefont {W.}~\bibnamefont
  {Altmannshofer}}, \bibinfo {author} {\bibfnamefont {A.~J.}\ \bibnamefont
  {Buras}}, \bibinfo {author} {\bibfnamefont {S.}~\bibnamefont {Gori}},
  \bibinfo {author} {\bibfnamefont {P.}~\bibnamefont {Paradisi}}, \ and\
  \bibinfo {author} {\bibfnamefont {D.~M.}\ \bibnamefont {Straub}},\ }\href
  {\doibase 10.1016/j.nuclphysb.2009.12.019} {\bibfield  {journal} {\bibinfo
  {journal} {Nucl. Phys.}\ }\textbf {\bibinfo {volume} {B830}},\ \bibinfo
  {pages} {17} (\bibinfo {year} {2010})},\ \Eprint
  {http://arxiv.org/abs/0909.1333} {arXiv:0909.1333 [hep-ph]} \BibitemShut
  {NoStop}%
\bibitem [{\citenamefont {Isidori}\ \emph {et~al.}(2010)\citenamefont
  {Isidori}, \citenamefont {Nir},\ and\ \citenamefont
  {Perez}}]{Isidori:2010kg}%
  \BibitemOpen
  \bibfield  {author} {\bibinfo {author} {\bibfnamefont {G.}~\bibnamefont
  {Isidori}}, \bibinfo {author} {\bibfnamefont {Y.}~\bibnamefont {Nir}}, \ and\
  \bibinfo {author} {\bibfnamefont {G.}~\bibnamefont {Perez}},\ }\href
  {\doibase 10.1146/annurev.nucl.012809.104534} {\bibfield  {journal} {\bibinfo
   {journal} {Ann. Rev. Nucl. Part. Sci.}\ }\textbf {\bibinfo {volume} {60}},\
  \bibinfo {pages} {355} (\bibinfo {year} {2010})},\ \Eprint
  {http://arxiv.org/abs/1002.0900} {arXiv:1002.0900 [hep-ph]} \BibitemShut
  {NoStop}%
\bibitem [{\citenamefont {Altmannshofer}\ \emph
  {et~al.}(2013{\natexlab{a}})\citenamefont {Altmannshofer}, \citenamefont
  {Harnik},\ and\ \citenamefont {Zupan}}]{Altmannshofer:2013lfa}%
  \BibitemOpen
  \bibfield  {author} {\bibinfo {author} {\bibfnamefont {W.}~\bibnamefont
  {Altmannshofer}}, \bibinfo {author} {\bibfnamefont {R.}~\bibnamefont
  {Harnik}}, \ and\ \bibinfo {author} {\bibfnamefont {J.}~\bibnamefont
  {Zupan}},\ }\href {\doibase 10.1007/JHEP11(2013)202} {\bibfield  {journal}
  {\bibinfo  {journal} {JHEP}\ }\textbf {\bibinfo {volume} {11}},\ \bibinfo
  {pages} {202} (\bibinfo {year} {2013}{\natexlab{a}})},\ \Eprint
  {http://arxiv.org/abs/1308.3653} {arXiv:1308.3653 [hep-ph]} \BibitemShut
  {NoStop}%
\bibitem [{\citenamefont {Arana-Catania}\ \emph {et~al.}(2013)\citenamefont
  {Arana-Catania}, \citenamefont {Heinemeyer},\ and\ \citenamefont
  {Herrero}}]{Arana-Catania:2013ggc}%
  \BibitemOpen
  \bibfield  {author} {\bibinfo {author} {\bibfnamefont {M.}~\bibnamefont
  {Arana-Catania}}, \bibinfo {author} {\bibfnamefont {S.}~\bibnamefont
  {Heinemeyer}}, \ and\ \bibinfo {author} {\bibfnamefont {M.~J.}\ \bibnamefont
  {Herrero}},\ }\href {\doibase 10.1103/PhysRevD.88.015026} {\bibfield
  {journal} {\bibinfo  {journal} {Phys. Rev.}\ }\textbf {\bibinfo {volume}
  {D88}},\ \bibinfo {pages} {015026} (\bibinfo {year} {2013})},\ \Eprint
  {http://arxiv.org/abs/1304.2783} {arXiv:1304.2783 [hep-ph]} \BibitemShut
  {NoStop}%
\bibitem [{\citenamefont {Arana-Catania}\ \emph {et~al.}(2014)\citenamefont
  {Arana-Catania}, \citenamefont {Heinemeyer},\ and\ \citenamefont
  {Herrero}}]{Arana-Catania:2014ooa}%
  \BibitemOpen
  \bibfield  {author} {\bibinfo {author} {\bibfnamefont {M.}~\bibnamefont
  {Arana-Catania}}, \bibinfo {author} {\bibfnamefont {S.}~\bibnamefont
  {Heinemeyer}}, \ and\ \bibinfo {author} {\bibfnamefont {M.~J.}\ \bibnamefont
  {Herrero}},\ }\href {\doibase 10.1103/PhysRevD.90.075003} {\bibfield
  {journal} {\bibinfo  {journal} {Phys. Rev.}\ }\textbf {\bibinfo {volume}
  {D90}},\ \bibinfo {pages} {075003} (\bibinfo {year} {2014})},\ \Eprint
  {http://arxiv.org/abs/1405.6960} {arXiv:1405.6960 [hep-ph]} \BibitemShut
  {NoStop}%
\bibitem [{\citenamefont {Ciuchini}\ \emph {et~al.}(2007)\citenamefont
  {Ciuchini}, \citenamefont {Masiero}, \citenamefont {Paradisi}, \citenamefont
  {Silvestrini}, \citenamefont {Vempati},\ and\ \citenamefont
  {Vives}}]{Ciuchini:2007ha}%
  \BibitemOpen
  \bibfield  {author} {\bibinfo {author} {\bibfnamefont {M.}~\bibnamefont
  {Ciuchini}}, \bibinfo {author} {\bibfnamefont {A.}~\bibnamefont {Masiero}},
  \bibinfo {author} {\bibfnamefont {P.}~\bibnamefont {Paradisi}}, \bibinfo
  {author} {\bibfnamefont {L.}~\bibnamefont {Silvestrini}}, \bibinfo {author}
  {\bibfnamefont {S.~K.}\ \bibnamefont {Vempati}}, \ and\ \bibinfo {author}
  {\bibfnamefont {O.}~\bibnamefont {Vives}},\ }\href {\doibase
  10.1016/j.nuclphysb.2007.05.032} {\bibfield  {journal} {\bibinfo  {journal}
  {Nucl. Phys.}\ }\textbf {\bibinfo {volume} {B783}},\ \bibinfo {pages} {112}
  (\bibinfo {year} {2007})},\ \Eprint {http://arxiv.org/abs/hep-ph/0702144}
  {arXiv:hep-ph/0702144 [HEP-PH]} \BibitemShut {NoStop}%
\bibitem [{\citenamefont {Hisano}\ \emph {et~al.}(1998)\citenamefont {Hisano},
  \citenamefont {Nomura},\ and\ \citenamefont {Yanagida}}]{Hisano:1997tc}%
  \BibitemOpen
  \bibfield  {author} {\bibinfo {author} {\bibfnamefont {J.}~\bibnamefont
  {Hisano}}, \bibinfo {author} {\bibfnamefont {D.}~\bibnamefont {Nomura}}, \
  and\ \bibinfo {author} {\bibfnamefont {T.}~\bibnamefont {Yanagida}},\ }\href
  {\doibase 10.1016/S0370-2693(98)00929-0} {\bibfield  {journal} {\bibinfo
  {journal} {Phys. Lett.}\ }\textbf {\bibinfo {volume} {B437}},\ \bibinfo
  {pages} {351} (\bibinfo {year} {1998})},\ \Eprint
  {http://arxiv.org/abs/hep-ph/9711348} {arXiv:hep-ph/9711348 [hep-ph]}
  \BibitemShut {NoStop}%
\bibitem [{\citenamefont {Casas}\ and\ \citenamefont
  {Ibarra}(2001)}]{Casas:2001sr}%
  \BibitemOpen
  \bibfield  {author} {\bibinfo {author} {\bibfnamefont {J.~A.}\ \bibnamefont
  {Casas}}\ and\ \bibinfo {author} {\bibfnamefont {A.}~\bibnamefont {Ibarra}},\
  }\href {\doibase 10.1016/S0550-3213(01)00475-8} {\bibfield  {journal}
  {\bibinfo  {journal} {Nucl. Phys.}\ }\textbf {\bibinfo {volume} {B618}},\
  \bibinfo {pages} {171} (\bibinfo {year} {2001})},\ \Eprint
  {http://arxiv.org/abs/hep-ph/0103065} {arXiv:hep-ph/0103065 [hep-ph]}
  \BibitemShut {NoStop}%
\bibitem [{\citenamefont {Ellis}\ \emph
  {et~al.}(2002{\natexlab{a}})\citenamefont {Ellis}, \citenamefont {Hisano},
  \citenamefont {Lola},\ and\ \citenamefont {Raidal}}]{Ellis:2001xt}%
  \BibitemOpen
  \bibfield  {author} {\bibinfo {author} {\bibfnamefont {J.~R.}\ \bibnamefont
  {Ellis}}, \bibinfo {author} {\bibfnamefont {J.}~\bibnamefont {Hisano}},
  \bibinfo {author} {\bibfnamefont {S.}~\bibnamefont {Lola}}, \ and\ \bibinfo
  {author} {\bibfnamefont {M.}~\bibnamefont {Raidal}},\ }\href {\doibase
  10.1016/S0550-3213(01)00583-1} {\bibfield  {journal} {\bibinfo  {journal}
  {Nucl. Phys.}\ }\textbf {\bibinfo {volume} {B621}},\ \bibinfo {pages} {208}
  (\bibinfo {year} {2002}{\natexlab{a}})},\ \Eprint
  {http://arxiv.org/abs/hep-ph/0109125} {arXiv:hep-ph/0109125 [hep-ph]}
  \BibitemShut {NoStop}%
\bibitem [{\citenamefont {Ellis}\ \emph
  {et~al.}(2002{\natexlab{b}})\citenamefont {Ellis}, \citenamefont {Hisano},
  \citenamefont {Raidal},\ and\ \citenamefont {Shimizu}}]{Ellis:2002fe}%
  \BibitemOpen
  \bibfield  {author} {\bibinfo {author} {\bibfnamefont {J.~R.}\ \bibnamefont
  {Ellis}}, \bibinfo {author} {\bibfnamefont {J.}~\bibnamefont {Hisano}},
  \bibinfo {author} {\bibfnamefont {M.}~\bibnamefont {Raidal}}, \ and\ \bibinfo
  {author} {\bibfnamefont {Y.}~\bibnamefont {Shimizu}},\ }\href {\doibase
  10.1103/PhysRevD.66.115013} {\bibfield  {journal} {\bibinfo  {journal} {Phys.
  Rev.}\ }\textbf {\bibinfo {volume} {D66}},\ \bibinfo {pages} {115013}
  (\bibinfo {year} {2002}{\natexlab{b}})},\ \Eprint
  {http://arxiv.org/abs/hep-ph/0206110} {arXiv:hep-ph/0206110 [hep-ph]}
  \BibitemShut {NoStop}%
\bibitem [{\citenamefont {Antusch}\ \emph {et~al.}(2006)\citenamefont
  {Antusch}, \citenamefont {Arganda}, \citenamefont {Herrero},\ and\
  \citenamefont {Teixeira}}]{Antusch:2006vw}%
  \BibitemOpen
  \bibfield  {author} {\bibinfo {author} {\bibfnamefont {S.}~\bibnamefont
  {Antusch}}, \bibinfo {author} {\bibfnamefont {E.}~\bibnamefont {Arganda}},
  \bibinfo {author} {\bibfnamefont {M.~J.}\ \bibnamefont {Herrero}}, \ and\
  \bibinfo {author} {\bibfnamefont {A.~M.}\ \bibnamefont {Teixeira}},\ }\href
  {\doibase 10.1088/1126-6708/2006/11/090} {\bibfield  {journal} {\bibinfo
  {journal} {JHEP}\ }\textbf {\bibinfo {volume} {11}},\ \bibinfo {pages} {090}
  (\bibinfo {year} {2006})},\ \Eprint {http://arxiv.org/abs/hep-ph/0607263}
  {arXiv:hep-ph/0607263 [hep-ph]} \BibitemShut {NoStop}%
\bibitem [{\citenamefont {Arganda}\ \emph {et~al.}(2007)\citenamefont
  {Arganda}, \citenamefont {Herrero},\ and\ \citenamefont
  {Teixeira}}]{Arganda:2007jw}%
  \BibitemOpen
  \bibfield  {author} {\bibinfo {author} {\bibfnamefont {E.}~\bibnamefont
  {Arganda}}, \bibinfo {author} {\bibfnamefont {M.~J.}\ \bibnamefont
  {Herrero}}, \ and\ \bibinfo {author} {\bibfnamefont {A.~M.}\ \bibnamefont
  {Teixeira}},\ }\href {\doibase 10.1088/1126-6708/2007/10/104} {\bibfield
  {journal} {\bibinfo  {journal} {JHEP}\ }\textbf {\bibinfo {volume} {10}},\
  \bibinfo {pages} {104} (\bibinfo {year} {2007})},\ \Eprint
  {http://arxiv.org/abs/0707.2955} {arXiv:0707.2955 [hep-ph]} \BibitemShut
  {NoStop}%
\bibitem [{\citenamefont {Mohapatra}\ and\ \citenamefont
  {Smirnov}(2006)}]{Mohapatra:2006gs}%
  \BibitemOpen
  \bibfield  {author} {\bibinfo {author} {\bibfnamefont {R.~N.}\ \bibnamefont
  {Mohapatra}}\ and\ \bibinfo {author} {\bibfnamefont {A.~Y.}\ \bibnamefont
  {Smirnov}},\ }\bibfield  {booktitle} {\emph {\bibinfo {booktitle}
  {{Elementary particle physics. Proceedings, Corfu Summer Institute,
  CORFU2005, Corfu, Greece, September 4-26, 2005}}},\ }\href {\doibase
  10.1146/annurev.nucl.56.080805.140534} {\bibfield  {journal} {\bibinfo
  {journal} {Ann. Rev. Nucl. Part. Sci.}\ }\textbf {\bibinfo {volume} {56}},\
  \bibinfo {pages} {569} (\bibinfo {year} {2006})},\ \Eprint
  {http://arxiv.org/abs/hep-ph/0603118} {arXiv:hep-ph/0603118 [hep-ph]}
  \BibitemShut {NoStop}%
\bibitem [{\citenamefont {Olive}\ \emph {et~al.}(2014)\citenamefont {Olive}
  \emph {et~al.}}]{Agashe:2014kda}%
  \BibitemOpen
  \bibfield  {author} {\bibinfo {author} {\bibfnamefont {K.~A.}\ \bibnamefont
  {Olive}} \emph {et~al.} (\bibinfo {collaboration} {Particle Data Group}),\
  }\href {\doibase 10.1088/1674-1137/38/9/090001} {\bibfield  {journal}
  {\bibinfo  {journal} {Chin. Phys.}\ }\textbf {\bibinfo {volume} {C38}},\
  \bibinfo {pages} {090001} (\bibinfo {year} {2014})}\BibitemShut {NoStop}%
\bibitem [{\citenamefont {Bai}\ \emph {et~al.}(2014)\citenamefont {Bai},
  \citenamefont {Christ}, \citenamefont {Izubuchi}, \citenamefont {Sachrajda},
  \citenamefont {Soni},\ and\ \citenamefont {Yu}}]{Bai:2014cva}%
  \BibitemOpen
  \bibfield  {author} {\bibinfo {author} {\bibfnamefont {Z.}~\bibnamefont
  {Bai}}, \bibinfo {author} {\bibfnamefont {N.~H.}\ \bibnamefont {Christ}},
  \bibinfo {author} {\bibfnamefont {T.}~\bibnamefont {Izubuchi}}, \bibinfo
  {author} {\bibfnamefont {C.~T.}\ \bibnamefont {Sachrajda}}, \bibinfo {author}
  {\bibfnamefont {A.}~\bibnamefont {Soni}}, \ and\ \bibinfo {author}
  {\bibfnamefont {J.}~\bibnamefont {Yu}},\ }\href {\doibase
  10.1103/PhysRevLett.113.112003} {\bibfield  {journal} {\bibinfo  {journal}
  {Phys. Rev. Lett.}\ }\textbf {\bibinfo {volume} {113}},\ \bibinfo {pages}
  {112003} (\bibinfo {year} {2014})},\ \Eprint {http://arxiv.org/abs/1406.0916}
  {arXiv:1406.0916 [hep-lat]} \BibitemShut {NoStop}%
\bibitem [{\citenamefont {Artuso}\ \emph {et~al.}(2015)\citenamefont {Artuso},
  \citenamefont {Borissov},\ and\ \citenamefont {Lenz}}]{Artuso:2015swg}%
  \BibitemOpen
  \bibfield  {author} {\bibinfo {author} {\bibfnamefont {M.}~\bibnamefont
  {Artuso}}, \bibinfo {author} {\bibfnamefont {G.}~\bibnamefont {Borissov}}, \
  and\ \bibinfo {author} {\bibfnamefont {A.}~\bibnamefont {Lenz}},\ }\href@noop
  {} {\  (\bibinfo {year} {2015})},\ \Eprint {http://arxiv.org/abs/1511.09466}
  {arXiv:1511.09466 [hep-ph]} \BibitemShut {NoStop}%
\bibitem [{\citenamefont {Amhis}\ \emph {et~al.}(2014)\citenamefont {Amhis}
  \emph {et~al.}}]{Amhis:2014hma}%
  \BibitemOpen
  \bibfield  {author} {\bibinfo {author} {\bibfnamefont {Y.}~\bibnamefont
  {Amhis}} \emph {et~al.} (\bibinfo {collaboration} {Heavy Flavor Averaging
  Group (HFAG)}),\ }\href@noop {} {\  (\bibinfo {year} {2014})},\ \Eprint
  {http://arxiv.org/abs/1412.7515} {arXiv:1412.7515 [hep-ex]} \BibitemShut
  {NoStop}%
\bibitem [{\citenamefont {Charles}\ \emph {et~al.}(2005)\citenamefont
  {Charles}, \citenamefont {Hocker}, \citenamefont {Lacker}, \citenamefont
  {Laplace}, \citenamefont {Le~Diberder}, \citenamefont {Malcles},
  \citenamefont {Ocariz}, \citenamefont {Pivk},\ and\ \citenamefont
  {Roos}}]{Charles:2004jd}%
  \BibitemOpen
  \bibfield  {author} {\bibinfo {author} {\bibfnamefont {J.}~\bibnamefont
  {Charles}}, \bibinfo {author} {\bibfnamefont {A.}~\bibnamefont {Hocker}},
  \bibinfo {author} {\bibfnamefont {H.}~\bibnamefont {Lacker}}, \bibinfo
  {author} {\bibfnamefont {S.}~\bibnamefont {Laplace}}, \bibinfo {author}
  {\bibfnamefont {F.~R.}\ \bibnamefont {Le~Diberder}}, \bibinfo {author}
  {\bibfnamefont {J.}~\bibnamefont {Malcles}}, \bibinfo {author} {\bibfnamefont
  {J.}~\bibnamefont {Ocariz}}, \bibinfo {author} {\bibfnamefont
  {M.}~\bibnamefont {Pivk}}, \ and\ \bibinfo {author} {\bibfnamefont
  {L.}~\bibnamefont {Roos}} (\bibinfo {collaboration} {CKMfitter Group}),\
  }\href {\doibase 10.1140/epjc/s2005-02169-1} {\bibfield  {journal} {\bibinfo
  {journal} {Eur. Phys. J.}\ }\textbf {\bibinfo {volume} {C41}},\ \bibinfo
  {pages} {1} (\bibinfo {year} {2005})},\ \Eprint
  {http://arxiv.org/abs/hep-ph/0406184} {arXiv:hep-ph/0406184 [hep-ph]}
  \BibitemShut {NoStop}%
\bibitem [{\citenamefont {Baldini}\ \emph {et~al.}(2016)\citenamefont {Baldini}
  \emph {et~al.}}]{TheMEG:2016wtm}%
  \BibitemOpen
  \bibfield  {author} {\bibinfo {author} {\bibfnamefont {A.~M.}\ \bibnamefont
  {Baldini}} \emph {et~al.} (\bibinfo {collaboration} {MEG}),\ }\href@noop {}
  {\  (\bibinfo {year} {2016})},\ \Eprint {http://arxiv.org/abs/1605.05081}
  {arXiv:1605.05081 [hep-ex]} \BibitemShut {NoStop}%
\bibitem [{\citenamefont {Baldini}\ \emph {et~al.}(2013)\citenamefont {Baldini}
  \emph {et~al.}}]{Baldini:2013ke}%
  \BibitemOpen
  \bibfield  {author} {\bibinfo {author} {\bibfnamefont {A.~M.}\ \bibnamefont
  {Baldini}} \emph {et~al.},\ }\href@noop {} {\  (\bibinfo {year} {2013})},\
  \Eprint {http://arxiv.org/abs/1301.7225} {arXiv:1301.7225 [physics.ins-det]}
  \BibitemShut {NoStop}%
\bibitem [{\citenamefont {Aubert}\ \emph {et~al.}(2010)\citenamefont {Aubert}
  \emph {et~al.}}]{Aubert:2009ag}%
  \BibitemOpen
  \bibfield  {author} {\bibinfo {author} {\bibfnamefont {B.}~\bibnamefont
  {Aubert}} \emph {et~al.} (\bibinfo {collaboration} {BaBar}),\ }\href
  {\doibase 10.1103/PhysRevLett.104.021802} {\bibfield  {journal} {\bibinfo
  {journal} {Phys. Rev. Lett.}\ }\textbf {\bibinfo {volume} {104}},\ \bibinfo
  {pages} {021802} (\bibinfo {year} {2010})},\ \Eprint
  {http://arxiv.org/abs/0908.2381} {arXiv:0908.2381 [hep-ex]} \BibitemShut
  {NoStop}%
\bibitem [{\citenamefont {Hayasaka}(2013)}]{Hayasaka:2013dsa}%
  \BibitemOpen
  \bibfield  {author} {\bibinfo {author} {\bibfnamefont {K.}~\bibnamefont
  {Hayasaka}} (\bibinfo {collaboration} {Belle, Belle-II}),\ }\bibfield
  {booktitle} {\emph {\bibinfo {booktitle} {{Proceedings, 13th International
  Workshop on Neutrino Factories, Superbeams and Beta beams (NuFact11)}}},\
  }\href {\doibase 10.1088/1742-6596/408/1/012069} {\bibfield  {journal}
  {\bibinfo  {journal} {J. Phys. Conf. Ser.}\ }\textbf {\bibinfo {volume}
  {408}},\ \bibinfo {pages} {012069} (\bibinfo {year} {2013})}\BibitemShut
  {NoStop}%
\bibitem [{\citenamefont {Bertl}\ \emph {et~al.}(2006)\citenamefont {Bertl}
  \emph {et~al.}}]{Bertl:2006up}%
  \BibitemOpen
  \bibfield  {author} {\bibinfo {author} {\bibfnamefont {W.~H.}\ \bibnamefont
  {Bertl}} \emph {et~al.} (\bibinfo {collaboration} {SINDRUM II}),\ }\href
  {\doibase 10.1140/epjc/s2006-02582-x} {\bibfield  {journal} {\bibinfo
  {journal} {Eur. Phys. J.}\ }\textbf {\bibinfo {volume} {C47}},\ \bibinfo
  {pages} {337} (\bibinfo {year} {2006})}\BibitemShut {NoStop}%
\bibitem [{\citenamefont {Abrams}\ \emph {et~al.}(2012)\citenamefont {Abrams}
  \emph {et~al.}}]{Abrams:2012er}%
  \BibitemOpen
  \bibfield  {author} {\bibinfo {author} {\bibfnamefont {R.~J.}\ \bibnamefont
  {Abrams}} \emph {et~al.} (\bibinfo {collaboration} {Mu2e}),\ }\href@noop {}
  {\  (\bibinfo {year} {2012})},\ \Eprint {http://arxiv.org/abs/1211.7019}
  {arXiv:1211.7019 [physics.ins-det]} \BibitemShut {NoStop}%
\bibitem [{\citenamefont {Bellgardt}\ \emph {et~al.}(1988)\citenamefont
  {Bellgardt} \emph {et~al.}}]{Bellgardt:1987du}%
  \BibitemOpen
  \bibfield  {author} {\bibinfo {author} {\bibfnamefont {U.}~\bibnamefont
  {Bellgardt}} \emph {et~al.} (\bibinfo {collaboration} {SINDRUM}),\ }\href
  {\doibase 10.1016/0550-3213(88)90462-2} {\bibfield  {journal} {\bibinfo
  {journal} {Nucl. Phys.}\ }\textbf {\bibinfo {volume} {B299}},\ \bibinfo
  {pages} {1} (\bibinfo {year} {1988})}\BibitemShut {NoStop}%
\bibitem [{\citenamefont {Blondel}\ \emph {et~al.}(2013)\citenamefont {Blondel}
  \emph {et~al.}}]{Blondel:2013ia}%
  \BibitemOpen
  \bibfield  {author} {\bibinfo {author} {\bibfnamefont {A.}~\bibnamefont
  {Blondel}} \emph {et~al.},\ }\href@noop {} {\  (\bibinfo {year} {2013})},\
  \Eprint {http://arxiv.org/abs/1301.6113} {arXiv:1301.6113 [physics.ins-det]}
  \BibitemShut {NoStop}%
\bibitem [{\citenamefont {Berger}(2014)}]{Berger:2014vba}%
  \BibitemOpen
  \bibfield  {author} {\bibinfo {author} {\bibfnamefont {N.}~\bibnamefont
  {Berger}} (\bibinfo {collaboration} {Mu3e}),\ }\bibfield  {booktitle} {\emph
  {\bibinfo {booktitle} {{Proceedings, 1st International Conference on Charged
  Lepton Flavor Violation (CLFV)}}},\ }\href {\doibase
  10.1016/j.nuclphysbps.2014.02.007} {\bibfield  {journal} {\bibinfo  {journal}
  {Nucl. Phys. Proc. Suppl.}\ }\textbf {\bibinfo {volume} {248-250}},\ \bibinfo
  {pages} {35} (\bibinfo {year} {2014})}\BibitemShut {NoStop}%
\bibitem [{\citenamefont {Demir}\ \emph {et~al.}(2004)\citenamefont {Demir},
  \citenamefont {Lebedev}, \citenamefont {Olive}, \citenamefont {Pospelov},\
  and\ \citenamefont {Ritz}}]{Demir:2003js}%
  \BibitemOpen
  \bibfield  {author} {\bibinfo {author} {\bibfnamefont {D.~A.}\ \bibnamefont
  {Demir}}, \bibinfo {author} {\bibfnamefont {O.}~\bibnamefont {Lebedev}},
  \bibinfo {author} {\bibfnamefont {K.~A.}\ \bibnamefont {Olive}}, \bibinfo
  {author} {\bibfnamefont {M.}~\bibnamefont {Pospelov}}, \ and\ \bibinfo
  {author} {\bibfnamefont {A.}~\bibnamefont {Ritz}},\ }\href {\doibase
  10.1016/j.nuclphysb.2003.12.026} {\bibfield  {journal} {\bibinfo  {journal}
  {Nucl. Phys.}\ }\textbf {\bibinfo {volume} {B680}},\ \bibinfo {pages} {339}
  (\bibinfo {year} {2004})},\ \Eprint {http://arxiv.org/abs/hep-ph/0311314}
  {arXiv:hep-ph/0311314 [hep-ph]} \BibitemShut {NoStop}%
\bibitem [{\citenamefont {Pospelov}\ and\ \citenamefont
  {Ritz}(2005)}]{Pospelov:2005pr}%
  \BibitemOpen
  \bibfield  {author} {\bibinfo {author} {\bibfnamefont {M.}~\bibnamefont
  {Pospelov}}\ and\ \bibinfo {author} {\bibfnamefont {A.}~\bibnamefont
  {Ritz}},\ }\href {\doibase 10.1016/j.aop.2005.04.002} {\bibfield  {journal}
  {\bibinfo  {journal} {Annals Phys.}\ }\textbf {\bibinfo {volume} {318}},\
  \bibinfo {pages} {119} (\bibinfo {year} {2005})},\ \Eprint
  {http://arxiv.org/abs/hep-ph/0504231} {arXiv:hep-ph/0504231 [hep-ph]}
  \BibitemShut {NoStop}%
\bibitem [{\citenamefont {Giudice}\ and\ \citenamefont
  {Romanino}(2006)}]{Giudice:2005rz}%
  \BibitemOpen
  \bibfield  {author} {\bibinfo {author} {\bibfnamefont {G.~F.}\ \bibnamefont
  {Giudice}}\ and\ \bibinfo {author} {\bibfnamefont {A.}~\bibnamefont
  {Romanino}},\ }\href {\doibase 10.1016/j.physletb.2006.01.027} {\bibfield
  {journal} {\bibinfo  {journal} {Phys. Lett.}\ }\textbf {\bibinfo {volume}
  {B634}},\ \bibinfo {pages} {307} (\bibinfo {year} {2006})},\ \Eprint
  {http://arxiv.org/abs/hep-ph/0510197} {arXiv:hep-ph/0510197 [hep-ph]}
  \BibitemShut {NoStop}%
\bibitem [{\citenamefont {Ellis}\ \emph {et~al.}(2008)\citenamefont {Ellis},
  \citenamefont {Lee},\ and\ \citenamefont {Pilaftsis}}]{Ellis:2008zy}%
  \BibitemOpen
  \bibfield  {author} {\bibinfo {author} {\bibfnamefont {J.~R.}\ \bibnamefont
  {Ellis}}, \bibinfo {author} {\bibfnamefont {J.~S.}\ \bibnamefont {Lee}}, \
  and\ \bibinfo {author} {\bibfnamefont {A.}~\bibnamefont {Pilaftsis}},\ }\href
  {\doibase 10.1088/1126-6708/2008/10/049} {\bibfield  {journal} {\bibinfo
  {journal} {JHEP}\ }\textbf {\bibinfo {volume} {10}},\ \bibinfo {pages} {049}
  (\bibinfo {year} {2008})},\ \Eprint {http://arxiv.org/abs/0808.1819}
  {arXiv:0808.1819 [hep-ph]} \BibitemShut {NoStop}%
\bibitem [{\citenamefont {McKeen}\ \emph {et~al.}(2013)\citenamefont {McKeen},
  \citenamefont {Pospelov},\ and\ \citenamefont {Ritz}}]{McKeen:2013dma}%
  \BibitemOpen
  \bibfield  {author} {\bibinfo {author} {\bibfnamefont {D.}~\bibnamefont
  {McKeen}}, \bibinfo {author} {\bibfnamefont {M.}~\bibnamefont {Pospelov}}, \
  and\ \bibinfo {author} {\bibfnamefont {A.}~\bibnamefont {Ritz}},\ }\href
  {\doibase 10.1103/PhysRevD.87.113002} {\bibfield  {journal} {\bibinfo
  {journal} {Phys. Rev.}\ }\textbf {\bibinfo {volume} {D87}},\ \bibinfo {pages}
  {113002} (\bibinfo {year} {2013})},\ \Eprint {http://arxiv.org/abs/1303.1172}
  {arXiv:1303.1172 [hep-ph]} \BibitemShut {NoStop}%
\bibitem [{\citenamefont {Ellis}\ and\ \citenamefont
  {Kane}(2016)}]{Ellis:2014tea}%
  \BibitemOpen
  \bibfield  {author} {\bibinfo {author} {\bibfnamefont {S.~A.~R.}\
  \bibnamefont {Ellis}}\ and\ \bibinfo {author} {\bibfnamefont {G.~L.}\
  \bibnamefont {Kane}},\ }\href {\doibase 10.1007/JHEP01(2016)077} {\bibfield
  {journal} {\bibinfo  {journal} {JHEP}\ }\textbf {\bibinfo {volume} {01}},\
  \bibinfo {pages} {077} (\bibinfo {year} {2016})},\ \Eprint
  {http://arxiv.org/abs/1405.7719} {arXiv:1405.7719 [hep-ph]} \BibitemShut
  {NoStop}%
\bibitem [{\citenamefont {Kane}\ \emph {et~al.}(2012)\citenamefont {Kane},
  \citenamefont {Kumar}, \citenamefont {Lu},\ and\ \citenamefont
  {Zheng}}]{Kane:2011kj}%
  \BibitemOpen
  \bibfield  {author} {\bibinfo {author} {\bibfnamefont {G.}~\bibnamefont
  {Kane}}, \bibinfo {author} {\bibfnamefont {P.}~\bibnamefont {Kumar}},
  \bibinfo {author} {\bibfnamefont {R.}~\bibnamefont {Lu}}, \ and\ \bibinfo
  {author} {\bibfnamefont {B.}~\bibnamefont {Zheng}},\ }\href {\doibase
  10.1103/PhysRevD.85.075026} {\bibfield  {journal} {\bibinfo  {journal} {Phys.
  Rev.}\ }\textbf {\bibinfo {volume} {D85}},\ \bibinfo {pages} {075026}
  (\bibinfo {year} {2012})},\ \Eprint {http://arxiv.org/abs/1112.1059}
  {arXiv:1112.1059 [hep-ph]} \BibitemShut {NoStop}%
\bibitem [{\citenamefont {Giudice}\ and\ \citenamefont
  {Strumia}(2012)}]{Giudice:2011cg}%
  \BibitemOpen
  \bibfield  {author} {\bibinfo {author} {\bibfnamefont {G.~F.}\ \bibnamefont
  {Giudice}}\ and\ \bibinfo {author} {\bibfnamefont {A.}~\bibnamefont
  {Strumia}},\ }\href {\doibase 10.1016/j.nuclphysb.2012.01.001} {\bibfield
  {journal} {\bibinfo  {journal} {Nucl. Phys.}\ }\textbf {\bibinfo {volume}
  {B858}},\ \bibinfo {pages} {63} (\bibinfo {year} {2012})},\ \Eprint
  {http://arxiv.org/abs/1108.6077} {arXiv:1108.6077 [hep-ph]} \BibitemShut
  {NoStop}%
\bibitem [{\citenamefont {Draper}\ \emph {et~al.}(2014)\citenamefont {Draper},
  \citenamefont {Lee},\ and\ \citenamefont {Wagner}}]{Draper:2013oza}%
  \BibitemOpen
  \bibfield  {author} {\bibinfo {author} {\bibfnamefont {P.}~\bibnamefont
  {Draper}}, \bibinfo {author} {\bibfnamefont {G.}~\bibnamefont {Lee}}, \ and\
  \bibinfo {author} {\bibfnamefont {C.~E.~M.}\ \bibnamefont {Wagner}},\ }\href
  {\doibase 10.1103/PhysRevD.89.055023} {\bibfield  {journal} {\bibinfo
  {journal} {Phys. Rev.}\ }\textbf {\bibinfo {volume} {D89}},\ \bibinfo {pages}
  {055023} (\bibinfo {year} {2014})},\ \Eprint {http://arxiv.org/abs/1312.5743}
  {arXiv:1312.5743 [hep-ph]} \BibitemShut {NoStop}%
\bibitem [{\citenamefont {Arvanitaki}\ \emph {et~al.}(2013)\citenamefont
  {Arvanitaki}, \citenamefont {Craig}, \citenamefont {Dimopoulos},\ and\
  \citenamefont {Villadoro}}]{Arvanitaki:2012ps}%
  \BibitemOpen
  \bibfield  {author} {\bibinfo {author} {\bibfnamefont {A.}~\bibnamefont
  {Arvanitaki}}, \bibinfo {author} {\bibfnamefont {N.}~\bibnamefont {Craig}},
  \bibinfo {author} {\bibfnamefont {S.}~\bibnamefont {Dimopoulos}}, \ and\
  \bibinfo {author} {\bibfnamefont {G.}~\bibnamefont {Villadoro}},\ }\href
  {\doibase 10.1007/JHEP02(2013)126} {\bibfield  {journal} {\bibinfo  {journal}
  {JHEP}\ }\textbf {\bibinfo {volume} {02}},\ \bibinfo {pages} {126} (\bibinfo
  {year} {2013})},\ \Eprint {http://arxiv.org/abs/1210.0555} {arXiv:1210.0555
  [hep-ph]} \BibitemShut {NoStop}%
\bibitem [{\citenamefont {Wells}(2003)}]{Wells:2003tf}%
  \BibitemOpen
  \bibfield  {author} {\bibinfo {author} {\bibfnamefont {J.~D.}\ \bibnamefont
  {Wells}},\ }in\ \href@noop {} {\emph {\bibinfo {booktitle} {{11th
  International Conference on Supersymmetry and the Unification of Fundamental
  Interactions (SUSY 2003) Tucson, Arizona, June 5-10, 2003}}}}\ (\bibinfo
  {year} {2003})\ \Eprint {http://arxiv.org/abs/hep-ph/0306127}
  {arXiv:hep-ph/0306127 [hep-ph]} \BibitemShut {NoStop}%
\bibitem [{\citenamefont {Pierce}(2004)}]{Pierce:2004mk}%
  \BibitemOpen
  \bibfield  {author} {\bibinfo {author} {\bibfnamefont {A.}~\bibnamefont
  {Pierce}},\ }\href {\doibase 10.1103/PhysRevD.70.075006} {\bibfield
  {journal} {\bibinfo  {journal} {Phys. Rev.}\ }\textbf {\bibinfo {volume}
  {D70}},\ \bibinfo {pages} {075006} (\bibinfo {year} {2004})},\ \Eprint
  {http://arxiv.org/abs/hep-ph/0406144} {arXiv:hep-ph/0406144 [hep-ph]}
  \BibitemShut {NoStop}%
\bibitem [{\citenamefont {Giudice}\ and\ \citenamefont
  {Romanino}(2004)}]{Giudice:2004tc}%
  \BibitemOpen
  \bibfield  {author} {\bibinfo {author} {\bibfnamefont {G.~F.}\ \bibnamefont
  {Giudice}}\ and\ \bibinfo {author} {\bibfnamefont {A.}~\bibnamefont
  {Romanino}},\ }\href {\doibase 10.1016/j.nuclphysb.2004.11.048} {\bibfield
  {journal} {\bibinfo  {journal} {Nucl. Phys.}\ }\textbf {\bibinfo {volume}
  {B699}},\ \bibinfo {pages} {65} (\bibinfo {year} {2004})},\ \bibinfo {note}
  {[Erratum: Nucl. Phys.B706,65(2005)]},\ \Eprint
  {http://arxiv.org/abs/hep-ph/0406088} {arXiv:hep-ph/0406088 [hep-ph]}
  \BibitemShut {NoStop}%
\bibitem [{\citenamefont {Bertone}\ \emph {et~al.}(2013)\citenamefont {Bertone}
  \emph {et~al.}}]{Bertone:2012cu}%
  \BibitemOpen
  \bibfield  {author} {\bibinfo {author} {\bibfnamefont {V.}~\bibnamefont
  {Bertone}} \emph {et~al.} (\bibinfo {collaboration} {ETM}),\ }\href {\doibase
  10.1007/JHEP07(2013)143, 10.1007/JHEP03(2013)089} {\bibfield  {journal}
  {\bibinfo  {journal} {JHEP}\ }\textbf {\bibinfo {volume} {03}},\ \bibinfo
  {pages} {089} (\bibinfo {year} {2013})},\ \bibinfo {note} {[Erratum:
  JHEP07,143(2013)]},\ \Eprint {http://arxiv.org/abs/1207.1287}
  {arXiv:1207.1287 [hep-lat]} \BibitemShut {NoStop}%
\bibitem [{\citenamefont {Bazavov}\ \emph {et~al.}(2010)\citenamefont {Bazavov}
  \emph {et~al.}}]{Bazavov:2010hj}%
  \BibitemOpen
  \bibfield  {author} {\bibinfo {author} {\bibfnamefont {A.}~\bibnamefont
  {Bazavov}} \emph {et~al.} (\bibinfo {collaboration} {MILC}),\ }\bibfield
  {booktitle} {\emph {\bibinfo {booktitle} {{Proceedings, 28th International
  Symposium on Lattice field theory (Lattice 2010)}}},\ }\href@noop {}
  {\bibfield  {journal} {\bibinfo  {journal} {PoS}\ }\textbf {\bibinfo {volume}
  {LATTICE2010}},\ \bibinfo {pages} {074} (\bibinfo {year} {2010})},\ \Eprint
  {http://arxiv.org/abs/1012.0868} {arXiv:1012.0868 [hep-lat]} \BibitemShut
  {NoStop}%
\bibitem [{\citenamefont {Carrasco}\ \emph {et~al.}(2014)\citenamefont
  {Carrasco} \emph {et~al.}}]{Carrasco:2013zta}%
  \BibitemOpen
  \bibfield  {author} {\bibinfo {author} {\bibfnamefont {N.}~\bibnamefont
  {Carrasco}} \emph {et~al.} (\bibinfo {collaboration} {ETM}),\ }\href
  {\doibase 10.1007/JHEP03(2014)016} {\bibfield  {journal} {\bibinfo  {journal}
  {JHEP}\ }\textbf {\bibinfo {volume} {03}},\ \bibinfo {pages} {016} (\bibinfo
  {year} {2014})},\ \Eprint {http://arxiv.org/abs/1308.1851} {arXiv:1308.1851
  [hep-lat]} \BibitemShut {NoStop}%
\bibitem [{\citenamefont {Bagger}\ \emph {et~al.}(1997)\citenamefont {Bagger},
  \citenamefont {Matchev},\ and\ \citenamefont {Zhang}}]{Bagger:1997gg}%
  \BibitemOpen
  \bibfield  {author} {\bibinfo {author} {\bibfnamefont {J.~A.}\ \bibnamefont
  {Bagger}}, \bibinfo {author} {\bibfnamefont {K.~T.}\ \bibnamefont {Matchev}},
  \ and\ \bibinfo {author} {\bibfnamefont {R.-J.}\ \bibnamefont {Zhang}},\
  }\href {\doibase 10.1016/S0370-2693(97)00920-9} {\bibfield  {journal}
  {\bibinfo  {journal} {Phys. Lett.}\ }\textbf {\bibinfo {volume} {B412}},\
  \bibinfo {pages} {77} (\bibinfo {year} {1997})},\ \Eprint
  {http://arxiv.org/abs/hep-ph/9707225} {arXiv:hep-ph/9707225 [hep-ph]}
  \BibitemShut {NoStop}%
\bibitem [{\citenamefont {Aushev}\ \emph {et~al.}(2010)\citenamefont {Aushev}
  \emph {et~al.}}]{Aushev:2010bq}%
  \BibitemOpen
  \bibfield  {author} {\bibinfo {author} {\bibfnamefont {T.}~\bibnamefont
  {Aushev}} \emph {et~al.},\ }\href@noop {} {\  (\bibinfo {year} {2010})},\
  \Eprint {http://arxiv.org/abs/1002.5012} {arXiv:1002.5012 [hep-ex]}
  \BibitemShut {NoStop}%
\bibitem [{\citenamefont {Barbieri}\ and\ \citenamefont
  {Giudice}(1993)}]{Barbieri:1993av}%
  \BibitemOpen
  \bibfield  {author} {\bibinfo {author} {\bibfnamefont {R.}~\bibnamefont
  {Barbieri}}\ and\ \bibinfo {author} {\bibfnamefont {G.~F.}\ \bibnamefont
  {Giudice}},\ }\href {\doibase 10.1016/0370-2693(93)91508-K} {\bibfield
  {journal} {\bibinfo  {journal} {Phys. Lett.}\ }\textbf {\bibinfo {volume}
  {B309}},\ \bibinfo {pages} {86} (\bibinfo {year} {1993})},\ \Eprint
  {http://arxiv.org/abs/hep-ph/9303270} {arXiv:hep-ph/9303270 [hep-ph]}
  \BibitemShut {NoStop}%
\bibitem [{\citenamefont {Degrassi}\ \emph {et~al.}(2000)\citenamefont
  {Degrassi}, \citenamefont {Gambino},\ and\ \citenamefont
  {Giudice}}]{Degrassi:2000qf}%
  \BibitemOpen
  \bibfield  {author} {\bibinfo {author} {\bibfnamefont {G.}~\bibnamefont
  {Degrassi}}, \bibinfo {author} {\bibfnamefont {P.}~\bibnamefont {Gambino}}, \
  and\ \bibinfo {author} {\bibfnamefont {G.~F.}\ \bibnamefont {Giudice}},\
  }\href {\doibase 10.1088/1126-6708/2000/12/009} {\bibfield  {journal}
  {\bibinfo  {journal} {JHEP}\ }\textbf {\bibinfo {volume} {12}},\ \bibinfo
  {pages} {009} (\bibinfo {year} {2000})},\ \Eprint
  {http://arxiv.org/abs/hep-ph/0009337} {arXiv:hep-ph/0009337 [hep-ph]}
  \BibitemShut {NoStop}%
\bibitem [{\citenamefont {Carena}\ \emph {et~al.}(2001)\citenamefont {Carena},
  \citenamefont {Garcia}, \citenamefont {Nierste},\ and\ \citenamefont
  {Wagner}}]{Carena:2000uj}%
  \BibitemOpen
  \bibfield  {author} {\bibinfo {author} {\bibfnamefont {M.}~\bibnamefont
  {Carena}}, \bibinfo {author} {\bibfnamefont {D.}~\bibnamefont {Garcia}},
  \bibinfo {author} {\bibfnamefont {U.}~\bibnamefont {Nierste}}, \ and\
  \bibinfo {author} {\bibfnamefont {C.~E.~M.}\ \bibnamefont {Wagner}},\ }\href
  {\doibase 10.1016/S0370-2693(01)00009-0} {\bibfield  {journal} {\bibinfo
  {journal} {Phys. Lett.}\ }\textbf {\bibinfo {volume} {B499}},\ \bibinfo
  {pages} {141} (\bibinfo {year} {2001})},\ \Eprint
  {http://arxiv.org/abs/hep-ph/0010003} {arXiv:hep-ph/0010003 [hep-ph]}
  \BibitemShut {NoStop}%
\bibitem [{\citenamefont {Freitas}\ and\ \citenamefont
  {Haisch}(2008)}]{Freitas:2008vh}%
  \BibitemOpen
  \bibfield  {author} {\bibinfo {author} {\bibfnamefont {A.}~\bibnamefont
  {Freitas}}\ and\ \bibinfo {author} {\bibfnamefont {U.}~\bibnamefont
  {Haisch}},\ }\href {\doibase 10.1103/PhysRevD.77.093008} {\bibfield
  {journal} {\bibinfo  {journal} {Phys. Rev.}\ }\textbf {\bibinfo {volume}
  {D77}},\ \bibinfo {pages} {093008} (\bibinfo {year} {2008})},\ \Eprint
  {http://arxiv.org/abs/0801.4346} {arXiv:0801.4346 [hep-ph]} \BibitemShut
  {NoStop}%
\bibitem [{\citenamefont {Altmannshofer}\ \emph
  {et~al.}(2013{\natexlab{b}})\citenamefont {Altmannshofer}, \citenamefont
  {Carena}, \citenamefont {Shah},\ and\ \citenamefont
  {Yu}}]{Altmannshofer:2012ks}%
  \BibitemOpen
  \bibfield  {author} {\bibinfo {author} {\bibfnamefont {W.}~\bibnamefont
  {Altmannshofer}}, \bibinfo {author} {\bibfnamefont {M.}~\bibnamefont
  {Carena}}, \bibinfo {author} {\bibfnamefont {N.~R.}\ \bibnamefont {Shah}}, \
  and\ \bibinfo {author} {\bibfnamefont {F.}~\bibnamefont {Yu}},\ }\href
  {\doibase 10.1007/JHEP01(2013)160} {\bibfield  {journal} {\bibinfo  {journal}
  {JHEP}\ }\textbf {\bibinfo {volume} {01}},\ \bibinfo {pages} {160} (\bibinfo
  {year} {2013}{\natexlab{b}})},\ \Eprint {http://arxiv.org/abs/1211.1976}
  {arXiv:1211.1976 [hep-ph]} \BibitemShut {NoStop}%
\bibitem [{\citenamefont {Czakon}\ \emph {et~al.}(2015)\citenamefont {Czakon},
  \citenamefont {Fiedler}, \citenamefont {Huber}, \citenamefont {Misiak},
  \citenamefont {Schutzmeier},\ and\ \citenamefont
  {Steinhauser}}]{Czakon:2015exa}%
  \BibitemOpen
  \bibfield  {author} {\bibinfo {author} {\bibfnamefont {M.}~\bibnamefont
  {Czakon}}, \bibinfo {author} {\bibfnamefont {P.}~\bibnamefont {Fiedler}},
  \bibinfo {author} {\bibfnamefont {T.}~\bibnamefont {Huber}}, \bibinfo
  {author} {\bibfnamefont {M.}~\bibnamefont {Misiak}}, \bibinfo {author}
  {\bibfnamefont {T.}~\bibnamefont {Schutzmeier}}, \ and\ \bibinfo {author}
  {\bibfnamefont {M.}~\bibnamefont {Steinhauser}},\ }\href {\doibase
  10.1007/JHEP04(2015)168} {\bibfield  {journal} {\bibinfo  {journal} {JHEP}\
  }\textbf {\bibinfo {volume} {04}},\ \bibinfo {pages} {168} (\bibinfo {year}
  {2015})},\ \Eprint {http://arxiv.org/abs/1503.01791} {arXiv:1503.01791
  [hep-ph]} \BibitemShut {NoStop}%
\bibitem [{\citenamefont {Misiak}\ \emph {et~al.}(2015)\citenamefont {Misiak}
  \emph {et~al.}}]{Misiak:2015xwa}%
  \BibitemOpen
  \bibfield  {author} {\bibinfo {author} {\bibfnamefont {M.}~\bibnamefont
  {Misiak}} \emph {et~al.},\ }\href {\doibase 10.1103/PhysRevLett.114.221801}
  {\bibfield  {journal} {\bibinfo  {journal} {Phys. Rev. Lett.}\ }\textbf
  {\bibinfo {volume} {114}},\ \bibinfo {pages} {221801} (\bibinfo {year}
  {2015})},\ \Eprint {http://arxiv.org/abs/1503.01789} {arXiv:1503.01789
  [hep-ph]} \BibitemShut {NoStop}%
\bibitem [{\citenamefont {Agashe}\ \emph {et~al.}(2013)\citenamefont {Agashe}
  \emph {et~al.}}]{Agashe:2013hma}%
  \BibitemOpen
  \bibfield  {author} {\bibinfo {author} {\bibfnamefont {K.}~\bibnamefont
  {Agashe}} \emph {et~al.} (\bibinfo {collaboration} {Top Quark Working
  Group}),\ }in\ \href
  {http://inspirehep.net/record/1263763/files/arXiv:1311.2028.pdf} {\emph
  {\bibinfo {booktitle} {{Community Summer Study 2013: Snowmass on the
  Mississippi (CSS2013) Minneapolis, MN, USA, July 29-August 6, 2013}}}}\
  (\bibinfo {year} {2013})\ \Eprint {http://arxiv.org/abs/1311.2028}
  {arXiv:1311.2028 [hep-ph]} \BibitemShut {NoStop}%
\bibitem [{ATL(2013)}]{ATL-PHYS-PUB-2013-012}%
  \BibitemOpen
  \href {http://cds.cern.ch/record/1604506} {\emph {\bibinfo {title}
  {{Sensitivity of ATLAS at HL-LHC to flavour changing neutral currents in top
  quark decays t → cH, with H → γγ}}}},\ \bibinfo {type} {Tech. Rep.}\
  \bibinfo {number} {ATL-PHYS-PUB-2013-012}\ (\bibinfo  {institution} {CERN},\
  \bibinfo {address} {Geneva},\ \bibinfo {year} {2013})\BibitemShut {NoStop}%
\bibitem [{\citenamefont {Dedes}\ \emph {et~al.}(2014)\citenamefont {Dedes},
  \citenamefont {Paraskevas}, \citenamefont {Rosiek}, \citenamefont {Suxho},\
  and\ \citenamefont {Tamvakis}}]{Dedes:2014asa}%
  \BibitemOpen
  \bibfield  {author} {\bibinfo {author} {\bibfnamefont {A.}~\bibnamefont
  {Dedes}}, \bibinfo {author} {\bibfnamefont {M.}~\bibnamefont {Paraskevas}},
  \bibinfo {author} {\bibfnamefont {J.}~\bibnamefont {Rosiek}}, \bibinfo
  {author} {\bibfnamefont {K.}~\bibnamefont {Suxho}}, \ and\ \bibinfo {author}
  {\bibfnamefont {K.}~\bibnamefont {Tamvakis}},\ }\href {\doibase
  10.1007/JHEP11(2014)137} {\bibfield  {journal} {\bibinfo  {journal} {JHEP}\
  }\textbf {\bibinfo {volume} {11}},\ \bibinfo {pages} {137} (\bibinfo {year}
  {2014})},\ \Eprint {http://arxiv.org/abs/1409.6546} {arXiv:1409.6546
  [hep-ph]} \BibitemShut {NoStop}%
\bibitem [{\citenamefont {Arganda}\ \emph {et~al.}(2016)\citenamefont
  {Arganda}, \citenamefont {Herrero}, \citenamefont {Morales},\ and\
  \citenamefont {Szynkman}}]{Arganda:2015uca}%
  \BibitemOpen
  \bibfield  {author} {\bibinfo {author} {\bibfnamefont {E.}~\bibnamefont
  {Arganda}}, \bibinfo {author} {\bibfnamefont {M.~J.}\ \bibnamefont
  {Herrero}}, \bibinfo {author} {\bibfnamefont {R.}~\bibnamefont {Morales}}, \
  and\ \bibinfo {author} {\bibfnamefont {A.}~\bibnamefont {Szynkman}},\ }\href
  {\doibase 10.1007/JHEP03(2016)055} {\bibfield  {journal} {\bibinfo  {journal}
  {JHEP}\ }\textbf {\bibinfo {volume} {03}},\ \bibinfo {pages} {055} (\bibinfo
  {year} {2016})},\ \Eprint {http://arxiv.org/abs/1510.04685} {arXiv:1510.04685
  [hep-ph]} \BibitemShut {NoStop}%
\bibitem [{\citenamefont {Aloni}\ \emph {et~al.}(2016)\citenamefont {Aloni},
  \citenamefont {Nir},\ and\ \citenamefont {Stamou}}]{Aloni:2015wvn}%
  \BibitemOpen
  \bibfield  {author} {\bibinfo {author} {\bibfnamefont {D.}~\bibnamefont
  {Aloni}}, \bibinfo {author} {\bibfnamefont {Y.}~\bibnamefont {Nir}}, \ and\
  \bibinfo {author} {\bibfnamefont {E.}~\bibnamefont {Stamou}},\ }\href
  {\doibase 10.1007/JHEP04(2016)162} {\bibfield  {journal} {\bibinfo  {journal}
  {JHEP}\ }\textbf {\bibinfo {volume} {04}},\ \bibinfo {pages} {162} (\bibinfo
  {year} {2016})},\ \Eprint {http://arxiv.org/abs/1511.00979} {arXiv:1511.00979
  [hep-ph]} \BibitemShut {NoStop}%
\bibitem [{\citenamefont {Paradisi}(2005)}]{Paradisi:2005fk}%
  \BibitemOpen
  \bibfield  {author} {\bibinfo {author} {\bibfnamefont {P.}~\bibnamefont
  {Paradisi}},\ }\href {\doibase 10.1088/1126-6708/2005/10/006} {\bibfield
  {journal} {\bibinfo  {journal} {JHEP}\ }\textbf {\bibinfo {volume} {10}},\
  \bibinfo {pages} {006} (\bibinfo {year} {2005})},\ \Eprint
  {http://arxiv.org/abs/hep-ph/0505046} {arXiv:hep-ph/0505046 [hep-ph]}
  \BibitemShut {NoStop}%
\bibitem [{\citenamefont {Kitano}\ \emph {et~al.}(2002)\citenamefont {Kitano},
  \citenamefont {Koike},\ and\ \citenamefont {Okada}}]{Kitano:2002mt}%
  \BibitemOpen
  \bibfield  {author} {\bibinfo {author} {\bibfnamefont {R.}~\bibnamefont
  {Kitano}}, \bibinfo {author} {\bibfnamefont {M.}~\bibnamefont {Koike}}, \
  and\ \bibinfo {author} {\bibfnamefont {Y.}~\bibnamefont {Okada}},\ }\href
  {\doibase 10.1103/PhysRevD.76.059902, 10.1103/PhysRevD.66.096002} {\bibfield
  {journal} {\bibinfo  {journal} {Phys. Rev.}\ }\textbf {\bibinfo {volume}
  {D66}},\ \bibinfo {pages} {096002} (\bibinfo {year} {2002})},\ \bibinfo
  {note} {[Erratum: Phys. Rev.D76,059902(2007)]},\ \Eprint
  {http://arxiv.org/abs/hep-ph/0203110} {arXiv:hep-ph/0203110 [hep-ph]}
  \BibitemShut {NoStop}%
\bibitem [{\citenamefont {Arkani-Hamed}\ and\ \citenamefont
  {Dimopoulos}(2005)}]{ArkaniHamed:2004fb}%
  \BibitemOpen
  \bibfield  {author} {\bibinfo {author} {\bibfnamefont {N.}~\bibnamefont
  {Arkani-Hamed}}\ and\ \bibinfo {author} {\bibfnamefont {S.}~\bibnamefont
  {Dimopoulos}},\ }\href {\doibase 10.1088/1126-6708/2005/06/073} {\bibfield
  {journal} {\bibinfo  {journal} {JHEP}\ }\textbf {\bibinfo {volume} {06}},\
  \bibinfo {pages} {073} (\bibinfo {year} {2005})},\ \Eprint
  {http://arxiv.org/abs/hep-th/0405159} {arXiv:hep-th/0405159 [hep-th]}
  \BibitemShut {NoStop}%
\bibitem [{\citenamefont {Arkani-Hamed}\ \emph {et~al.}(2005)\citenamefont
  {Arkani-Hamed}, \citenamefont {Dimopoulos}, \citenamefont {Giudice},\ and\
  \citenamefont {Romanino}}]{ArkaniHamed:2004yi}%
  \BibitemOpen
  \bibfield  {author} {\bibinfo {author} {\bibfnamefont {N.}~\bibnamefont
  {Arkani-Hamed}}, \bibinfo {author} {\bibfnamefont {S.}~\bibnamefont
  {Dimopoulos}}, \bibinfo {author} {\bibfnamefont {G.~F.}\ \bibnamefont
  {Giudice}}, \ and\ \bibinfo {author} {\bibfnamefont {A.}~\bibnamefont
  {Romanino}},\ }\href {\doibase 10.1016/j.nuclphysb.2004.12.026} {\bibfield
  {journal} {\bibinfo  {journal} {Nucl. Phys.}\ }\textbf {\bibinfo {volume}
  {B709}},\ \bibinfo {pages} {3} (\bibinfo {year} {2005})},\ \Eprint
  {http://arxiv.org/abs/hep-ph/0409232} {arXiv:hep-ph/0409232 [hep-ph]}
  \BibitemShut {NoStop}%
\bibitem [{\citenamefont {Wells}(2005)}]{Wells:2004di}%
  \BibitemOpen
  \bibfield  {author} {\bibinfo {author} {\bibfnamefont {J.~D.}\ \bibnamefont
  {Wells}},\ }\href {\doibase 10.1103/PhysRevD.71.015013} {\bibfield  {journal}
  {\bibinfo  {journal} {Phys. Rev.}\ }\textbf {\bibinfo {volume} {D71}},\
  \bibinfo {pages} {015013} (\bibinfo {year} {2005})},\ \Eprint
  {http://arxiv.org/abs/hep-ph/0411041} {arXiv:hep-ph/0411041 [hep-ph]}
  \BibitemShut {NoStop}%
\bibitem [{\citenamefont {Acharya}\ \emph {et~al.}(2007)\citenamefont
  {Acharya}, \citenamefont {Bobkov}, \citenamefont {Kane}, \citenamefont
  {Kumar},\ and\ \citenamefont {Shao}}]{Acharya:2007rc}%
  \BibitemOpen
  \bibfield  {author} {\bibinfo {author} {\bibfnamefont {B.~S.}\ \bibnamefont
  {Acharya}}, \bibinfo {author} {\bibfnamefont {K.}~\bibnamefont {Bobkov}},
  \bibinfo {author} {\bibfnamefont {G.~L.}\ \bibnamefont {Kane}}, \bibinfo
  {author} {\bibfnamefont {P.}~\bibnamefont {Kumar}}, \ and\ \bibinfo {author}
  {\bibfnamefont {J.}~\bibnamefont {Shao}},\ }\href {\doibase
  10.1103/PhysRevD.76.126010} {\bibfield  {journal} {\bibinfo  {journal} {Phys.
  Rev.}\ }\textbf {\bibinfo {volume} {D76}},\ \bibinfo {pages} {126010}
  (\bibinfo {year} {2007})},\ \Eprint {http://arxiv.org/abs/hep-th/0701034}
  {arXiv:hep-th/0701034 [hep-th]} \BibitemShut {NoStop}%
\bibitem [{\citenamefont {Acharya}\ \emph {et~al.}(2008)\citenamefont
  {Acharya}, \citenamefont {Bobkov}, \citenamefont {Kane}, \citenamefont
  {Shao},\ and\ \citenamefont {Kumar}}]{Acharya:2008zi}%
  \BibitemOpen
  \bibfield  {author} {\bibinfo {author} {\bibfnamefont {B.~S.}\ \bibnamefont
  {Acharya}}, \bibinfo {author} {\bibfnamefont {K.}~\bibnamefont {Bobkov}},
  \bibinfo {author} {\bibfnamefont {G.~L.}\ \bibnamefont {Kane}}, \bibinfo
  {author} {\bibfnamefont {J.}~\bibnamefont {Shao}}, \ and\ \bibinfo {author}
  {\bibfnamefont {P.}~\bibnamefont {Kumar}},\ }\href {\doibase
  10.1103/PhysRevD.78.065038} {\bibfield  {journal} {\bibinfo  {journal} {Phys.
  Rev.}\ }\textbf {\bibinfo {volume} {D78}},\ \bibinfo {pages} {065038}
  (\bibinfo {year} {2008})},\ \Eprint {http://arxiv.org/abs/0801.0478}
  {arXiv:0801.0478 [hep-ph]} \BibitemShut {NoStop}%
\bibitem [{\citenamefont {Moroi}\ and\ \citenamefont
  {Nagai}(2013)}]{Moroi:2013sfa}%
  \BibitemOpen
  \bibfield  {author} {\bibinfo {author} {\bibfnamefont {T.}~\bibnamefont
  {Moroi}}\ and\ \bibinfo {author} {\bibfnamefont {M.}~\bibnamefont {Nagai}},\
  }\href {\doibase 10.1016/j.physletb.2013.04.049} {\bibfield  {journal}
  {\bibinfo  {journal} {Phys. Lett.}\ }\textbf {\bibinfo {volume} {B723}},\
  \bibinfo {pages} {107} (\bibinfo {year} {2013})},\ \Eprint
  {http://arxiv.org/abs/1303.0668} {arXiv:1303.0668 [hep-ph]} \BibitemShut
  {NoStop}%
\bibitem [{\citenamefont {Ellis}\ and\ \citenamefont
  {Kane}(2015)}]{Ellis:2015dra}%
  \BibitemOpen
  \bibfield  {author} {\bibinfo {author} {\bibfnamefont {S.~A.~R.}\
  \bibnamefont {Ellis}}\ and\ \bibinfo {author} {\bibfnamefont {G.~L.}\
  \bibnamefont {Kane}},\ }\href@noop {} {\  (\bibinfo {year} {2015})},\ \Eprint
  {http://arxiv.org/abs/1505.04191} {arXiv:1505.04191 [hep-ph]} \BibitemShut
  {NoStop}%
\end{thebibliography}%

\end{document}